\documentclass[a4paper,11pt]{article}
\pdfoutput=1 

\usepackage{jinstpub} 

\usepackage{lineno}

\title{The BiPo-3 detector for the measurement of ultra low natural radioactivities of thin materials}

\collaboration[]{The SuperNEMO collaboration}

\author[a]{A.S.~Barabash,}
\author[b,c]{A.~Basharina-Freshville,} 
\author[c]{E.~Birdsall,}
\author[d]{S.~Blondel,}
\author[c]{S.~Blot,}
\author[d]{M.~Bongrand,}
\author[d]{D.~Boursette,}
\author[e,v]{V.~Brudanin,}
\author[f]{J.~Busto,}
\author[g]{A.J.~Caffrey,}
\author[d]{S.~Calvez,}
\author[b]{M.~Cascella,}
\author[w]{S.~Cebri\'an,}
\author[h]{C.~Cerna,}
\author[m]{J.P ~Cesar,}
\author[h]{E.~Chauveau,} 
\author[b]{A.~Chopra,}
\author[w]{T.~Dafn\'i,}
\author[c]{S.~De~Capua,}
\author[i]{D.~Duchesneau,}
\author[k]{D.~Durand,}
\author[e]{V.~Egorov,}
\author[d,b]{G.~Eurin,} 
\author[c]{J.J.~Evans,} 
\author[j]{L.~Fajt,} 
\author[e]{D.~Filosofov,}
\author[b]{R.~Flack,} 
\author[d]{X.~Garrido,}
\author[d,1]{H.~G\'omez\note{Corresponding author},}
\author[k]{B.~Guillon,}
\author[c]{P.~Guzowski,}
\author[l]{K.~Hol\'{y},} 
\author[j]{R.~Hod\'{a}k,}
\author[h]{A.~Huber,}
\author[h]{C.~Hugon,}
\author[w]{F.J.~Iguaz,}
\author[w]{I.G.~Irastorza,}
\author[i]{A.~Jeremie,}
\author[d]{S.~Jullian,}
\author[b]{M.~Kauer,} 
\author[e]{A.~Klimenko,}
\author[e]{O.~Kochetov,}
\author[a]{S.I.~Konovalov,} 
\author[e]{V.~Kovalenko,}
\author[m]{K.~Lang,} 
\author[k]{Y.~Lemi\`ere,}
\author[i]{T.~Le Noblet,}
\author[m]{Z.~Liptak,}
\author[b]{X.R.~Liu,} 
\author[d]{P.~Loaiza,}
\author[h]{G.~Lutter,}
\author[w]{G.~Luz\'on,}
\author[l]{M.~Macko,}
\author[j]{F.~Mamedov,}
\author[h]{C.~Marquet,} 
\author[k]{F.~Mauger,}
\author[n]{B.~Morgan,}
\author[b]{J.~Mott,} 
\author[e]{I.~Nemchenok,}
\author[o]{M.~Nomachi,}
\author[m]{F.~Nova,}
\author[p]{H.~Ohsumi,} 
\author[k]{G.~Olivi\'ero} 
\author[w]{A.~Ortiz de Sol\'orzano,} 
\author[m]{R.B.~Pahlka,}
\author[c]{J.~Pater,}
\author[h]{F.~Perrot,}
\author[h,q]{F.~Piquemal,} 
\author[l]{P.~Povinec,}
\author[j]{P.~P\v{r}idal,}
\author[n]{Y.A.~Ramachers,} 
\author[i]{A.~Remoto,}
\author[b]{B.~Richards,}
\author[g]{C.L.~Riddle,}
\author[j]{E.~Rukhadze,} 
\author[b]{R.~Saakyan,} 
\author[m]{R.~Salazar,} 
\author[d]{X.~Sarazin,} 
\author[e,r]{Yu.~Shitov,}
\author[d,s]{L.~Simard,}
\author[l]{F.~\v{S}imkovic,} 
\author[j]{A.~Smetana,}
\author[j]{K.~Smolek,} 
\author[e]{A.~Smolnikov,} 
\author[c]{S.~S\"oldner-Rembold,}
\author[h]{B.~Soul\'e,}
\author[j]{I.~\v{S}tekl,} 
\author[b]{J.~Thomas,} 
\author[e]{V.~Timkin,} 
\author[b]{S.~Torre,}
\author[t]{Vl.I.~Tretyak,} 
\author[e]{V.I.~Tretyak,}
\author[a]{V.I.~Umatov,}
\author[b]{C.~Vilela,} 
\author[u]{V.~Vorobel,}
\author[b]{D.~Waters,}
\author[u]{A.~\v{Z}ukauskas}

\affiliation[a]{NRC "Kurchatov Institute", ITEP, 117218 Moscow, Russia}
\affiliation[b]{University College London, London WC1E 6BT, United Kingdom}
\affiliation[c]{University of Manchester, Manchester M13 9PL, United Kingdom}
\affiliation[d]{LAL, Universit\'e Paris-Sud, CNRS/IN2P3, Universit\'e Paris-Saclay, F-91405 Orsay, France}
\affiliation[e]{JINR, 141980 Dubna, Russia}
\affiliation[f]{Aix Marseille Univ., CNRS, CPPM, Marseille, France}
\affiliation[g]{Idaho National Laboratory, Idaho Falls, ID 83415, United States}
\affiliation[h]{CENBG, Universit\'e de Bordeaux, CNRS/IN2P3, F-33175 Gradignan, France}
\affiliation[i]{LAPP, Universit\'e de Savoie, CNRS/IN2P3, F-74941 Annecy-le-Vieux, France}
\affiliation[j]{Institute of Experimental and Applied Physics, Czech Technical University in Prague, CZ-12800 Prague, Czech Republic}
\affiliation[k]{Normandie Univ., ENSICAEN, UNICAEN, CNRS/IN2P3, LPC Caen, \ F-14000 Caen, France}
\affiliation[l]{FMFI,~Comenius~University,~SK-842~48~Bratislava, Slovakia}
\affiliation[m]{University of Texas at Austin,  Austin, TX 78712, United States}
\affiliation[n]{University of Warwick, Coventry CV4 7AL, United Kingdom}
\affiliation[o]{Osaka University, 1-1 Machikaney arna Toyonaka, Osaka 560-0043, Japan}
\affiliation[p]{Saga University, Saga 840-8502, Japan}
\affiliation[q]{Laboratoire Souterrain de Modane, F-73500 Modane, France}
\affiliation[r]{Imperial College London, London SW7 2AZ, United Kingdom}
\affiliation[s]{Institut Universitaire de France, F-75005 Paris, France}
\affiliation[t]{Institute for Nuclear Research, MSP 03680 Kyiv, Ukraine}
\affiliation[u]{Charles University, Prague, Faculty of Mathematics and Physics, CZ-12116 Prague, Czech Republic}
\affiliation[v]{National Research Nuclear University MEPhI, 115409, Moscow, Russia}
\affiliation[w]{Grupo de F\'isica Nuclear y Astropart\'iculas, Departamento de  F\'isica Te\'orica, Universidad de Zaragoza, C/P Cerbuna 12 50009, Zaragoza, Spain}

\emailAdd{hgomez@apc.univ-paris7.fr}

\abstract{The BiPo-3 detector, running at the Canfranc Underground Laboratory (Laboratorio Subterr\'aneo de Canfranc, LSC, Spain) since 2013, is a low-radioactivity detector dedicated to measuring ultra low natural radionuclide contaminations of $^{208}$Tl ($^{232}$Th chain) and $^{214}$Bi ($^{238}$U chain) in thin materials. The total sensitive surface area of the detector is 3.6~m$^2$. The detector has been developed to measure the radiopurity of the selenium double $\beta$-decay source foils of the SuperNEMO experiment. In this paper the design and performance of the detector, and results of the background measurements in $^{208}$Tl and $^{214}$Bi, are presented, and the validation of the BiPo-3 measurement with a calibrated aluminium foil is discussed. Results of the $^{208}$Tl and $^{214}$Bi activity measurements of the first enriched $^{82}$Se foils of the double $\beta$-decay SuperNEMO experiment are reported. The sensitivity of the BiPo-3 detector for the measurement of the SuperNEMO $^{82}$Se foils is $\mathcal{A}$($^{208}$Tl)~$<2$~$\mu$Bq/kg (90\%~C.L.) and $\mathcal{A}$($^{214}$Bi)~$<140$~$\mu$Bq/kg (90\%~C.L.) after 6 months of measurement.}


\begin{document}
\maketitle
\flushbottom

\section{Introduction}
\label{sec:introduction}

The BiPo-3 detector, running at the Canfranc Underground Laboratory (Laboratorio Subterr\'aneo de Canfranc, LSC, Spain) since 2013, has been initially developed to measure ultra low natural radionuclide contaminations of $^{208}$Tl and $^{214}$Bi in the SuperNEMO double $\beta$-decay source foils. The goal of the SuperNEMO experiment is to search for the neutrinoless double $\beta$-decay ($\beta\beta 0\nu$)~\cite{supernemo} as an experimental proof of the non-conservation of the total lepton number (L), as postulated in the standard model (SM) of Particle Physics, and the Majorana nature of neutrino, i.e. the identity between neutrino and antineutrino. As baseline, SuperNEMO will measure 100~kg of $^{82}$Se double $\beta$-decay isotope, with a corresponding transition energy of Q$_{\beta\beta}$~=~2995~keV, with a sensitivity of $T_{1/2}(\beta\beta 0\nu)>10^{26}$~years after 5 year measurement. Other isotopes with higher Q$_{\beta\beta}$, as it is the case of $^{150}$Nd (Q$_{\beta\beta}$~=~3368~keV), are also considered. The isotopes are in powder form mixed with polyvinyl alcohol (PVA) glue and deposited between Mylar foils with a total surface area of 250~m$^2$. Possible contaminations of $^{208}$Tl and $^{214}$Bi, produced inside the double $\beta$-decay source foils by the natural radioactive decay chains of $^{232}$Th and $^{238}$U respectively, represent two of the main sources of background for SuperNEMO, since their corresponding endpoint energies, Q$_{\beta}$~=~4.99~MeV\footnote{This value as well as all the other $\beta$ endpoints and $\alpha$ energies have been extracted from \cite{nudat}} (3.27~MeV) for the case of $^{208}$Tl ($^{214}$Bi), are above the Q$_{\beta\beta}$ value of $^{82}$Se. The required radiopurities of the SuperNEMO double $\beta$-decay foils are $\mathcal{A}$($^{208}$Tl)~$<$~2~$\mathrm{\mu}$Bq/kg and $\mathcal{A}$($^{214}$Bi)~$<$~10~$\mathrm{\mu}$Bq/kg in order to achieve the desired sensitivity~\cite{supernemo}. This is more than one order of magnitude lower than the sensitivity that can be reached by non-destructive radiopurity measurement techniques, as $\gamma$-spectrometry with ultra low background High Purity Germanium (HPGe) detectors~\cite{nemo3-bkg}.

BiPo prototypes~\cite{bipo1} were developed first in order to demonstrate that a sensitivity of a few $\mathrm{\mu}$Bq/kg in $^{208}$Tl was reachable using the proposed BiPo experimental technique. In 2009, the SuperNEMO collaboration decided to build a larger BiPo detector named BiPo-3, with a total sensitive surface area of 3.6 m$^2$. The detector size was driven by the requirements of a relatively small detector and sufficient sensitivity to measure the radiopurity of the selenium double $\beta$-decay source foils planned for the SuperNEMO demonstrator. We show in this paper that the BiPo-3 detector is a generic low-radioactivity planar detector, which can measure the natural radioactivity in $^{208}$Tl and $^{214}$Bi of thin materials (surface density less than about 50~mg/cm$^2$) with an unprecedented sensitivity.

\section{The BiPo-3 detector}

\subsection{BiPo-3 measurement principle}
\label{sec:principle}

\begin{figure}[!]
  \centering
  \includegraphics[width=0.44\textwidth]{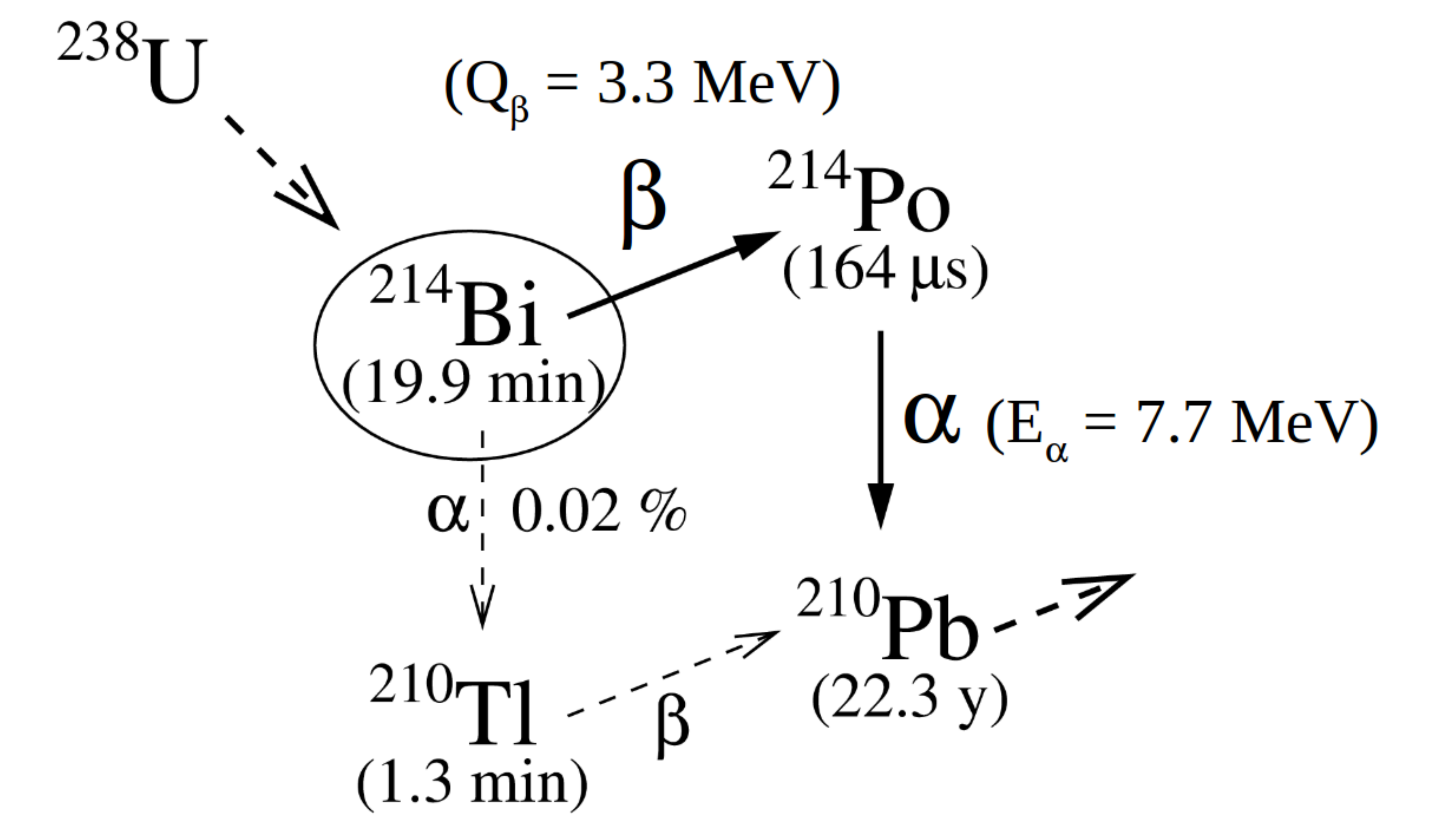}
  \hspace{1.5cm}
  \includegraphics[width=0.44\textwidth]{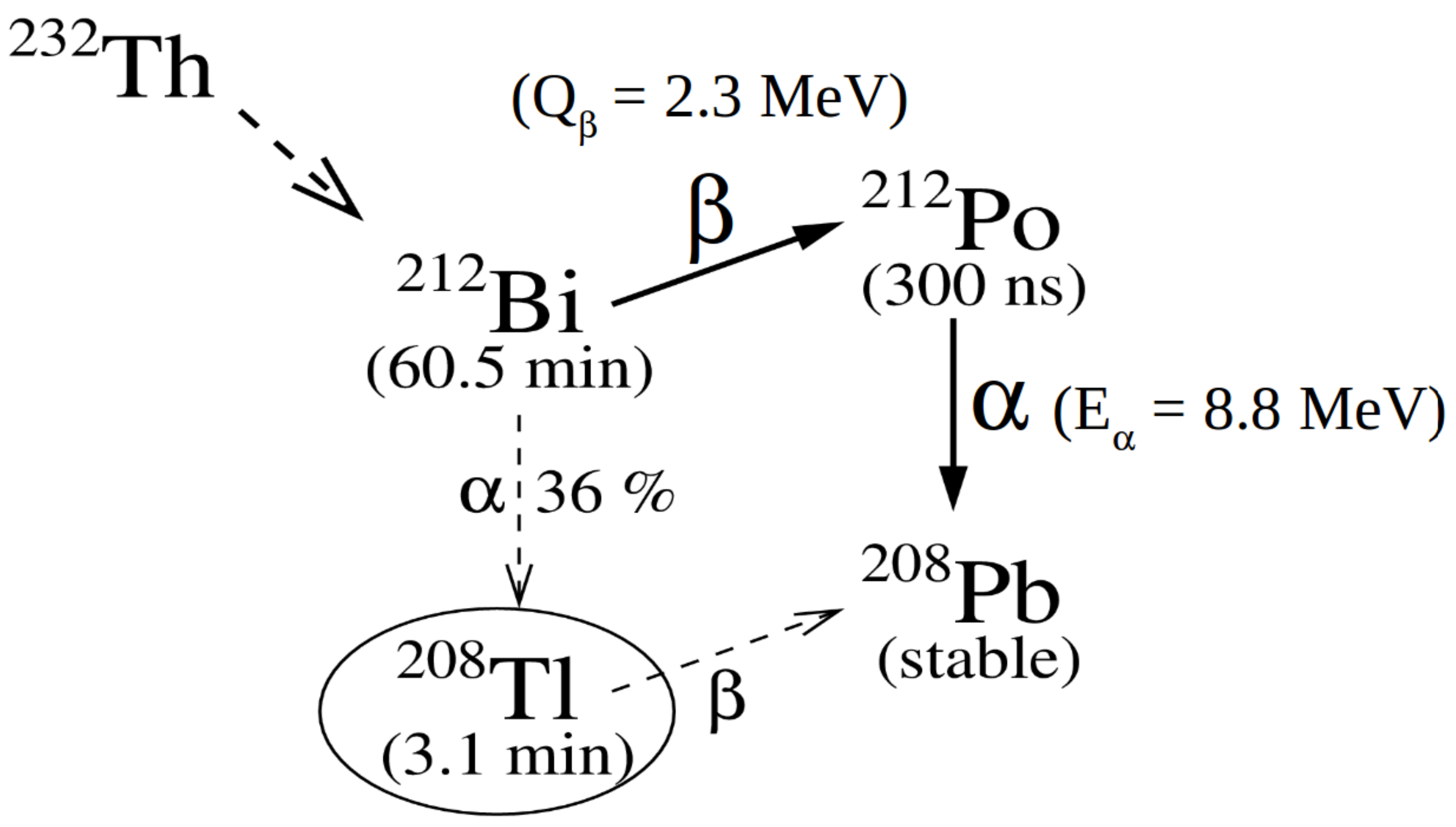}
  \caption{The $^{214}$Bi$\rightarrow^{214}$Po and $^{212}$Bi$\rightarrow^{212}$Po cascades used for the $^{214}$Bi and $^{208}$Tl measurements.}
  \label{fig:bipo-proc}
\end{figure}

In order to measure $^{208}$Tl and $^{214}$Bi contaminations, the underlying concept of the BiPo-3 detector is to observe the so-called BiPo process, which corresponds to the detection of an electron followed by a delayed $\alpha$ particle. The $^{214}$Bi isotope is a ($\beta$,$\gamma$) emitter (Q$_{\beta}$ = 3.27 MeV) decaying to $^{214}$Po, which is an $\alpha$ emitter ($E_{\alpha}$ = 7.69 MeV)  with a half-life of 164~$\mathrm{\mu}$s. The $^{208}$Tl isotope is measured by detecting its parent,  $^{212}$Bi. Here $^{212}$Bi decays with a branching ratio of 64\% via a $\beta$ emission (Q$_{\beta}$ = 2.25 MeV) to $^{212}$Po, which is a pure $\alpha$ emitter ($E_{\alpha}$ = 8.78 MeV) with a short half-life of 300 ns. Corresponding decay schemes are showed in Figure \ref{fig:bipo-proc}.

\begin{figure}[!]
  \centering
  \includegraphics[scale=0.7]{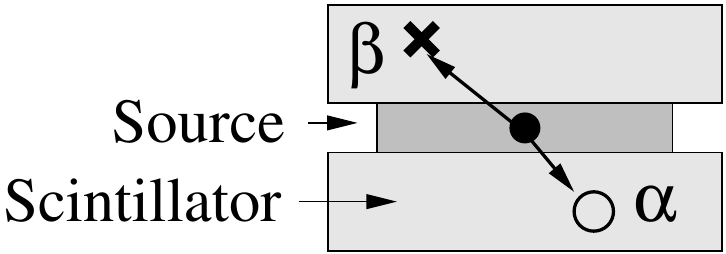}
  \hspace{1.5cm}
  \includegraphics[scale=0.4]{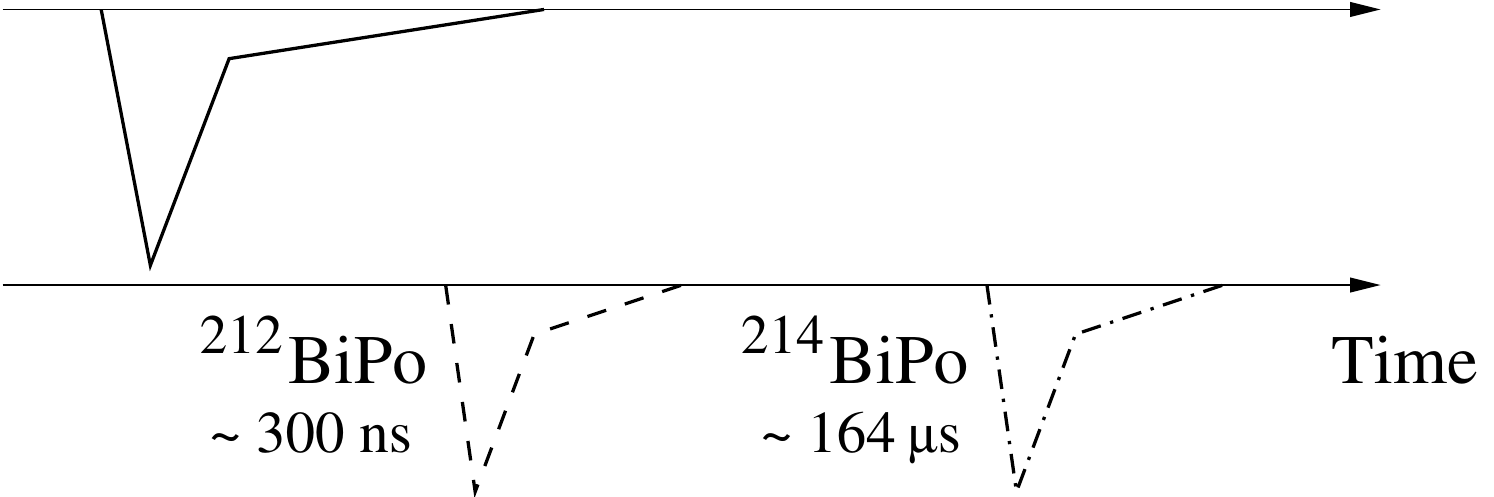}
  \caption{Schematic view of the BiPo detection technique with the source foil inserted between two plastic scintillator plates (left). The dot represents the contamination, the cross and open circle represent energy depositions in the scintillators by the prompt $\beta$ and the delayed $\alpha$ signal respectively. The prompt $\beta$ signal and the delayed $\alpha$ signal  observed by the top and bottom scintillators respectively are schematically illustrated (right). The timing of the delayed signal depends on the isotope to be measured.}
  \label{fig:bipo-event}
\end{figure}

The BiPo-3 experimental technique consists in installing the foil of interest between two thin ultra radiopure organic plastic scintillators, as illustrated in Figure \ref{fig:bipo-event}. The $^{212}$Bi ($^{208}$Tl) and $^{214}$Bi contaminations inside the foil are measured by detecting the $\beta$-decay as an energy deposition in one scintillator and no coincidence signal from the opposite side, and the delayed $\alpha$ particle as a delayed signal in the second opposite scintillator without a coincidence in the first one. Such a BiPo event is identified as a {\it back-to-back} event since the $\beta$ and $\alpha$ particles enter different scintillators on opposite sides of the foil. The timing of the delayed $\alpha$ particle depends on the isotope to be measured. The energy of the delayed $\alpha$ particle provides information on whether the contamination is on the surface or in the bulk of the foil. A second topology of BiPo events can in principle be used as well. This involves the {\it same-side} BiPo events for which the prompt $\beta$ signal and the delayed $\alpha$ signal are detected in the same scintillator without a coincidence signal in the scintillator on the opposite side. The detection of the {\it same-side} events would increase the BiPo-3 efficiency. However, the level of background is much higher than the one measured in the {\it back-to-back} topology because bulk contamination inside scintillators can mimic {\it same-side} events, reducing the sensitivity of the measurement. For this reason, only the {\it back-to-back} topology will be used for the rest of the analysis presented.

\subsection{Description of the BiPo-3 detector}
\label{sec:detector}

\begin{figure}[!]
  \centering
  \includegraphics[scale=0.5]{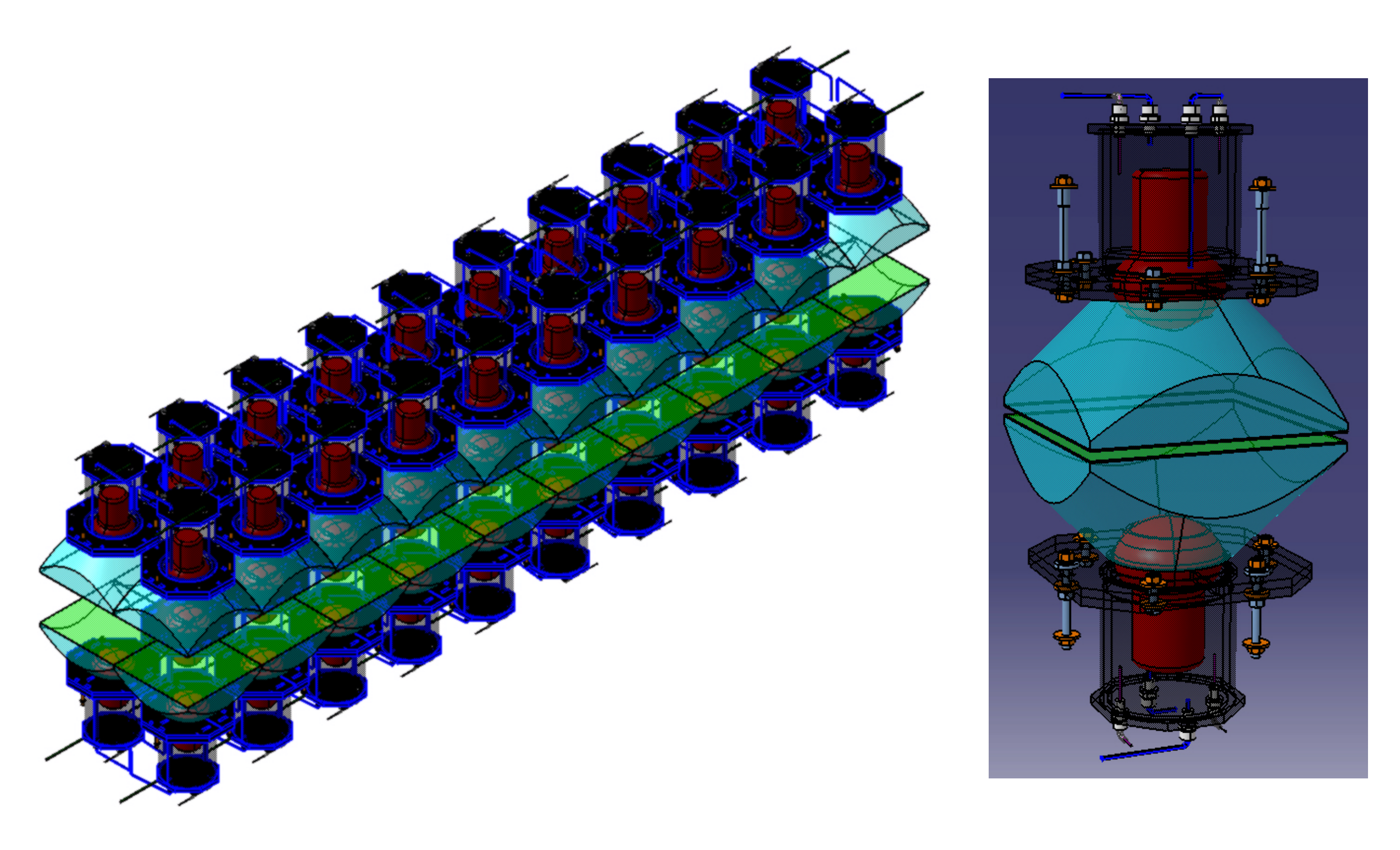}
  \caption{Details of the assembly of the 40 optical sub-modules inside a BiPo-3 module. On the right, a pair of optical sub-modules with the two thin scintillators (in green) face-to-face, coupled with a PMMA optical guide (blue) to a low-radioactivity 5-inch PMT (red).}
  \label{fig:bipo3-design}
\end{figure}

The BiPo-3 detector is composed of two modules. Each module (see Figure \ref{fig:bipo3-design}) consists of 20 pairs of optical sub-modules, positioned in two rows. Each optical sub-module consists of a polystyrene-based scintillator plate coupled to a 5-inch low-radioactivity photomultiplier tube (PMT, model Hamamatsu R6594-MOD) through a poly-methyl methacrylate (PMMA) optical guide. The sub-modules are arranged face-to-face to form a pair. The size of each scintillator is 300$\times$300$\times$2 mm$^3$.  The BiPo-3 detector corresponds to a total of 80 optical sub-modules and a total detector surface of 3.6 m$^2$.

The scintillators are prepared with a mono-diamond tool from raw and radiopure polystyrene scintillator blocks produced by the Joint Institute for Nuclear Research (JINR, Russia). The surface of the scintillators facing the source foil is covered with a 200 nm thick layer of evaporated ultra radiopure aluminium in order to isolate optically each scintillator from its neighbour, and to improve the light collection efficiency. The radiopurity of the aluminium has been measured by HPGe spectrometry before evaporation, obtaining an upper limit of $\mathcal{A}(^{208}\mathrm{Tl})<0.2$ mBq/kg. The entrance surfaces of the scintillators are carefully cleaned before and after the aluminium deposition using the following cleaning sequence: acetic acid, ultra pure water, isopropanol, and finally a second ultra pure water rinse. The light guide is first wrapped in a Tyvek layer in order to diffuse and collect the light into the PMT. Then it is covered with a black polyethylene film to avoid any optical crosstalk between sub-modules.

\begin{figure}[!]
  \centering
  \includegraphics[scale=0.5]{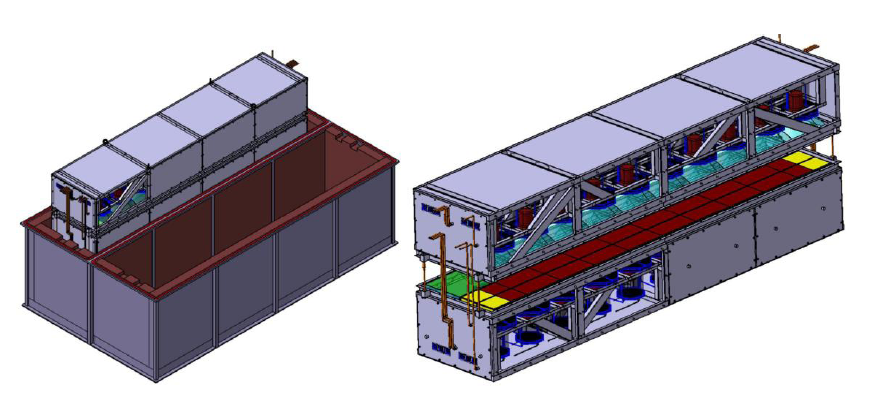}
  \caption{(Left) View of one of the two BiPo-3 modules inside the shield. The dimensions of the tank enclosing the shield are 3.9 $\times$ 2.1 $\times$ 1.4 m; (right) view of a BiPo-3 module with its upper part open for source installation (lateral iron plates have been removed for illustration purposes).}
  \label{fig:bipo3-design-3}
\end{figure}

The mechanical structure supporting the optical sub-modules (see Figure \ref{fig:bipo3-design-3}) is a welded pure iron structure closed with pure iron plates (2 cm thick). The radon tightness of the BiPo-3 modules is provided by silicone seals. All material used for the detector has been selected by HPGe measurements ensuring high radiopurity. The total mass of one BiPo-3 module is approximately 700 kg.

\begin{figure}[!]
  \centering
  \includegraphics[scale=0.5]{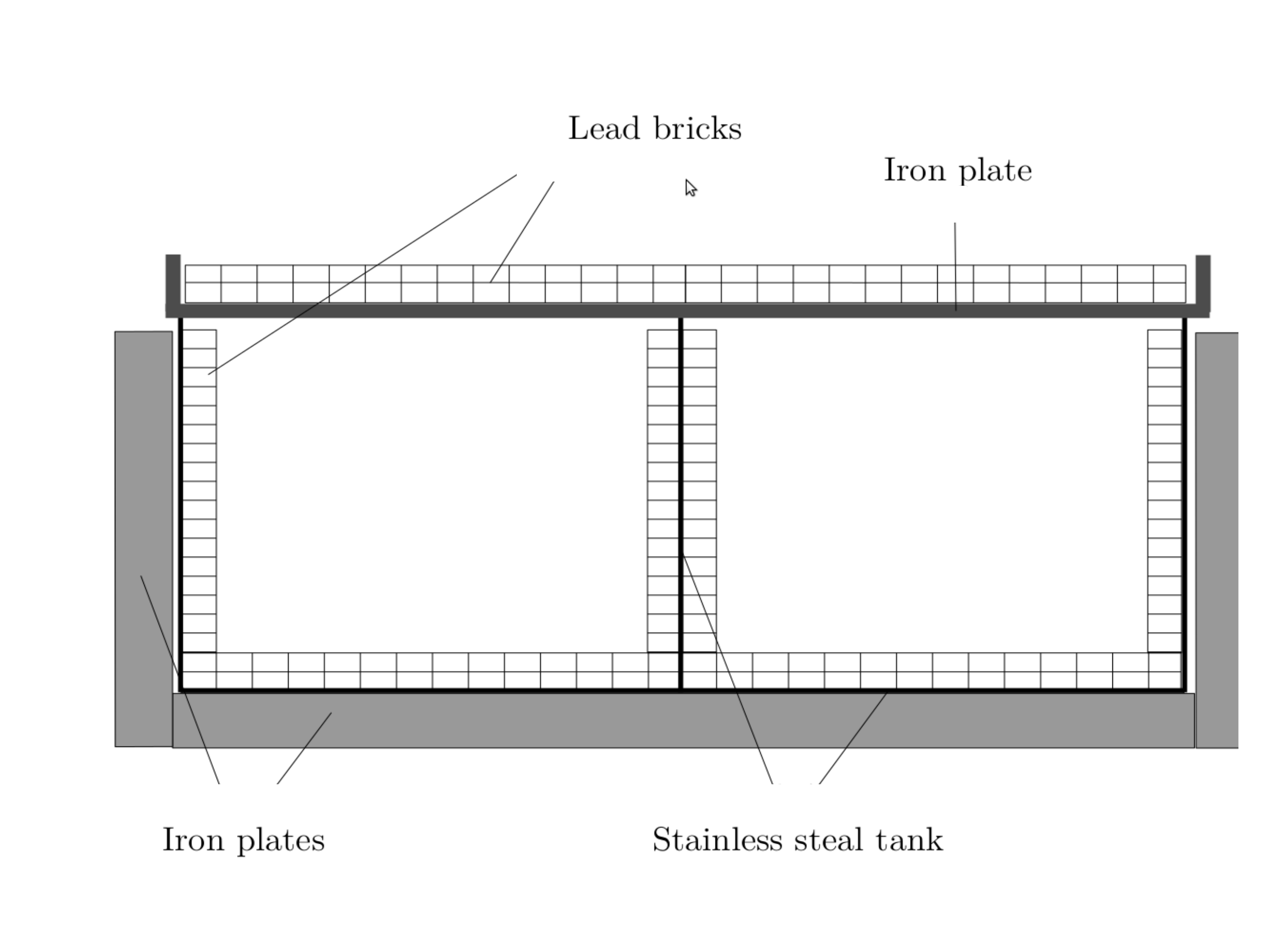}
  \caption{Schematic view of the low-radioactivity shield for the two BiPo-3 modules.}
  \label{fig:bipo3-design-shield}
\end{figure}

The two BiPo-3 modules are installed inside a low-radioactivity shield (see Figure \ref{fig:bipo3-design-shield}). The shield is built out of a radon-tight stainless steel tank with the upper part composed of a pure iron lid (2~cm thick). Low-activity lead bricks are assembled inside the tank and above the upper iron plate for a total thickness of 10~cm. Pure iron plates (18~cm thick) are added under the tank and on its lateral sides. 

Note that a preliminary shield has been used for initial measurements with Module 1. Measurements acquired with this shield will be indicated when needed. The preliminary shield consisted of pure iron plates, 18~cm thick, on the bottom and the top, and 10~cm lead bricks on the lateral side, and was wrapped in a radon-tight plastic foil.

Special attention is required for the radon suppression in order to avoid radon diffusion between the two scintillator faces. Pure nitrogen flushing of the whole detector in separate volumes reduces potential radon background. The most critical volume corresponds to the one surrounding the scintillators. It is isolated by a radon-tight  barrier film (ethyl vinyl alcohol copolymer and high density polyethylene - EVOH) which is placed around the light guides' collars near the PMTs. The PMTs are the main source of radon emanation and are enclosed in a gas-tight black box (made of black polyethylene). The mechanical structure of the two BiPo-3 modules and the shield tank are made gas-tight to avoid any external radon diffusion into the detector. The total volume of the BiPo-3 detector to be flushed is about 2~m$^3$. The seals are made of RTV 615 silicone mixed with pure selenium powder in order to obtain a black compound against light leaks. 

The insertion of the samples inside the BiPo-3 detector is performed in the LSC clean room. The corresponding module is opened by lifting the upper part with a manual hydraulic jack. The samples are placed directly on the surface of the lower scintillators and the upper scintillators are moved down until they are in contact with the samples.  

\subsection{Electronic readout and data acquisition}

PMTs signals are sampled with MATACQ VME digitizer boards~\cite{matacq}, which contain 4~readout channels with the following features: 2.5~$\mu$s time window, 1~GS/s sampling rate, 12-bit amplitude resolution, 1~V amplitude dynamic range and an electronic noise level of about 250~$\mathrm{\mu}$V (r.m.s.). The gain of each PMT is adjusted in order to have a signal amplitude of about 400~mV for a 1~MeV electron. The single photoelectron level is about 1~mV, corresponding to about 2~keV. The acquisition is triggered each time a PMT pulse exceeds 50~mV, corresponding to an energy threshold of about 130~keV. 

\begin{figure}[!]
  \centering
  \includegraphics[width=1.0\textwidth]{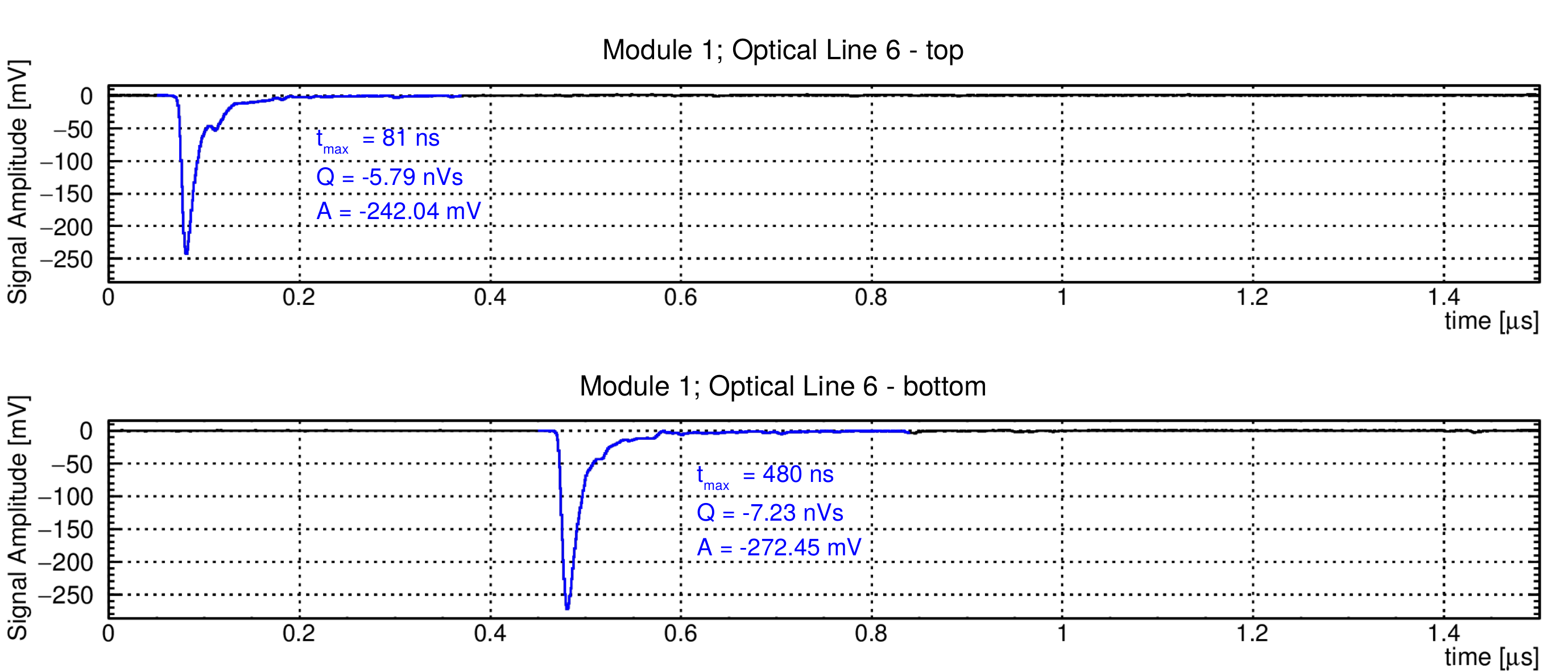}
  \caption{Display of a $^{212}$BiPo event, detected in the BiPo-3 detector during a background measurement. The prompt signal (top), and the delayed signal (bottom) are observed in opposite scintillators within a window of 1.5~$\mu$s.}
  \label{fig:display-bipo212}
\end{figure}

\begin{figure}[!]
  \centering
  \includegraphics[width=1.0\textwidth]{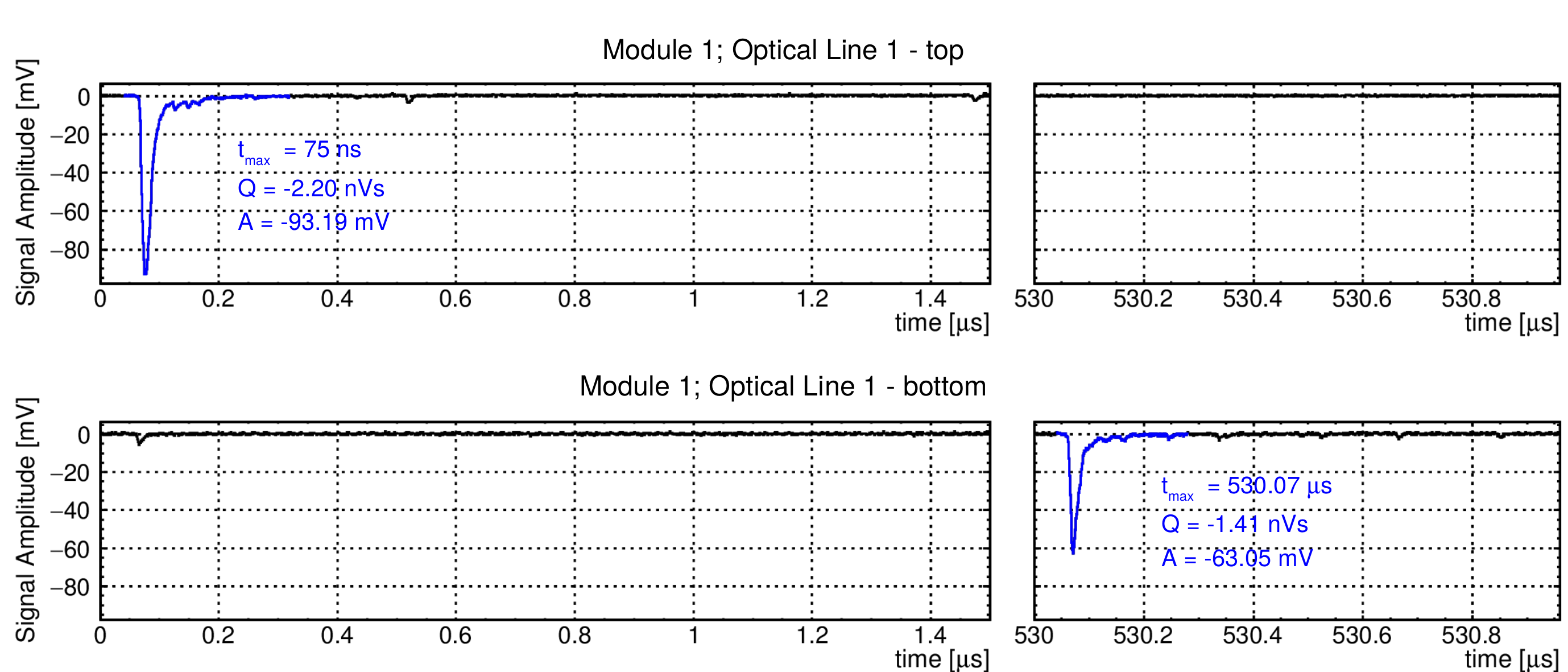}
  \caption{Display of a $^{214}$BiPo event, detected in the BiPo-3 detector during a background measurement. The prompt signal (top) starts the acquisition; no signal is observed in the opposite scintillator within the first 1.5~$\mu$s sampling. The delayed signal (bottom) is observed by the second delayed trigger.}
  \label{fig:display-bipo214}
\end{figure}

A dedicated trigger board has been developed in order to trigger both the $^{212}$Bi$\rightarrow^{212}$Po ($^{212}$BiPo) and the $^{214}$Bi$\rightarrow^{214}$Po ($^{214}$BiPo) cascades with a double sampling technique. The signals of both the triggered PMT and the PMT associated with the opposite scintillator are sampled for 1.5~$\mathrm{\mu}$s. After a 10~$\mathrm{\mu}$s dead time, a watchdog of 1~ms is started, waiting for a possible second delayed pulse above 30~mV (about 80~keV) in these two PMTs. If the second delayed trigger is validated, pulses of the two PMTs are sampled for 960~ns. Finally, the prompt and delayed samplings are stored on a computer. If no second delayed trigger is validated, only the first sampling is stored. The first trigger is used to tag the $^{212}$Bi$\rightarrow^{212}$Po cascade and the second delayed trigger is used to tag the $^{214}$Bi$\rightarrow^{214}$Po cascade. The recorded PMT signals of typical $^{212}$BiPo and $^{214}$BiPo events are displayed in Figures~\ref{fig:display-bipo212} and \ref{fig:display-bipo214}, respectively. 

Each BiPo-3 module has its own independent acquisition system allowing the manipulation of one module without disturbing the measurement of the second one. Four trigger boards, 20 MATACQ boards, and two controller boards are used for the complete acquisition of data from the two modules.

\subsection{BiPo-3 simulations}

Different Monte Carlo simulations have been done mainly to better understand the detector performance and to compute the detection efficiency of BiPo events depending on their origin (sample, scintillator surface or bulk volume), and in different configurations (background or sample measurement). They have been performed using the SuperNEMO simulation software, using DECAY0 as event generator~\cite{decay4}. This is a GEANT4-based framework which allows the precise detector definition, including the main features of the BiPo optical modules.

\subsection{Energy and timing calibration}

For each detected event, several pulses corresponding to different optical sub-modules are sampled. The arrival time of each pulse is determined estimating the amplitude mean value on a 10 ns sliding window. When an amplitude mean value higher than -5 mV is obtained, the arrival time is fixed and the region of interest (RoI) start is then set 10 ns before the arrival time. The amplitude mean value is also estimated over a sliding window to determine the end of the RoI. It is set 100 ns after the time where the amplitude mean value is lower than -1 mV. The integral of the pulse over this RoI, equivalent to the charge, is directly proportional to the energy deposited in the scintillator. A pedestal value is also calculated for each pulse, corresponding to the average value of the amplitude over the 100~ns period before the start of the pulse. Its value is subtracted from each sampling point for the computation of the charge. 

\begin{figure}[!]
  \centering
  \includegraphics[scale=0.5]{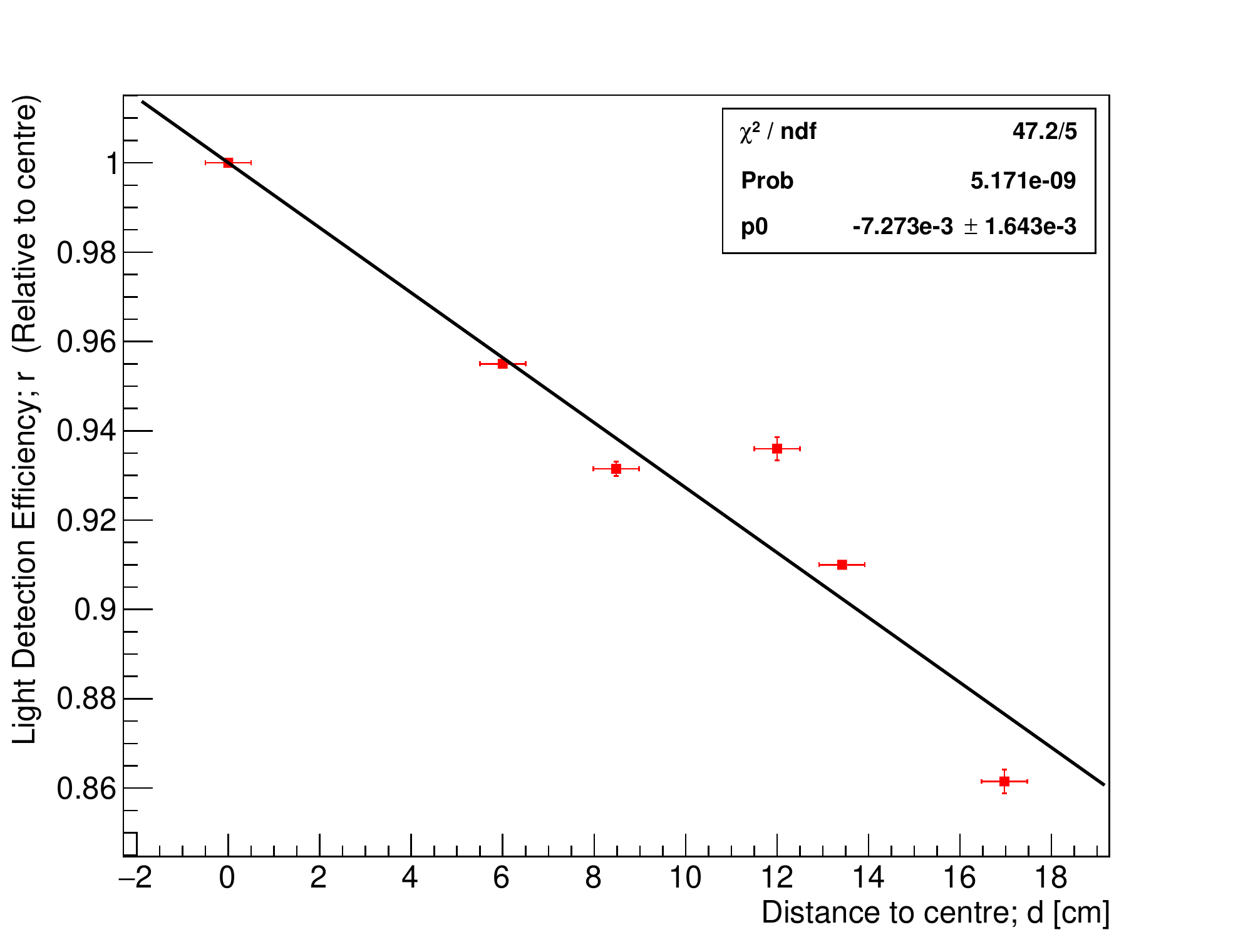}
  \caption{Average relative light detection efficiency $r$  (with respect to the centre) of the 80 scintillator blocks of the BiPo-3 detector, as a function of the distance $d$ to the centre of the entrance face of the scintillator, measured with an $^{241}$Am $\alpha$ source. The data are fitted by a linear function, and this response is taken into account in the BiPo-3 detector simulations. The equation of the best fit is $r= 1 - 7.273 \times 10^{-3} d$ (cm).}
  \label{fig:scanning-law}
\end{figure}

The linearity and the gain stability of all PMTs were initially measured on a PMT test bench. During the construction of the BiPo-3 detector, each optical sub-module was tested and pre-calibrated in energy with a dedicated calibration test bench. A $^{207}$Bi source, which provides internal conversion electrons with energies of $482$~keV and $976$~keV from the K-lines (with branching ratios of $1.5\%$ and $7.1\%$, respectively), was placed at the centre of the entrance face of the scintillator and used to define the PMT HV nominal value in order to equalize the gains of the 80 optical sub-modules. An $^{241}$Am source, emitting 5.5~MeV $\alpha$ particles, was used to measure the uniformity of the surface response of the scintillators. A scan of 25 points at the surface of each scintillator was performed. The $^{241}$Am source was covered by a 6~$\mathrm{\mu}$m thick polyethylene film in order to prevent the scintillator from being contaminated by the source. A decrease of about 15\% in the light collection was observed when the $\alpha$ source was moved from the centre of the entrance face of the scintillator to the edges. The average measured radial response, presented in Figure~\ref{fig:scanning-law}, is fitted by a linear function, and is taken into account in the BiPo-3 detector simulations. The dispersion of the averaged responses with respect to the linear fit induces a systematic uncertainty of 5\% on the BiPo detection efficiency, as discussed in section~\ref{sec:detection-efficiency}.

After the installation of a new sample, and once the corresponding module is installed inside the shield, an energy and timing calibration is performed before strating the data taking. To do this, five $^{54}$Mn sources, emitting $\gamma$'s of 835~keV, are positioned on the top of the corresponding module. After closing the upper shield lid, about 12 hours of calibration data are collected.

The energy calibration is performed measuring the single Compton edge at 639~keV for electrons fully contained in one scintillator. Events with a signal detected in one scintillator and no coincidence signal detected in the opposite scintillator are selected for the energy calibration. The simulated and measured energy spectra are fitted with a Kolmogorov statistical test. By selecting the highest probability of agreement returned by the test, two energy calibration parameters are obtained: the charge to energy conversion factor (at the centre of the entrance face of the scintillator) and the energy resolution of the scintillator. The test is performed between 500 and 750~keV, providing a precise result while optimizes the computing time.

\begin{figure}[!]
  \centering
  \includegraphics[scale=0.5]{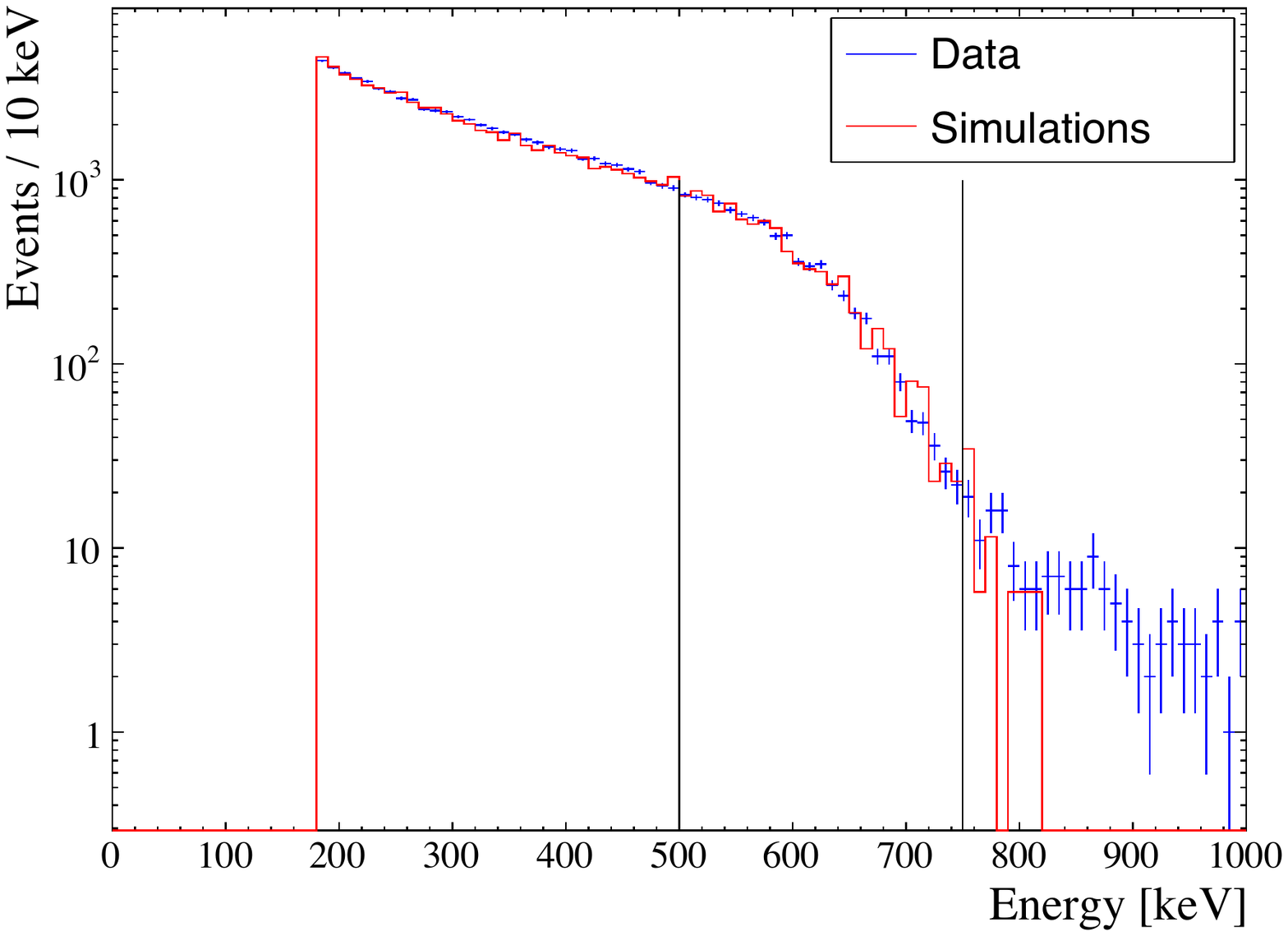}
  \vspace{-6cm}
  \caption{Example of $^{54}$Mn energy calibration of an optical sub-module: the simulated energy spectrum is fitted to the observed spectrum in the Compton edge $[500 - 750]$~keV energy window.  }
  \label{fig:calib-energy}
\end{figure}

An example of the result of the energy calibration measured for an optical sub-module is presented in Figure~\ref{fig:calib-energy}.  The Compton spectrum  and Compton edge are well reproduced by the simulations. The excess of background events observed above the Compton edge is due to the residual radon inside the shield during the calibration. The charge to energy conversion factors are typically around 100~keV/nVs and the energy resolution is $\sigma_E \sim 20\% / \sqrt{E(\mathrm{1 MeV})}$.
The energy resolution is estimated with a relative precision of about 20\% ($\sigma_E$ varies from 16\% to 24\%). 
The charge to energy conversion factor of a given scintillator varies by less than 10\% among different calibrations.
A 10\% variation of the charge to energy conversion factor induces a systematic uncertainty of 4\% on the BiPo detection efficiency, as discussed in section~\ref{sec:detection-efficiency}.

\begin{figure}[!]
  \centering
  \includegraphics[width=0.65\textwidth]{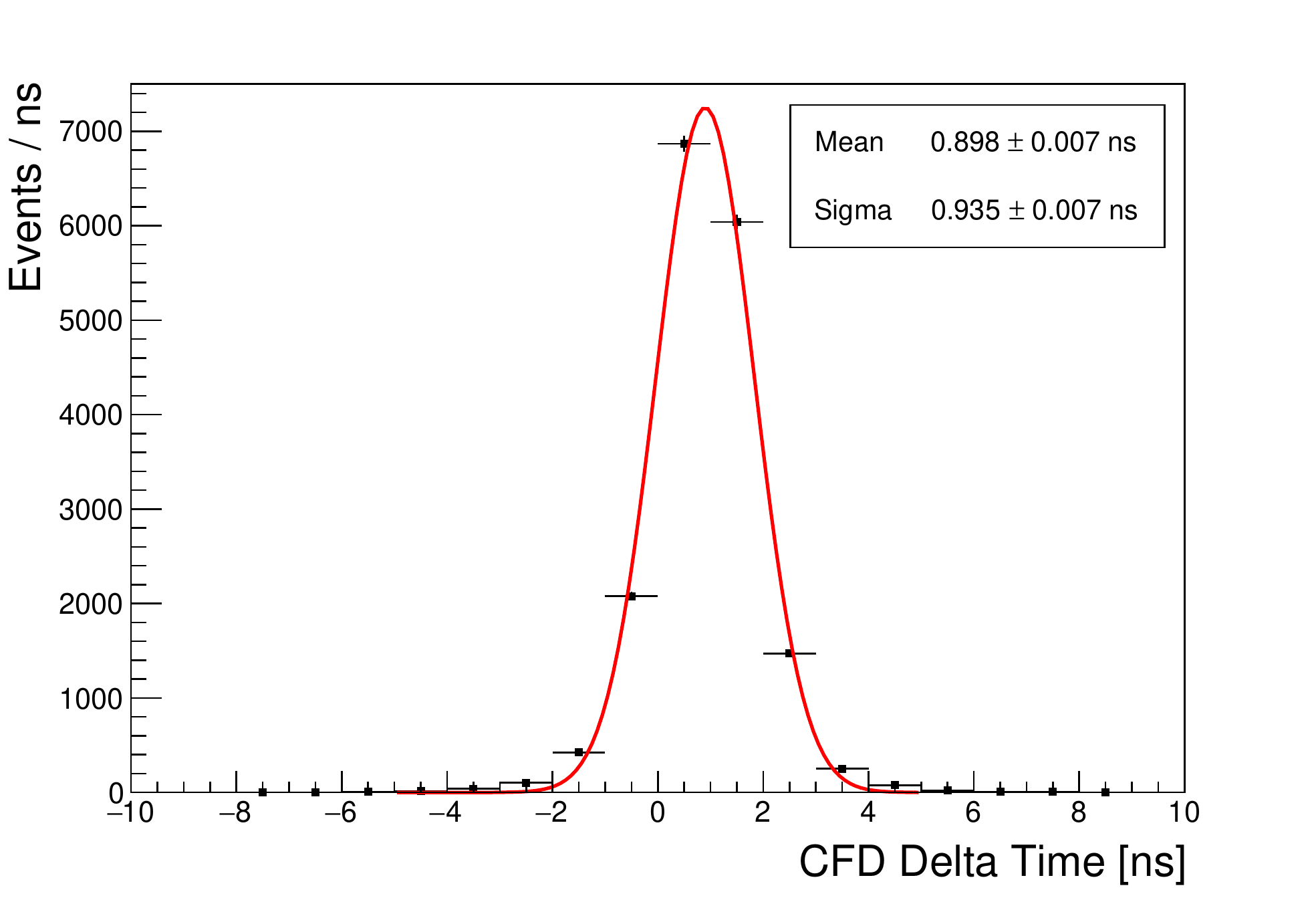}
  \caption{Example of a Gaussian fit of the constant fraction discrimination (CFD) time difference between two opposite optical sub-modules.}
  \label{fig:calib-time}
\end{figure}

The timing calibration is performed using the Compton electrons produced in one scintillator that crossed over into the opposite scintillator.
Events with a signal detected in coincidence in both opposite scintillators are selected for the timing calibration. The average time difference between the two PMT signals is measured and corrected in the  BiPo analysis software. An example of the time difference distribution measured during a calibration is presented in Figure~\ref{fig:calib-time}.
The results of the timing calibration are very stable from one calibration to the next one, with variations of less than 1~ns. Such variation is negligible for the BiPo measurements. 

\subsection{Scintillation quenching of $\alpha$ particles}
\label{sec:quenching}

\begin{figure}[!]
  \centering
  \includegraphics[width=0.75\textwidth]{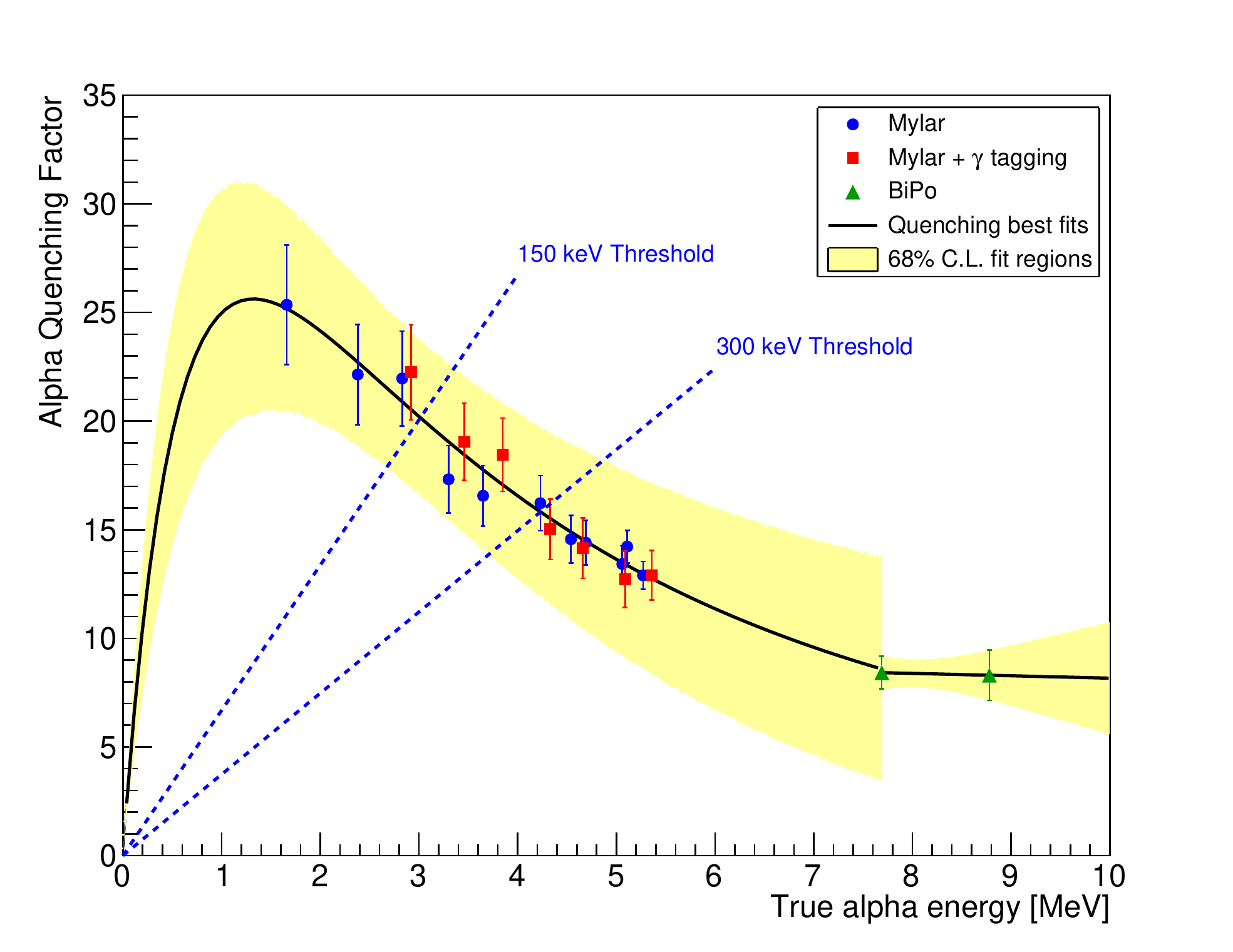}
  \caption{Quenching factor of an $\alpha$ particle in the BiPo-3 scintillator as a function of the energy deposited by the $\alpha$ particle on the scintillator. The blue dots are the measurements using an $^{241}$Am $\alpha$ source and different Mylar thickness. The red dots are the measurements obtained by detecting the low energy $\gamma$ in coincidence with the $\alpha$ particle to tag them. The green dots correspond to the measurements of the BiPo decays from the radon and thoron contaminations on the surface of the BiPo-3 scintillators. The curve corresponds to the fit of the different points, being the values used in the BiPo-3 simulations (see text for details). The 68\% C.L. band is also showed together with the energy thresholds of 150~keV (300~keV) equivalent electron, applied to the delayed $\alpha$ signal for the $^{212}$BiPo ($^{214}$BiPo) detection.}
  \label{fig:qf}
\end{figure}

The efficiency  of the BiPo-3 detector is limited by the capacity of an $\alpha$ particle to escape from the foil being measured, and then to deposit an amount of scintillation light above the detection threshold. 
The scintillation yield for $\alpha$ particles is quenched because of their very large stopping power. The amount of light produced by an $\alpha$ particle is smaller than that produced by an electron of the same energy. Moreover the quenching effect is not linear in energy. 
A dedicated measurement of the quenching factor $Q_{\alpha}$, defined as the ratio of the amount of light produced by an electron to the one produced by an $\alpha$ particle of the same energy, has been performed with the polystyrene-based plastic scintillators used in BiPo-3 for different $\alpha$ energies. 
The $\alpha$ particles of 5.5~MeV emitted by a $^{241}$Am source were used. Their energies have been reduced by adding successively thin Mylar foils between the source and the scintillator. The total thickness of the foils varies from 1.5 to 26.5~$\mathrm{\mu}$m. A simulation of $\alpha$ particles emitted by $^{241}$Am and crossing several Mylar foils has been performed in order to determine the expected spectrum of the energy deposited in the scintillator. Then the quenching factor is calculated. Internal conversion electrons from a $^{207}$Bi source are used for energy calibration. The results are presented in Figure~\ref{fig:qf}. This measurement is an improvement over the previous measurement performed for the BiPo-1 prototype~\cite{bipo1}: the energy loss fluctuations have been reduced thanks to a collimator, and the background has been reduced by using a smaller scintillator sample. Another set of measurements has been carried out with a NaI crystal by detecting the low energy $\gamma$ emitted in a delayed coincidence (half-life of about 60~ns) with the $\alpha$ particle, in order to suppress the external background. Results of this second measurement are presented also in Figure~\ref{fig:qf}.
The values of the quenching factor for $\alpha$ particles with energies of 7.69~MeV and 8.78~MeV have been obtained separately by analysing the BiPo decays from the $^{222}$Rn and $^{220}$Rn (usually referred as thoron) contaminations respectively, on the surface of the BiPo-3 scintillators, as discussed in section~\ref{sec:alu}.
The measured values of the quenching factor are fitted by an effective model given by: 

\begin{eqnarray}
\nonumber Q_{\alpha} & = & -a  \left( \frac{1}{(bE_{\alpha}+1)^c} - \frac{1}{(bE_{\alpha}+1)^{c/2}} \right) \ \mathrm{for} \ E_{\alpha} < 7.69 \  \mathrm{(MeV)} \\
\nonumber Q_{\alpha} & = & d  - e E_{\alpha} \ \mathrm{for} \ E_{\alpha} > 7.69 \ \mathrm{(MeV)}
\end{eqnarray}

\parindent=0cm where $E_{\alpha}$ is given in MeV and the fitted parameters are $a=99.82 \pm 12.92$, $b=0.316 \pm 0.250$, $c=3.836 \pm 1.397$, $d=12.64 \pm 10.17$ and $e=0.539 \pm 1.267$.
This effective model is used in the BiPo-3 Monte Carlo simulations.
The measured values are about 20\% higher than the theoretical values calculated on the basis of the Birk's model~\cite{tretyak}.
A 20\% uncertainty on the  quenching factor for $\alpha$ particles corresponds to a systematic error of the BiPo detection efficiency of 5\%  as discussed in section ~\ref{sec:detection-efficiency}. 
The energy threshold applied to the delayed $\alpha$ scintillation signal is 150~keV (300~keV) equivalent electron for the $^{212}$BiPo ($^{214}$BiPo) detection, as presented in section~\ref{sec:bkg}. It corresponds to an energy threshold on the $\alpha$ energy of 2.2~MeV (4.5~MeV).

\subsection{Pulse shape analysis and noise rejection}

\begin{figure}[!]
  \centering
  \includegraphics[scale=0.4]{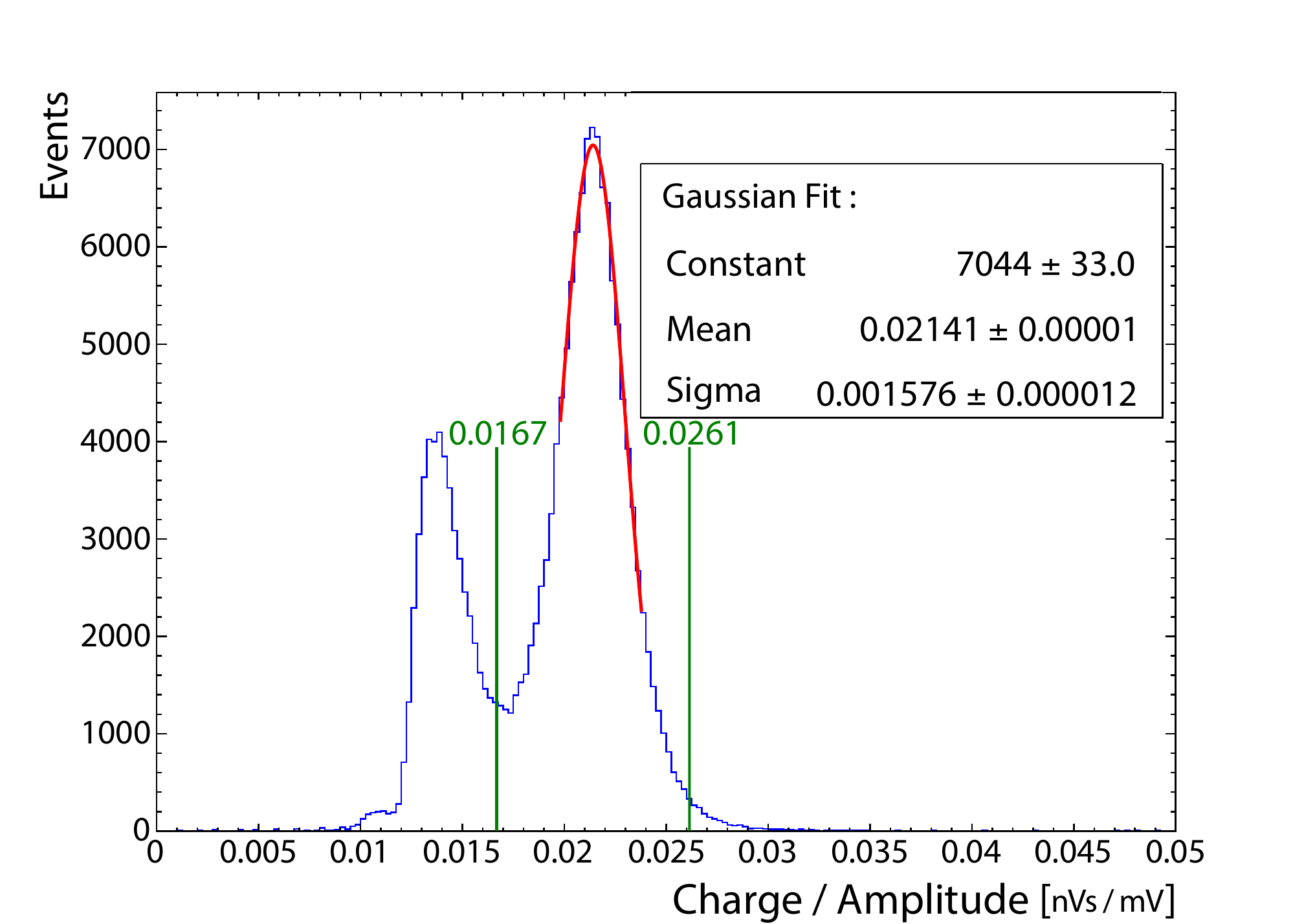}
  \includegraphics[width=1.0\textwidth]{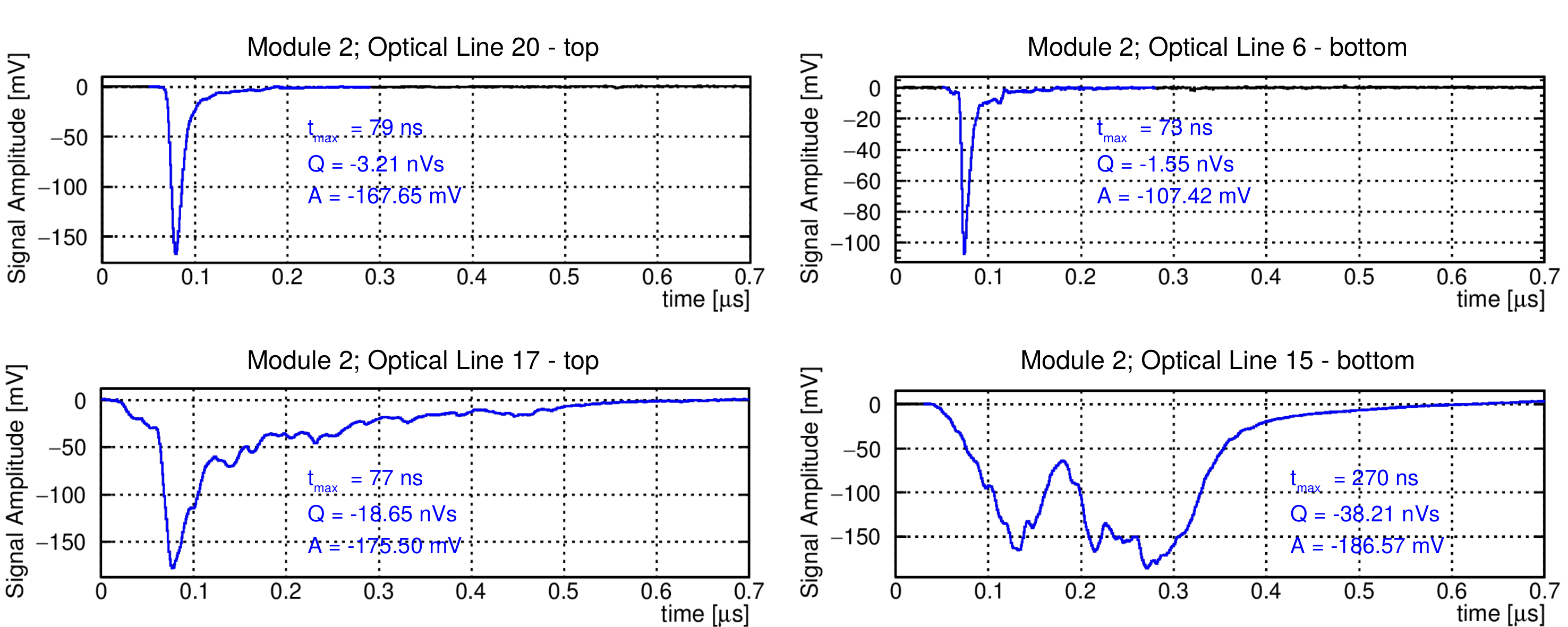}
  \caption{Upper panel: charge-over-amplitude ratio ($Q/A$) distribution measured for one optical sub-module. Lower panel: different examples of typical PMT signals for different $Q/A$ values: a good scintillation signal (top left), a signal emitted inside the optical guide (low $Q/A$ ratio, top right), and two noise signals (large $Q/A$, bottom).}
  \label{fig:charge-amplitude}
\end{figure}

A pulse shape analysis, based on the charge over amplitude ratio ($Q/A$), is applied on the recorded pulses in order to reject noise. Given that both the charge and the amplitude of the signal are proportional to the energy, the $Q/A$ ratio must be constant.
A typical distribution of the $Q/A$ ratio of an optical sub-module is represented in Figure~\ref{fig:charge-amplitude}. Two distinct distributions of PMT signals are well identified. The main distribution, fitted by a Gaussian, corresponds to PMT signals induced by the scintillation in the plastic scintillators. 
The second distribution is understood as scintillation light background emitted inside the optical guide. Rare pulses with $Q/A$ ratio above the signal peak are also observed. They correspond to sporadic noise in the PMT. Examples of pulses with typical $Q/A$ ratio are displayed in Figure~\ref{fig:charge-amplitude}. 
Scintillation signals are selected by requiring a $Q/A$ ratio within 3$\sigma$ of the main signal distribution. 

\section{Background measurement}
\label{sec:bkg}

\subsection{Background sources}

\begin{figure}[!]
  \centering
  \includegraphics[scale=0.4]{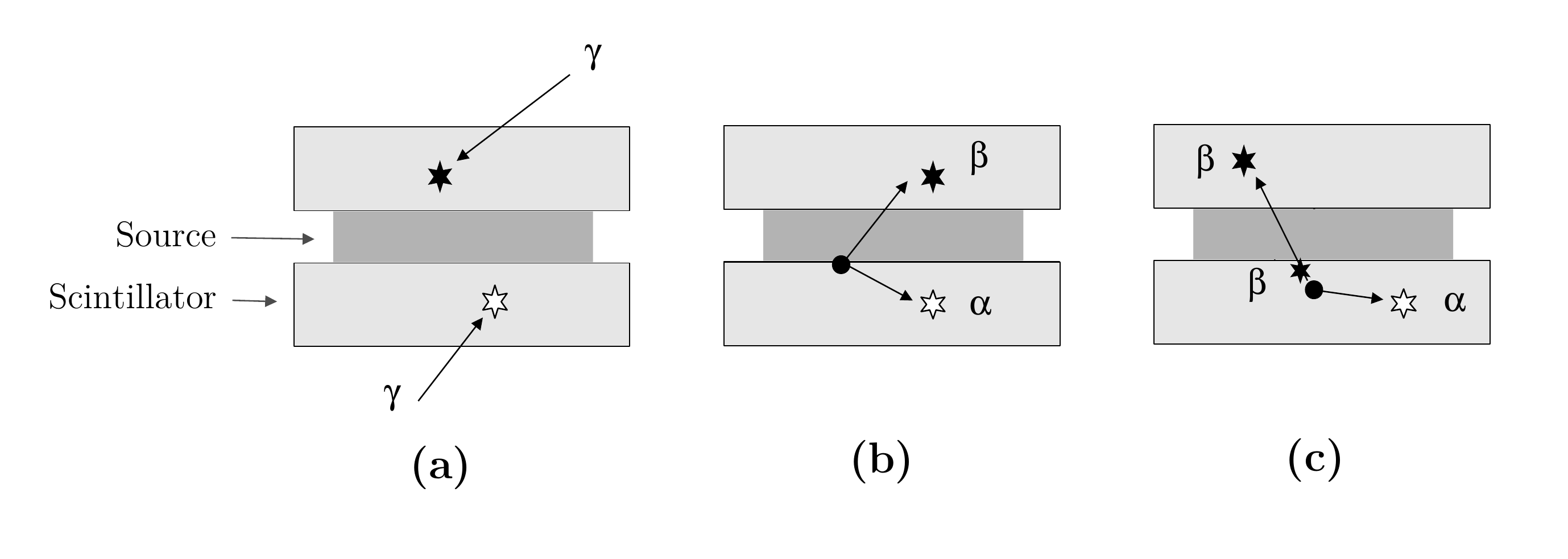}
  \caption{Illustration of the possible sources of background: (a) random coincidences due to $\gamma$-rays, (b) $^{212}$Bi or $^{214}$Bi contamination on the surface of the scintillators, (c) $^{212}$Bi or $^{214}$Bi  contamination in the volume of the scintillators. The dots correspond to the contamination, the black stars to the prompt signal and the white stars to the delayed signal.}
  \label{fig:bkg-schema}
\end{figure}

The first source of background  is the rate of random coincidences between two opposite scintillators, within the delay time window, as illustrated in Figure~\ref{fig:bkg-schema}~(a). The delay time distribution of the random coincidences is flat and the energy distributions of both prompt and delayed signals are located at low energy, since the single counting rate is dominated by Compton electrons diffused by external $\gamma$-rays.

The second  source of background comes from $^{212}$Bi and $^{214}$Bi contaminations on the surface of the scintillator (the so-called surface background), in contact with the sample foil, as illustrated in Figure~\ref{fig:bkg-schema}~(b). The delayed $\alpha$ particle, emitted from the surface of the scintillator, deposits all its energy inside the scintillator, in contrast with a  BiPo decay inside the sample foil, where the $\alpha$ particle loses part of its energy in the foil before reaching the scintillator. Therefore the surface background can be rejected by requiring the $\alpha$ energy to be lower than a given value. 
In the case of a $^{212}$Bi or $^{214}$Bi bulk contamination inside the scintillator volume, the delayed $\alpha$ particle also deposits all its energy inside the scintillator. In contrast, the prompt electron first triggers one scintillator block before escaping and entering the opposite one, as illustrated in Figure~\ref{fig:bkg-schema} (c). Therefore two prompt signals are detected in coincidence in the two opposite scintillators, allowing the rejection of this class of background events, as discussed in section~\ref{sec:coinc-rejection}.
However, if the contamination is located very close to the surface, the prompt electron escapes the first scintillator without depositing enough energy to induce a trigger. The signal in coincidence with the prompt signal is thus not detected and this background is similar to a surface background event. The lowest signal detection threshold must be set in order to reduce the background coming from the scintillator contamination. 
We note that a thoron ($^{220}$Rn) or radon ($^{222}$Rn) deposition on the surface of the scintillators may also cause a surface background. 

Thus the background can be defined by two components: the random coincidences and the surface background.  
The energy spectrum of the delayed signal is the most sensitive observable to discriminate the two background components. 
For surface background, the delayed $\alpha$ particle deposits all its energy inside the scintillator, corresponding to a peak at about 800~keV for $^{214}$Bi and about 1~MeV for $^{212}$Bi, while random coincidence signals are dominant at low energy.

\subsection{Measurement conditions}

The background is measured by closing the detector, without any sample between the scintillators. Opposite scintillators are directly in contact. 
The background has been measured separately for each of the two BiPo-3 modules. 

The background of the Module 1 has been measured from July 2012 to September 2012 at the beginning of the commissioning with a preliminary shield, and from July 2013 to September 2013 with the final shield and after having introduced and measured several samples.
The background of Module 2 has been measured from February 2013 to May 2013, with the final shield.

From August 2014 to December 2014, one half of Module 1 was empty  during the measurement of the first enriched $^{82}$Se foils (see Section~\ref{sec:sn-foil-measurement}), and from December 2014 until February 2015, about half of Module 2 was empty during the measurement of reflecting foils used for bolometric double $\beta$-decay experiments. These two measurements are used to control the stability and reliability of the two previous dedicated background measurements.
The $^{214}$Bi background measurement of Module 2 during the measurement of reflecting foils is unavailable because of a problem with the delayed $^{214}$BiPo trigger during that period.

\subsection{Event selection}
\label{sec:event-selection}

The selection procedure of the {\it back-to-back} BiPo events are described in the following.
The energy of the prompt signal is required to be greater than 200~keV. 
The energy of the delayed signal is required to be greater than 150~keV  for $^{212}$BiPo events, and greater than 300~keV for $^{214}$BiPo events. 
The higher energy threshold for the $^{214}$BiPo measurement is set in order to suppress the external radon background, as explained in section~\ref{sec:bkg-214}. 
The delay time between the prompt and the delayed signal is greater than 20~ns and lower than 1400~ns for $^{212}$BiPo events, and greater than 10~$\mathrm{\mu}$s and lower than 1000~$\mu$s for $^{214}$BiPo events. 
The charge over amplitude $Q/A$ ratio of the prompt and delayed signals is required to be contained within the scintillation signal range. 
If a signal greater than 3~mV (about 10~keV) is detected in coincidence with the prompt signal in the opposite scintillator, the BiPo event is recognized as a bulk contamination background event and is rejected. This coincidence rejection is discussed in section~\ref{sec:coinc-rejection}.

An excess of BiPo events is observed in the first days of background measurement. This excess is due to the deposition of thoron ($^{220}$Rn) and radon ($^{222}$Rn) on the surface of the scintillators when the detector is opened in the clean room, before starting the background run. 
The thoron decays successively to $^{216}$Po, then to $^{212}$Pb, which finally decays to $^{212}$Bi with a half-life of 10.6 hours. 
This results in an excess of $^{212}$Bi decays on the surface of the scintillators for several dozens of hours. 
The radon decays with a half-life of 3.8 days and produces an excess of $^{214}$Bi decays on the surface of the scintillators in the first days (see for instance Figure~\ref{fig:bulk-bkg-1}). 
In order to reject the thoron (radon) contamination, the 3 (15) first days of data are systematically removed for the $^{212}$BiPo ($^{214}$BiPo) events selection. 

\subsection{Detection efficiency}
\label{sec:detection-efficiency}

For all measurements performed with the BiPo-3 detectors, the corresponding energy spectra and detection efficiencies are evaluated by simulating $^{212}$BiPo and $^{214}$BiPo decay cascades, uniformly distributed on the sample volume and the surface of the scintillators. This provides the efficiencies of BiPo events depending on their origin. From simulated data, the efficiencies are obtained after the application sequentially of the different selection criteria described in section \ref{sec:event-selection}, except for the noise rejection. 

The associated systematics are computed from these simulations taking into account the following sources: energy calibration, quenching factor, maximum allowed delay time, noise rejection, non-uniformity on the light collection efficiency and sample geometry. For all these sources, their corresponding uncertainties and their contribution to the overall efficiency systematic are summarized in Table \ref{tab:eff-sys}. It is worth mentioning that for the case of the background measurements, the systematic contribution due to the uncertainty on the sample geometry does not need to be taken into account, while for further sample measurements it has been considered. Table \ref{tab:det-eff} shows the efficiencies after each subsequent selection cut for the background measurement, so without any sample in between the scintillators. The final obtained values are 30$\pm$2.7 \% for $^{212}$BiPo events and 26$\pm$2.3 \% for $^{214}$BiPo events. 

\begin{table}[htb]
\centering
\begin{tabular}{lcc}
 & & Efficiency \\
Parameter & Associated uncertainty & systematic contribution \\
\hline
Energy calibration & 10 \% & 4 \% \\
Quenching factor & 20 \%  & 5 \% \\
Maximum delay & 20 \% & 2 \% \\
Light collection efficiency & 10 \% & 5 \% \\
Noise rejection & 3 \% & 3 \% \\
Sample thickness & & 5 \% \\
\hline
Total systematic (without sample)& & 8.9 \% \\
Total systematic (with sample)& & 10.2 \% \\
\hline
\end{tabular}
\caption{Summary of the parameters contributing to the systematic on the detection efficiency, both for $^{212}$BiPo and $^{214}$BiPo decay cascades. The uncertainties of these parameters are quoted together with the contribution to the systematic.}
\label{tab:eff-sys}
\end{table}

\begin{table}[htb]
\centering
\begin{tabular}{ccc}
Applied selection & $^{212}$BiPo efficiency & $^{214}$BiPo efficiency\\
\hline
Back-to-back & 38 \% & 35 \% \\
E$_{prompt}$ > 200 keV & 32 \% & 28 \% \\
E$_{delay}$ > 150 keV / E$_{delay}$ > 300 keV & 32 \% & 28 \% \\
$\Delta$t & 30 \% & 26 \% \\
\hline
Final efficency & 30$\pm$2.7 \% & 26$\pm$2.3 \% \\
\hline
\end{tabular}
\caption{Summary of the detection efficiencies, both for $^{212}$BiPo and $^{214}$BiPo decay cascades, for background measurement after the sequential application of the different selection criteria. Final efficiency values are obtained once the latest criterion, the time difference between the prompt and delayed signals ($\Delta$t), is applied. The systematic unceratainty of these effciencies are estimated based on the contributions summarized in Table \ref{tab:eff-sys}.}
\label{tab:det-eff}
\end{table}

\subsection{Analysis method}
\label{sec:bkg-analysis-method}

The background activities (surface background and random coincidences) are compared to the observed data, by fitting simultaneously the expected energy spectra of the delayed signal of the two background components, using the likelihood method.
The energy spectra and the detection efficiencies of the $^{212}$Bi and $^{214}$Bi surface background are evaluated by simulating $^{212}$BiPo and $^{214}$BiPo decay cascades, uniformly distributed on the surface of the scintillators, without any sample in-between. The surface background detection efficiency is 30$\pm$2.7 \% for $^{212}$BiPo events and 26$\pm$2.3 \% for $^{214}$BiPo events. 
The energy spectrum of the random coincidence background is measured directly using the single counting events.
The criteria to select single counting events are as follows: (i) events with a single PMT signal (no peak in the opposite scintillator) are selected; (ii) the charge over amplitude $Q/A$ ratio of the signal is required to be contained within the scintillation signal range; (iii) if a signal greater than 3~mV is detected in coincidence in the opposite scintillator, the event is rejected.

The rate of random coincidences is also determined independently by measuring the single counting rate of the scintillator plates, using the single counting events. 
The single rate is calculated by using all the data available and by averaging over all the scintillators.
The expected number of random coincidences is equal to $2 \times r_p \times r_d \times \Delta T \times T_{obs}$,
where $r_p$ is the single rate measured by applying the prompt energy threshold (200~keV), 
$r_d$ is the single rate measured by applying the delayed energy threshold (150~keV for  $^{212}$Bi and 300~keV for $^{214}$Bi), 
$\Delta T$ is the time window (1380~ns for  $^{212}$Bi and 990~$\mu$s for $^{214}$Bi), and $T_{obs}$ is the duration of the measurement. 
Comparing the expected number of random coincidences with its fitted value gives a cross-check and a validation of the fitting procedure.

The delay time distribution is also used {\it a posteriori} to validate the proportion of surface background to random coincidences. This is done by fitting an exponential decay with a half-life set to the value of the polonium decay half-lives (300~ns for $^{212}$Po and 164~$\mathrm{\mu}$s for $^{214}$Po) plus a constant value for a flat random coincidence distribution. 

\subsection{Result of the background measurement in the $^{212}$BiPo channel}

The results of the background measurements in the $^{212}$BiPo channel are presented in Table~\ref{tab:bkg-bipo212}. 
The level of random coincidences, calculated by measuring the single rate, is about $6 \times 10^{-4}$~counts/day/m$^2$ of scintillator surface area, and is negligible. Therefore only the surface background contributes to the observed background.  
We note that the levels of surface background measured separately in the two BiPo-3 modules are equal, within the statistical uncertainty. 
It is worth mentioning that the background is stable after having opened several times the Module 1, and that no contamination has been introduced. 
The background level is also stable when a sample is placed in the other half of the BiPo-3 module (the first enriched $^{82}$Se foils for Module 1 or reflecting foils for Module 2).
It has been verified that the $^{212}$BiPo background events are uniformly distributed, both in space across the detector and in arrival time. 

\begin{table}[htb]
\centering
\begin{tabular}{c|c|c|c||c||c|c}
  & Module 1   & Module 1  & Module 2  &             & Module 1      & Module 2  \\
  & Temporary  & Final     & Final     & {\bf Combined} & $^{82}$Se foil & Reflect. foil   \\
  & shield  & shield & shield &             & Run           & Run       \\
\hline
\hline
Duration (days)           &  73.5 & 51.2  & 75.7  & 200.4  &  104.0 & 134.9 \\
\hline
Scint. surface  (m$^2$)   & 2.7   & 3.06  & 3.42  & 3.10   & 1.26  & 1.98 \\
\hline
Data events               & 9     & 8     & 12    & 29     & 5     & 10 \\ 
\hline
Surf. Bkg (fit)           & 7.4$\pm$2.9   & 8.0$\pm$3.4   & 12.0$\pm$3.5  & 27.7$\pm$5.4   & 5.0$\pm$2.1   & 11.6$\pm$3.4\\
Coinc. (fit)              & 1.6$\pm$1.5   & 0.0           &  0.0          &  1.3$\pm$1.4   & 0.0   & 0.5$\pm$0.6 \\
\hline
Coinc. (single rate)      & 0.20   & 0.10 & 0.14   & 0.44    & 0.06  & 0.14 \\ 
\hline
\hline
$\mathcal{A}(^{208}\mathrm{Tl})$ $\mathrm{\mu}$Bq/m$^2$ & $0.8 \pm 0.3$  & $1.0 \pm 0.4$  & $1.0 \pm 0.3$  & {\bf $0.9 \pm 0.2$}  & $1.0 \pm 0.5$ & $0.8 \pm 0.3$ \\
\end{tabular}
\caption{Results of the $^{212}$BiPo background measurements: separate and combined results of the three dedicated background measurements, and results of two additional background measurements carried out during the measurement of the first enriched $^{82}$Se foils (Module 1) and during the measurement of reflecting foils (Module 2). We report the duration of measurement, the active scintillator surface area, the number of detected $^{212}$BiPo events, the number of surface background events and random coincidences calculated by the fit of the delayed energy spectrum, the expected number of random coincidences calculated with the single rate measurement, and the activity in $^{208}$Tl on the surface of the scintillators.}
\label{tab:bkg-bipo212}
\end{table}

\begin{figure}[!]
  \centering
  \includegraphics[scale=0.37]{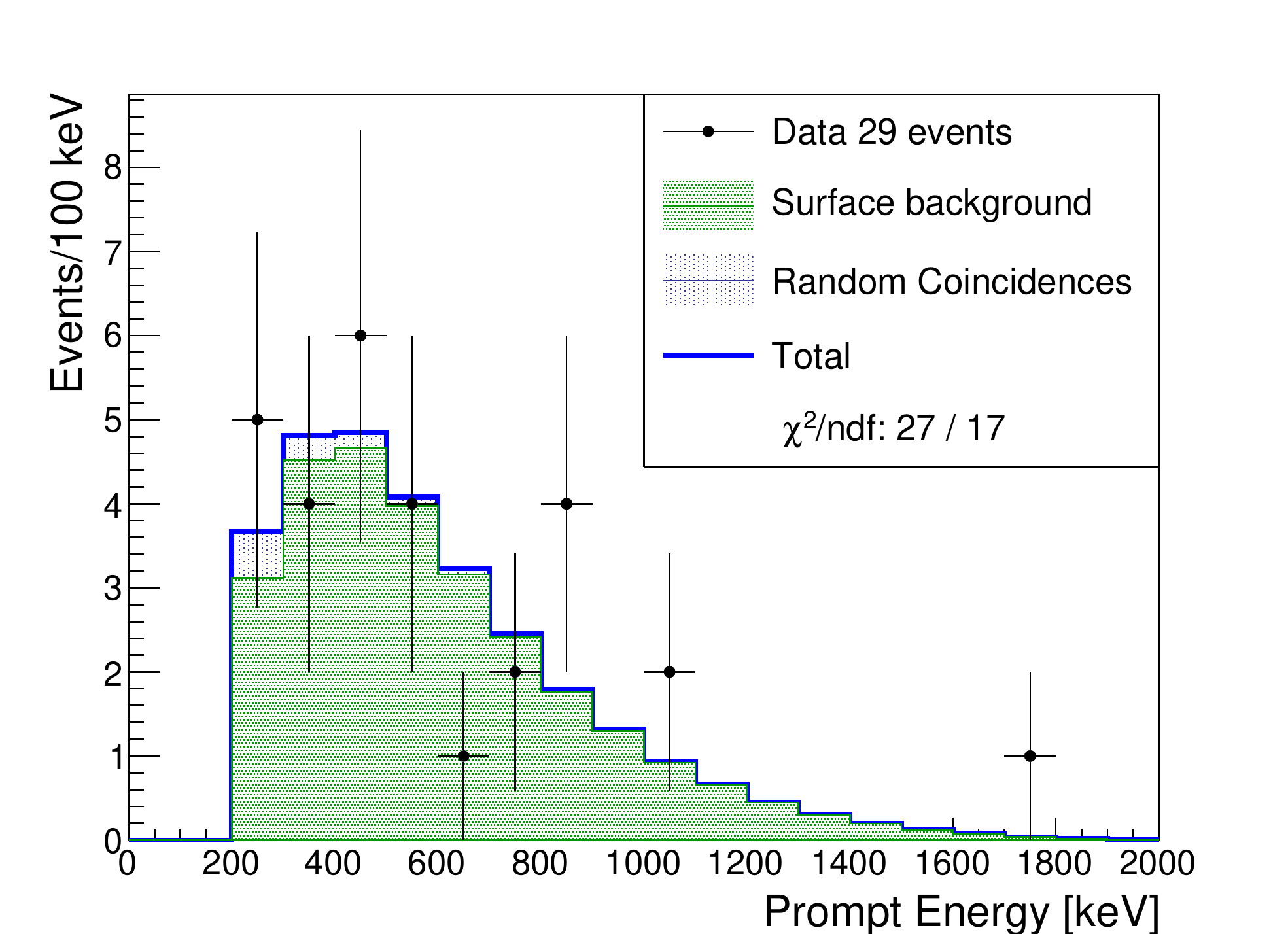}
  \includegraphics[scale=0.37]{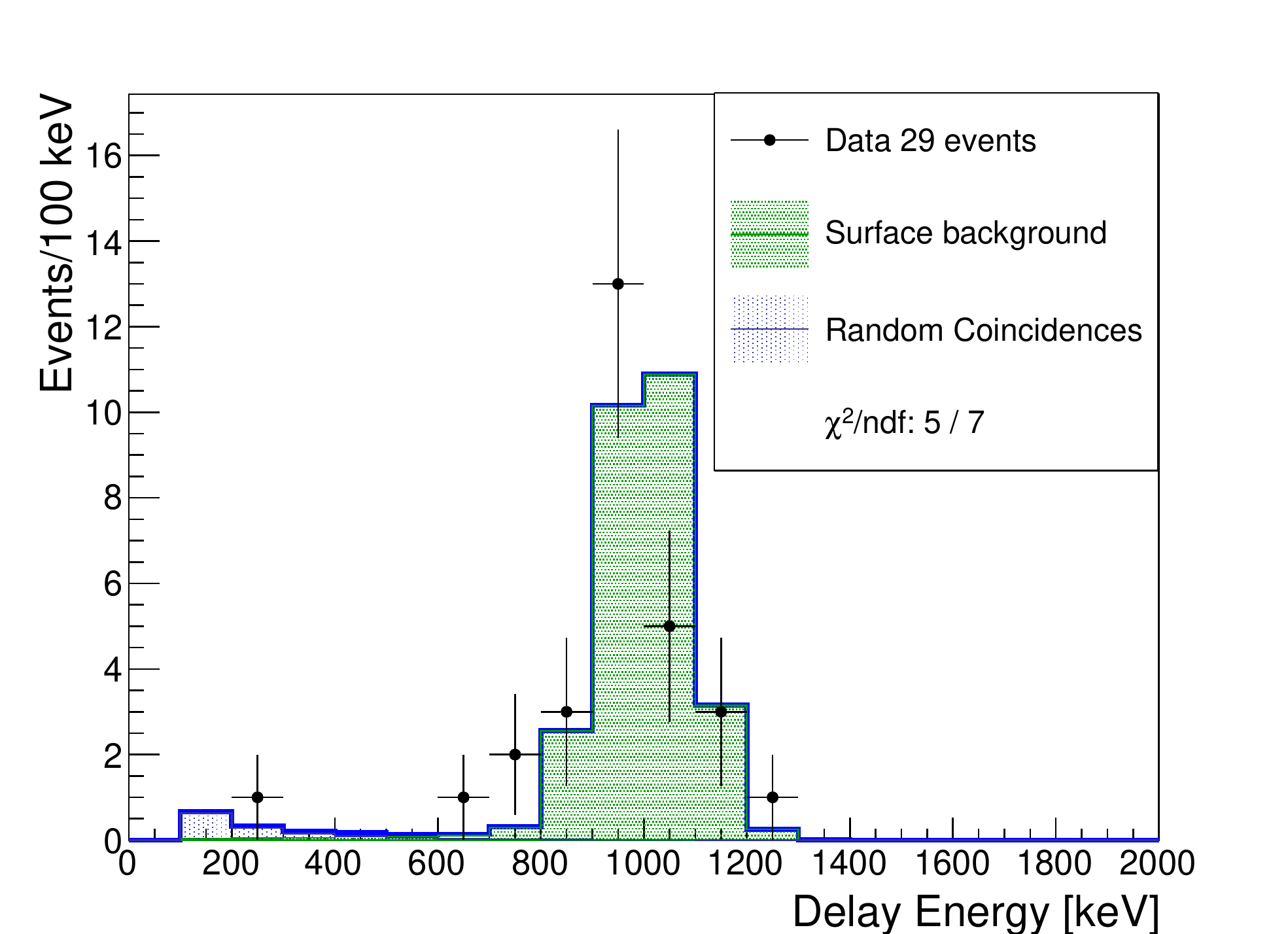}
  \includegraphics[width=0.50\textwidth]{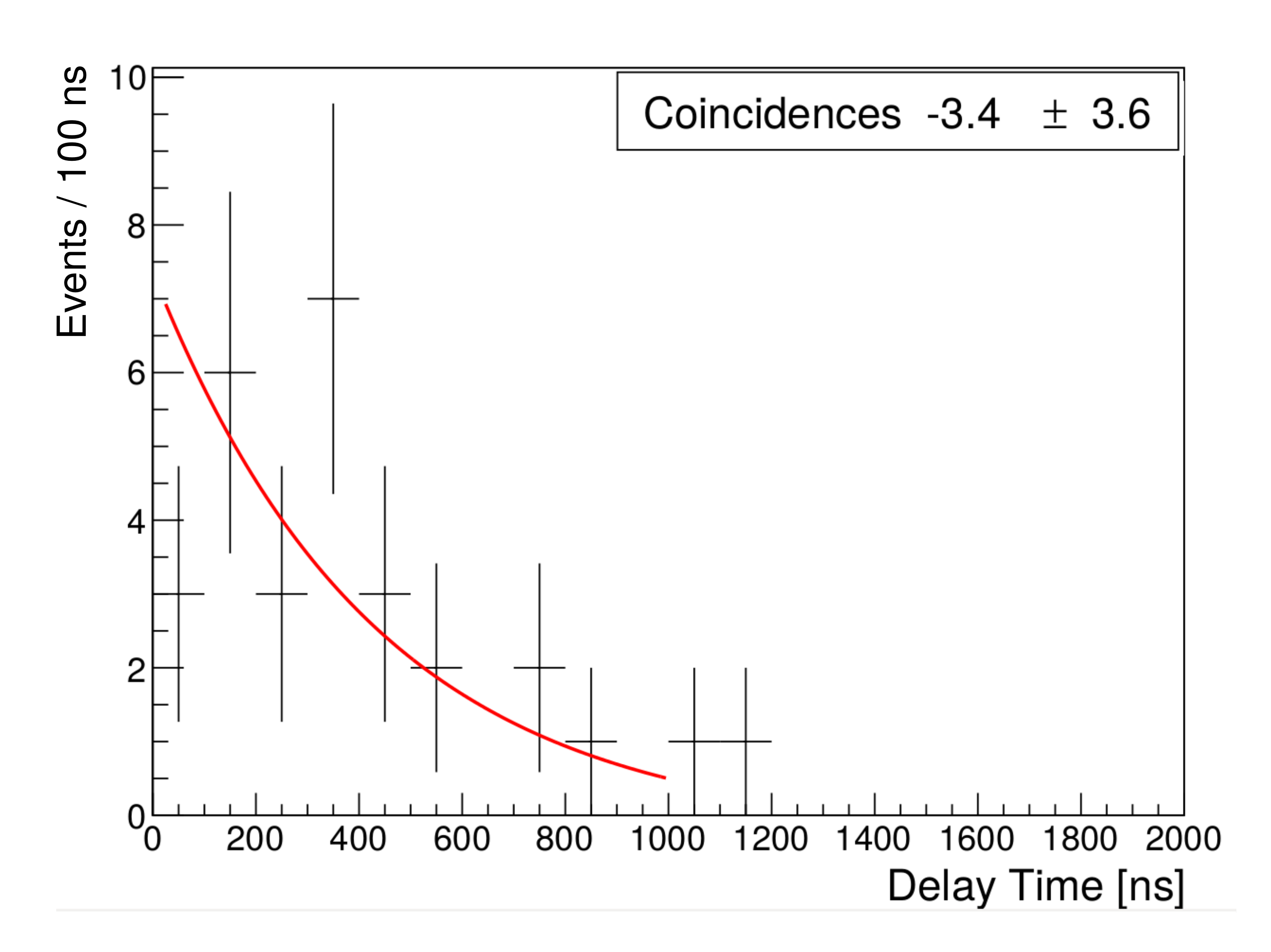}
  \caption{Distributions of the prompt energy (top left), the delayed energy (top right) and the delay time (bottom), for the $^{212}$BiPo background measurements, adding 	the three distinct sets of dedicated background measurements (Module 1 with preliminary and final shield, and Module 2 with final shield), corresponding to 200.4 days of data collection and an effective scintillator surface area of 3.10~m$^2$.
  The data are fitted by the expected background from the $^{212}$Bi contamination on the surface of the scintillators (green histogram) and from the random coincidences (blue histogram). The delay time distribution is fitted a posteriori by an exponential decay with a half-life set to the value of the $^{212}$Po decay half-life (300~ns) plus a constant value for a flat random coincidence contribution.}
  \label{fig:bkg-212}
\end{figure}

By adding the three distinct sets of dedicated background measurements (Module 1, with preliminary and final shield and Module 2 with final shield), corresponding to 200.4 days of data collection and a scintillator surface area of 3.10~m$^2$, 29 $^{212}$BiPo background events have been observed. 
The energy distributions of the prompt and delayed signals and the delay time between the two signals are presented in Figure~\ref{fig:bkg-212} for these 29 $^{212}$BiPo background events. The result of the fit of the two background components is also presented. 
The fitted $^{208}$Tl activity on the surface of the scintillators is $\mathcal{A}(^{208}\mathrm{Tl})=0.9 \pm 0.2$~$\mathrm{\mu}$Bq/m$^2$. 
The delay time distribution is fitted by a exponential decay with a half-life equal to the $^{212}$Po half-life of 300~ns. 
The observed delay time distribution is in agreement with the expected $^{212}$Po decay time distribution.

The ultra low $^{208}$Tl activity of $0.9 \pm 0.2~\mu$Bq/m$^2$ on the surface of the scintillators is similar to the one measured in the BiPo-1 prototype~\cite{bipo1}. 

\subsection{Result of the background measurement in the $^{214}$BiPo channel}
\label{sec:bkg-214}

\begin{figure}[!]
  \centering
  \includegraphics[width=0.48\textwidth]{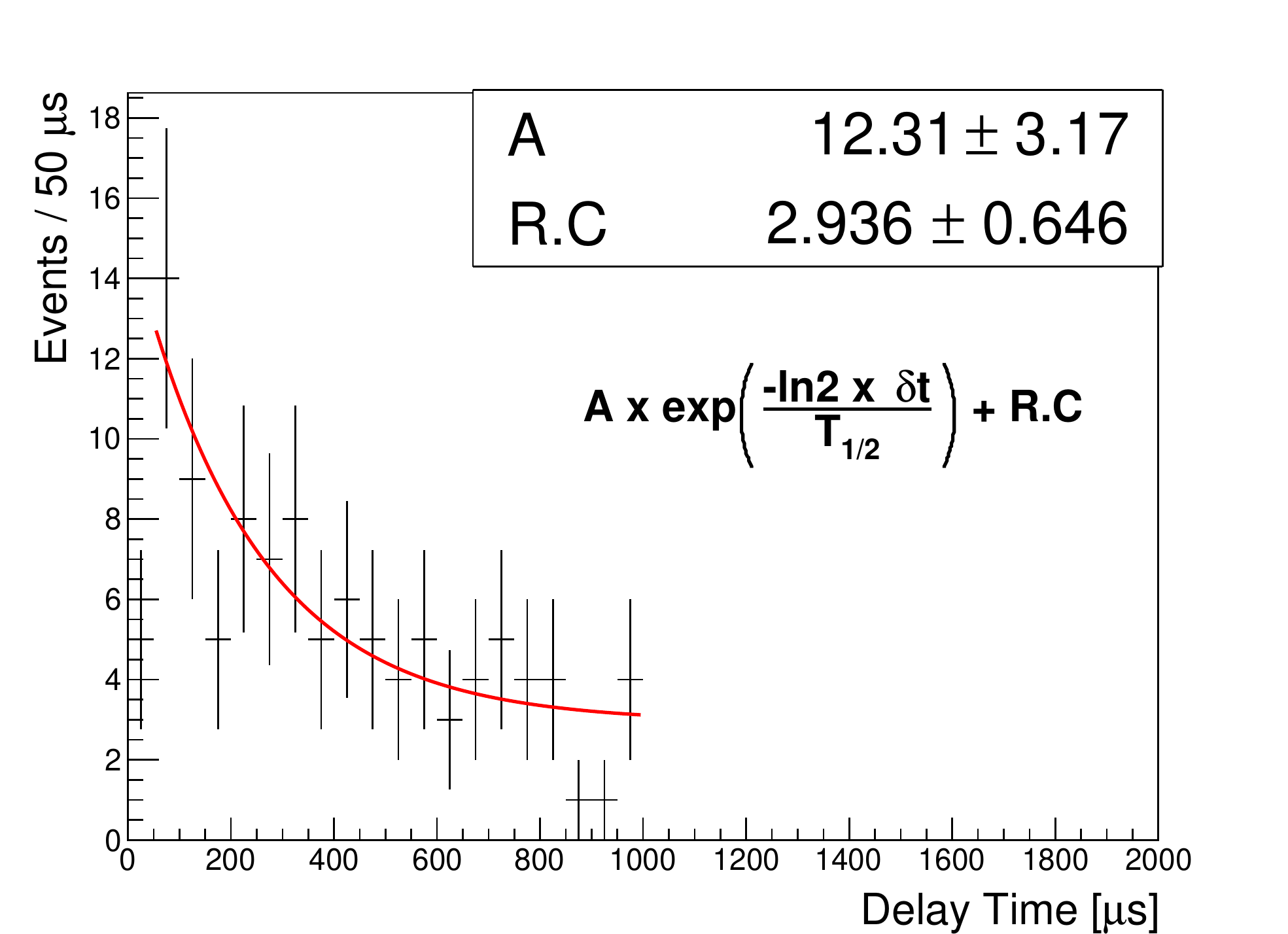}
  \includegraphics[width=0.48\textwidth]{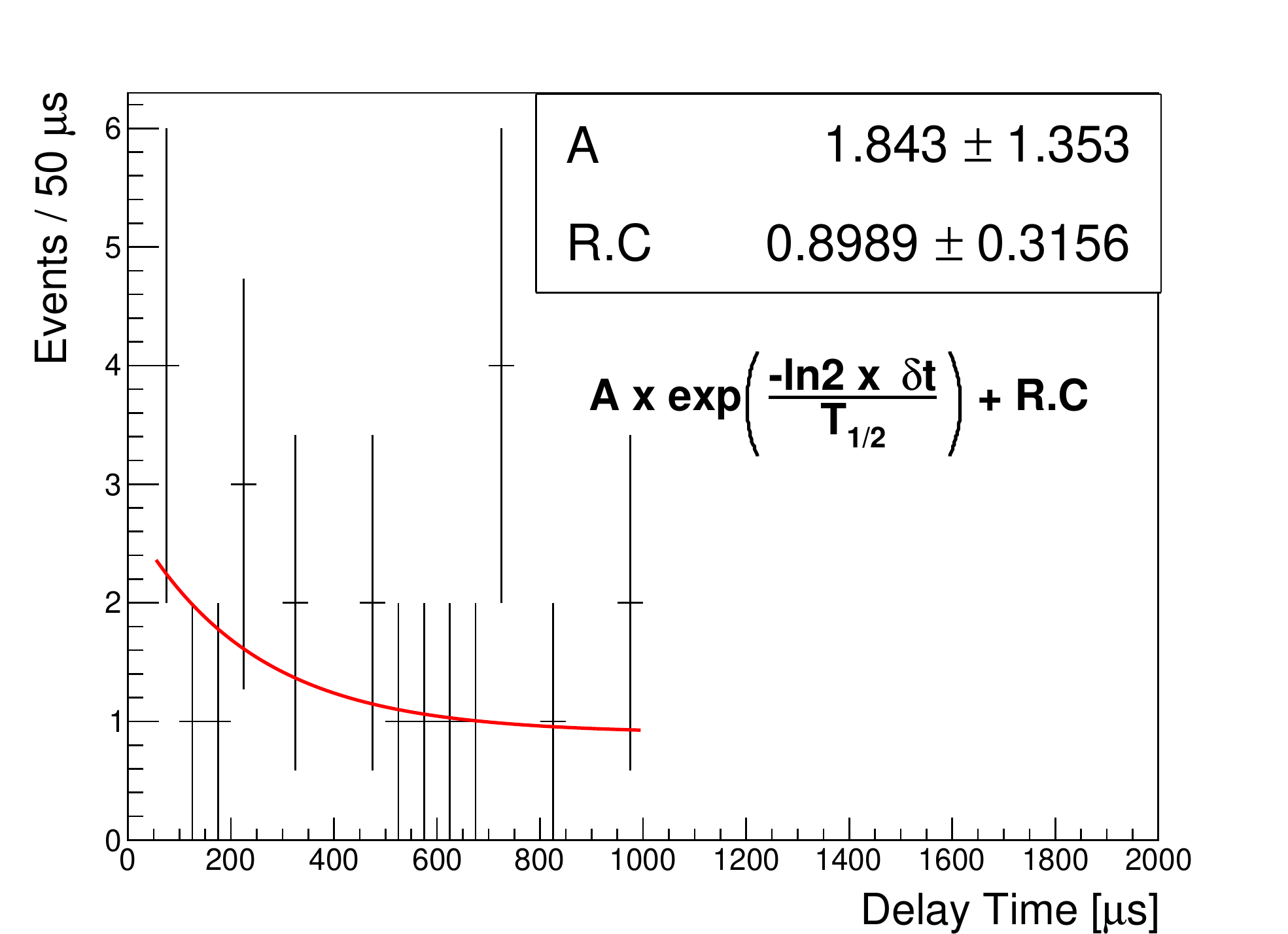}
  \caption{Delay time distribution between prompt and delayed signals for $^{214}$BiPo events selected from the three sets of dedicated background measurements (Module 1 with preliminary and final shield and Module 2 with final shield), with energy of the delayed signal between 100~keV and 600~keV (left), and between 300~keV and 600~keV (right).}
  \label{fig:bkg-ext-radon}
\end{figure}

For the  $^{214}$BiPo measurement, the random coincidence background becomes larger than the surface background due to the longer $^{214}$Po decay half-life, leading to BiPo events with preferentially low energy signals and a flat delay time distribution.
However, another source of background is also observed at low delayed energy when studying the delay time distribution of these events, as shown in Figure~\ref{fig:bkg-ext-radon}~(left). In addition to the flat distribution expected for random coincidences, an excess of events is observed at low delay time with an exponential distribution and a half-life in agreement with the $^{214}$Po half-life.
This means that this background is produced by a $^{214}$BiPo decay. However the decay does not occur on the scintillator surface since the delayed $\alpha$ particle is measured with a lower energy than expected for surface background. It indicates that the $^{214}$BiPo decay is not located directly on the scintillator surface, but at an unknown distance from the lateral sides of the scintillators. This may explain the energy loss of the delayed $\alpha$ particle. A possible origin is an external radon contamination on the lateral sides of the scintillators. 
This background is suppressed by applying an energy threshold of 300~keV to the delayed signal, as shown in Figure~\ref{fig:bkg-ext-radon}~(right) where the exponential component becomes negligible. 
Therefore an energy threshold of 300~keV on the delayed signal is systematically applied for the $^{214}$BiPo measurement. 
The counterpart is that the $^{214}$BiPo efficiency is reduced by about 60\% when measuring samples. 

\begin{figure}[!]
  \centering
  \includegraphics[scale=0.37]{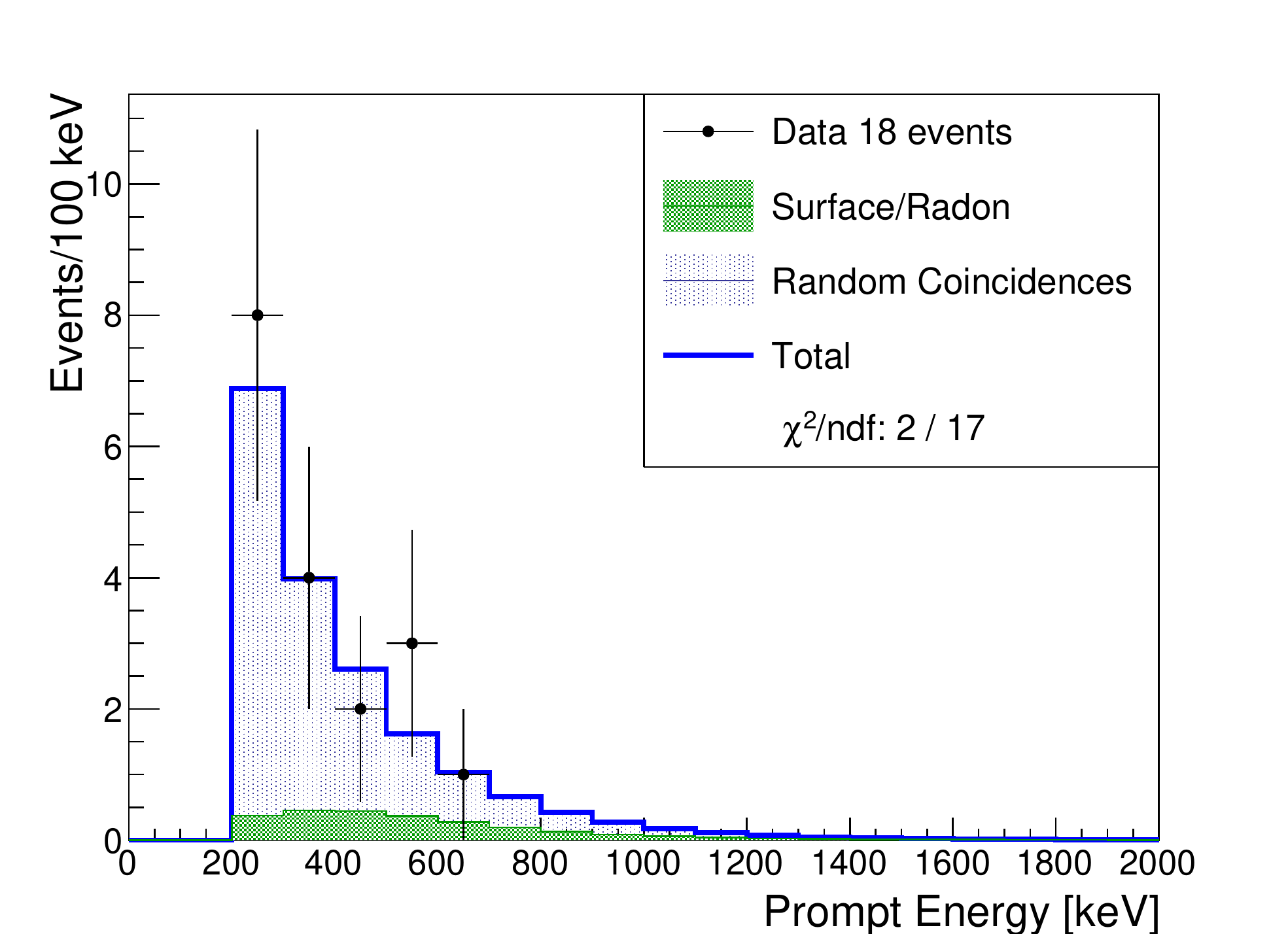}
  \includegraphics[scale=0.37]{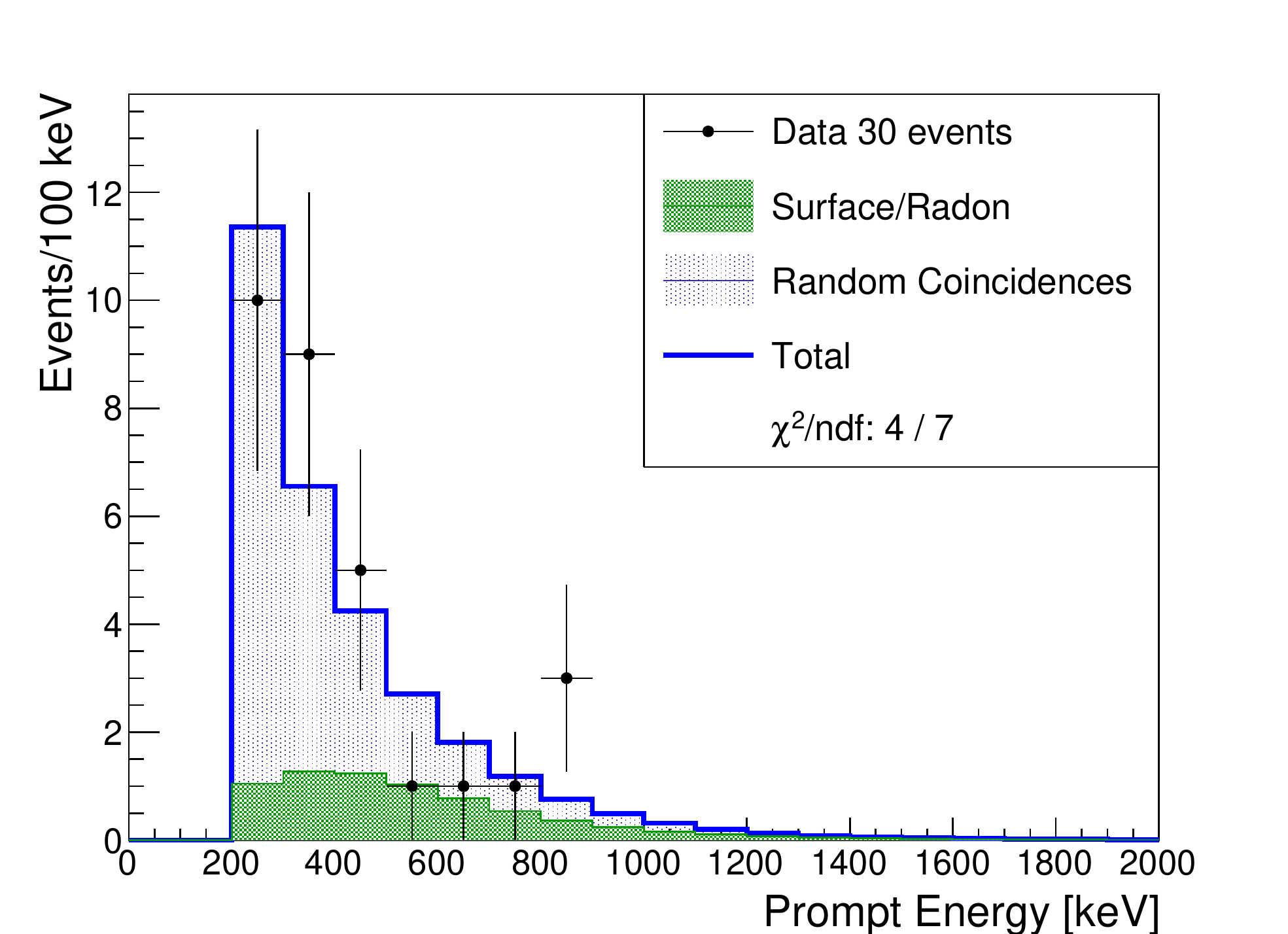}
  \includegraphics[scale=0.37]{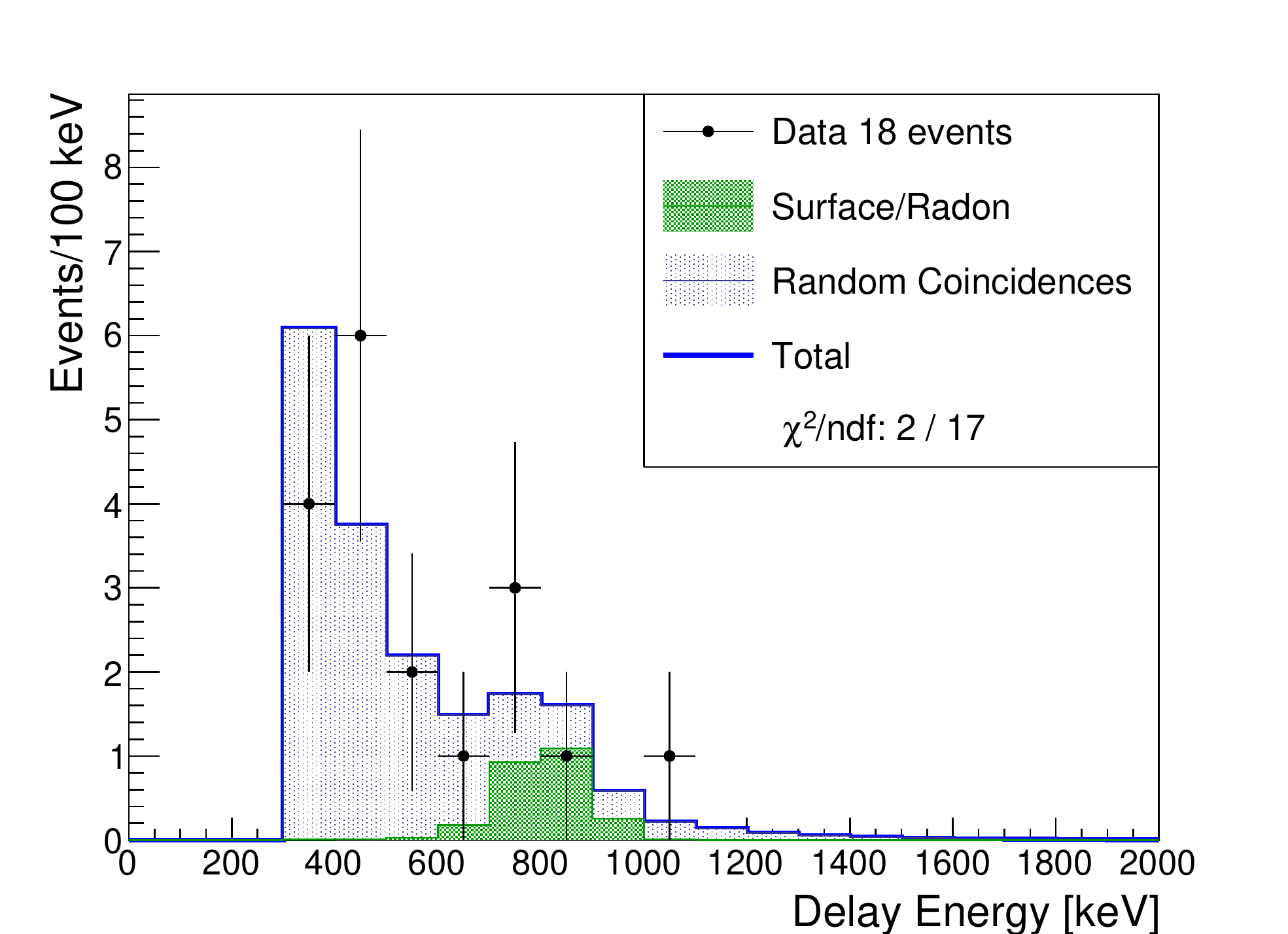}
  \includegraphics[scale=0.37]{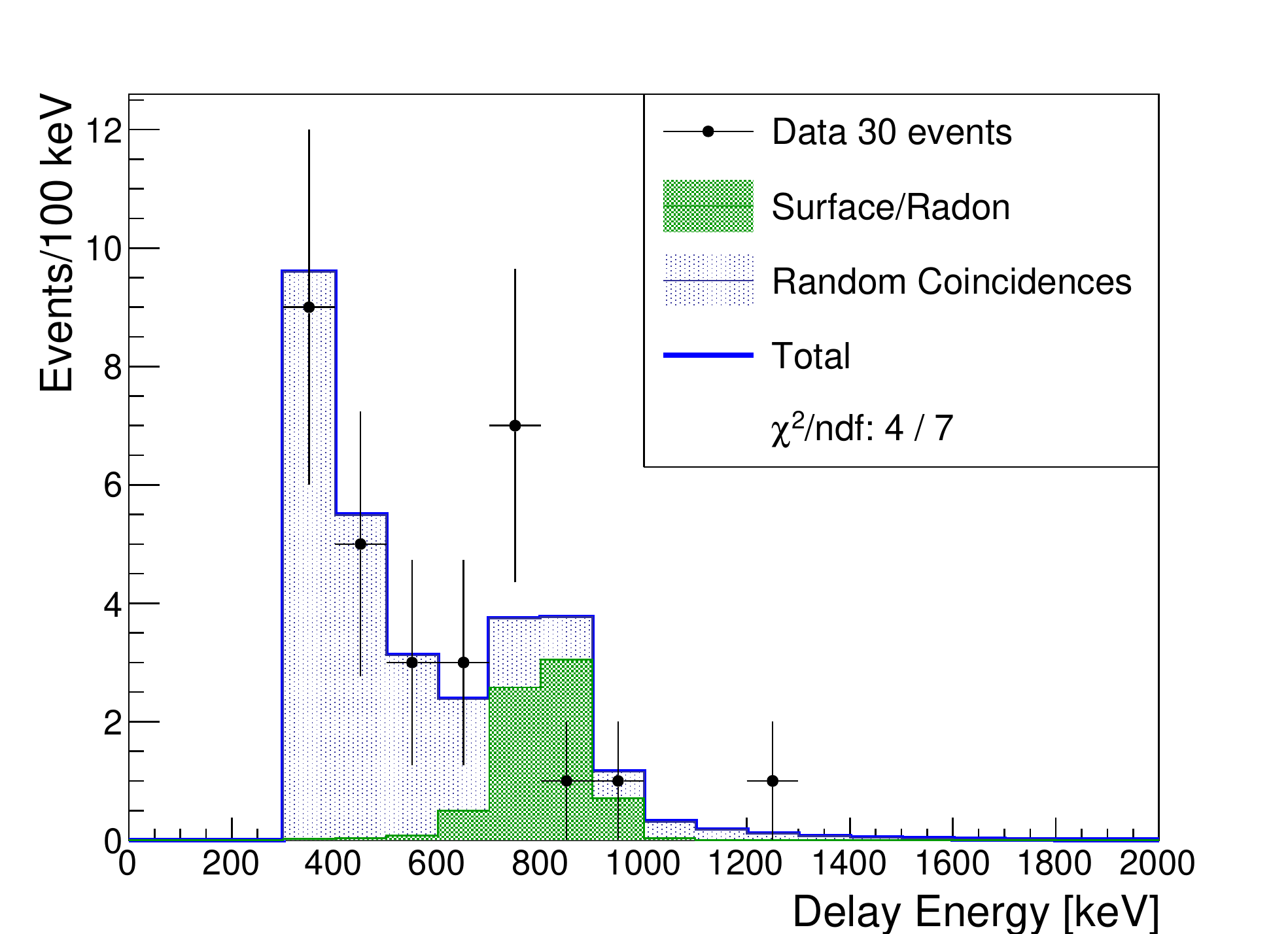}
  \includegraphics[scale=0.37]{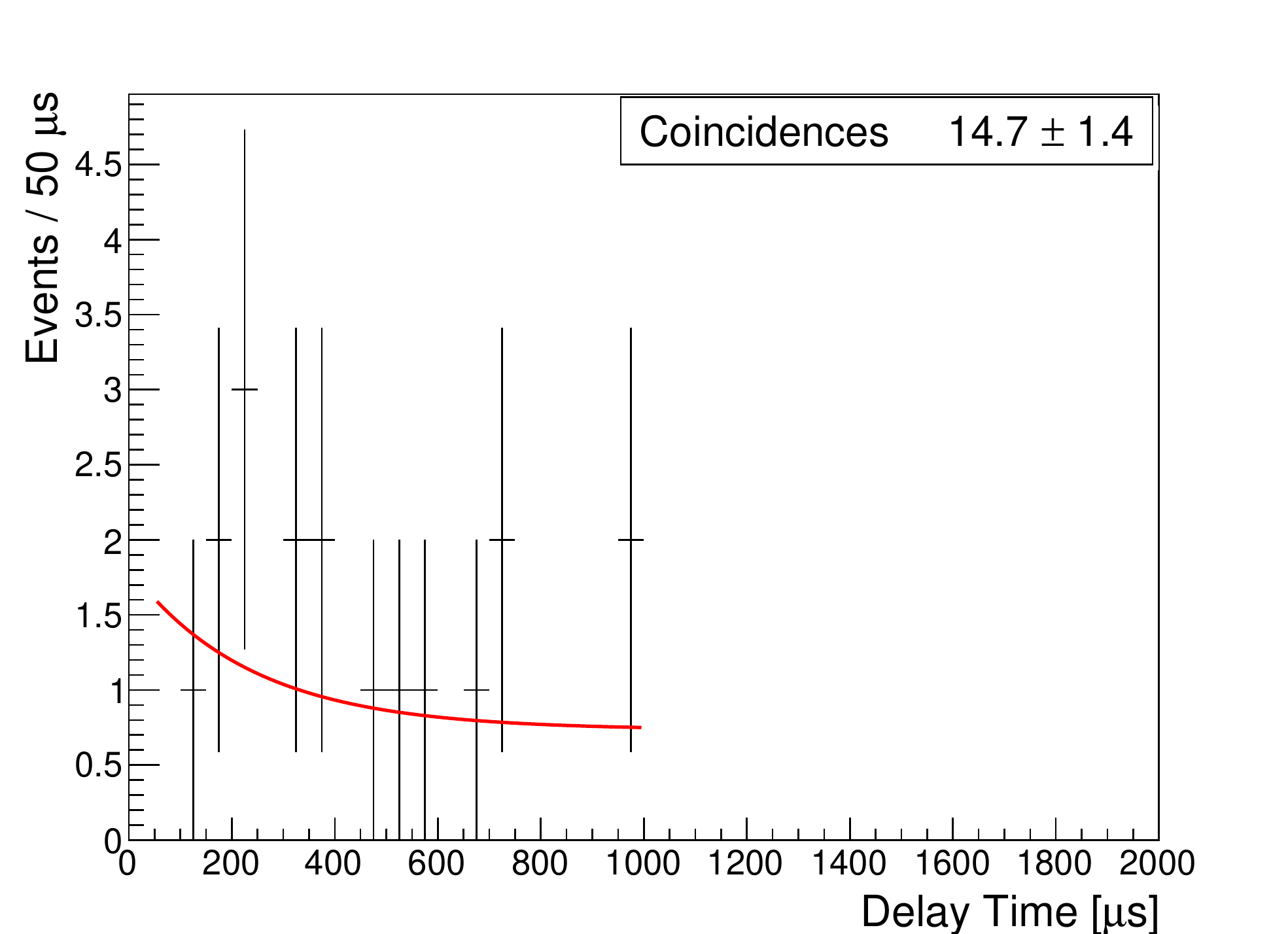}
  \includegraphics[scale=0.37]{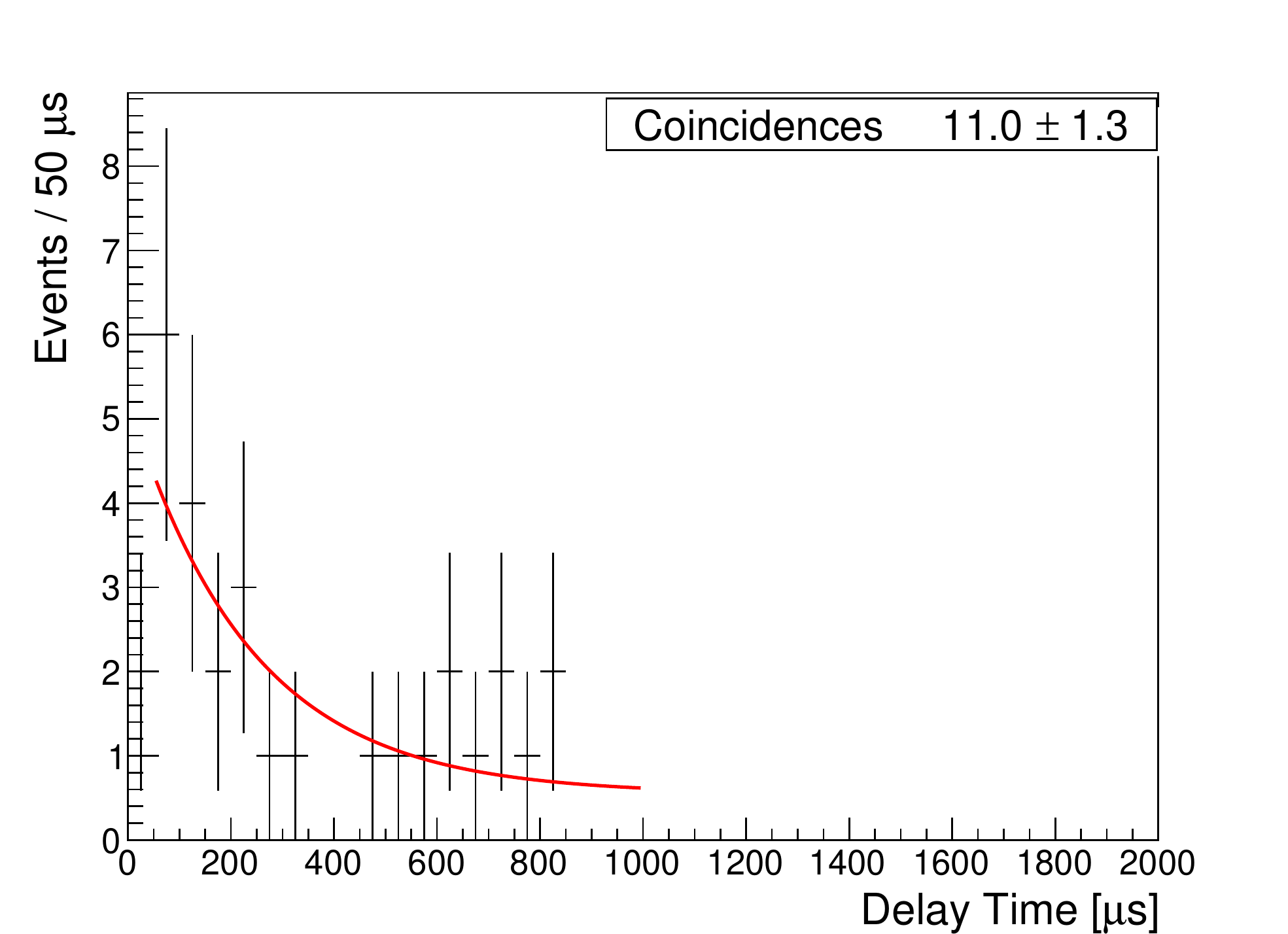}
  \caption{Distributions of the prompt energy (top), the delayed energy (middle) and the delay time (bottom), for the $^{214}$BiPo background measurements performed with the final shield. Left figures show the results for Module 1 with 36.2 days of measurement and an active scintillator surface area of 3.06~m$^2$. Right figures show the results for Module 2 with 75.7 days of measurement and an active scintillator surface area of 3.42~m$^2$. The data are fitted by the expected background from the $^{214}$Bi contamination on the surface of the scintillators (green histogram) and from the random coincidences (blue histogram). The delay time distribution is fitted a posteriori by an exponential decay with an half-life set to the value of the $^{214}$Po decay half-life (164~$\mu$s) plus a constant value for a flat random coincidence distribution.}
  \label{fig:bkg-214}
\end{figure}

The results of the $^{214}$BiPo background measurements are presented in Table~\ref{tab:bkg-bipo214}. 
The expected rate of random coincidences, calculated by measuring the single rate, is 0.13~counts/day/m$^2$ of scintillator surface area for Module 1 with the final shield, and 0.10~counts/day/m$^2$ for Module 2. 
The energy distributions of prompt and delayed signals and the delay time between the two signals are presented in Figure~\ref{fig:bkg-214}, separately for the two distinct sets of background measurements with the final shield. The contribution of the two background components (random coincidences and surface background) are fitted to the delayed energy spectra. 
As presented in  Table~\ref{tab:bkg-bipo214}, the number of random coincidences estimated by the fit is in agreement with the expected rate calculated from the single rates. It demonstrates the reliability both of the fit and of the estimated $^{214}$Bi on the surface of the scintillators. 
The levels of surface background measured separately in the two BiPo-3 modules with the final shield are equal within the statistical uncertainty. 
A higher surface background was observed with the temporary shield, due to a poor tightness against external radon. Therefore this measurement is not taken into account for the estimation of the detector surface background. 
Adding the two distinct sets of background measurements of the two modules with the final shield (corresponding to 111.9 days of data collection and a scintillator surface area of 3.24~m$^2$), the fitted $^{214}$Bi activity on the surface of the scintillators is $\mathcal{A}(^{214}\mathrm{Bi})=1.0 \pm 0.4$~$\mathrm{\mu}$Bq/m$^2$. 
It is worth mentioning that the background remains stable, within the statistical uncertainty, during the measurement of enriched $^{82}$Se foils placed in half of Module 1. 

\begin{table}[htb]
\centering
\begin{tabular}{c||c|c|c||c||c}
  &  Module 1   & Module 1  & Module 2  & {\bf Combined}   & Module 1       \\
  &  Temporary  & Final     & Final     & {\bf Final   }   & $^{82}$Se foil  \\
  &  shield     & shield    & shield    & {\bf shield  }   &  Run           \\
\hline
\hline
Duration (days)            & 73.5 & 36.2  & 75.7  & 111.9 & 91.9    \\
\hline
Scint. surf. area  (m$^2$) & 2.7 & 3.06  & 3.42  & 3.24  & 1.26    \\
\hline
Data events                & 27  & 18    & 30    & 48    & 17      \\ 
\hline
Surf. Bkg. (fit)           & 11.7$\pm$3.8 & 2.5$\pm$2.3   & 6.9$\pm$3.4   & 9.4$\pm$4.1   &  2.5$\pm$2.3  \\
Coinc. (fit)               & 15.3$\pm$4.3 & 15.5$\pm$4.2  & 23.1$\pm$5.3  & 38.5$\pm$7.6  & 14.5$\pm$4.0   \\
\hline
Coinc. (single rate)       & 18.2 & 14.3  & 25.0  & 39.3  & 12.9  \\ 
\hline
\hline
$\mathcal{A}(^{214}\mathrm{Bi})$ $\mathrm{\mu}$Bq/m$^2$ & $2.5 \pm 0.8$  & $1.0 \pm 0.9$  & $1.0 \pm 0.5$  & {\bf $1.0 \pm 0.4$}  & $0.9 \pm 0.6$ \\
\end{tabular}
\caption{Results of the $^{214}$BiPo background measurements: separate results of the three dedicated background measurements, combined result of the two dedicated background measurements with final shield, and results of an additional background measurement carried out during the measurement of the first enriched $^{82}$Se foils (Module 1). We report the duration of the measurement, the active scintillator surface area, the number of detected $^{214}$BiPo events, the number of surface background events and random coincidences calculated by the fit of the delayed energy spectrum, the expected number of random coincidences calculated with the single rate measurement, and the activity in $^{214}$Bi on the surface of the scintillators.}
\label{tab:bkg-bipo214}
\end{table}

\subsection{Rejection of bulk background from the scintillator volume}
\label{sec:coinc-rejection}

\begin{figure}[!]
  \centering
  \includegraphics[width=1.0\textwidth]{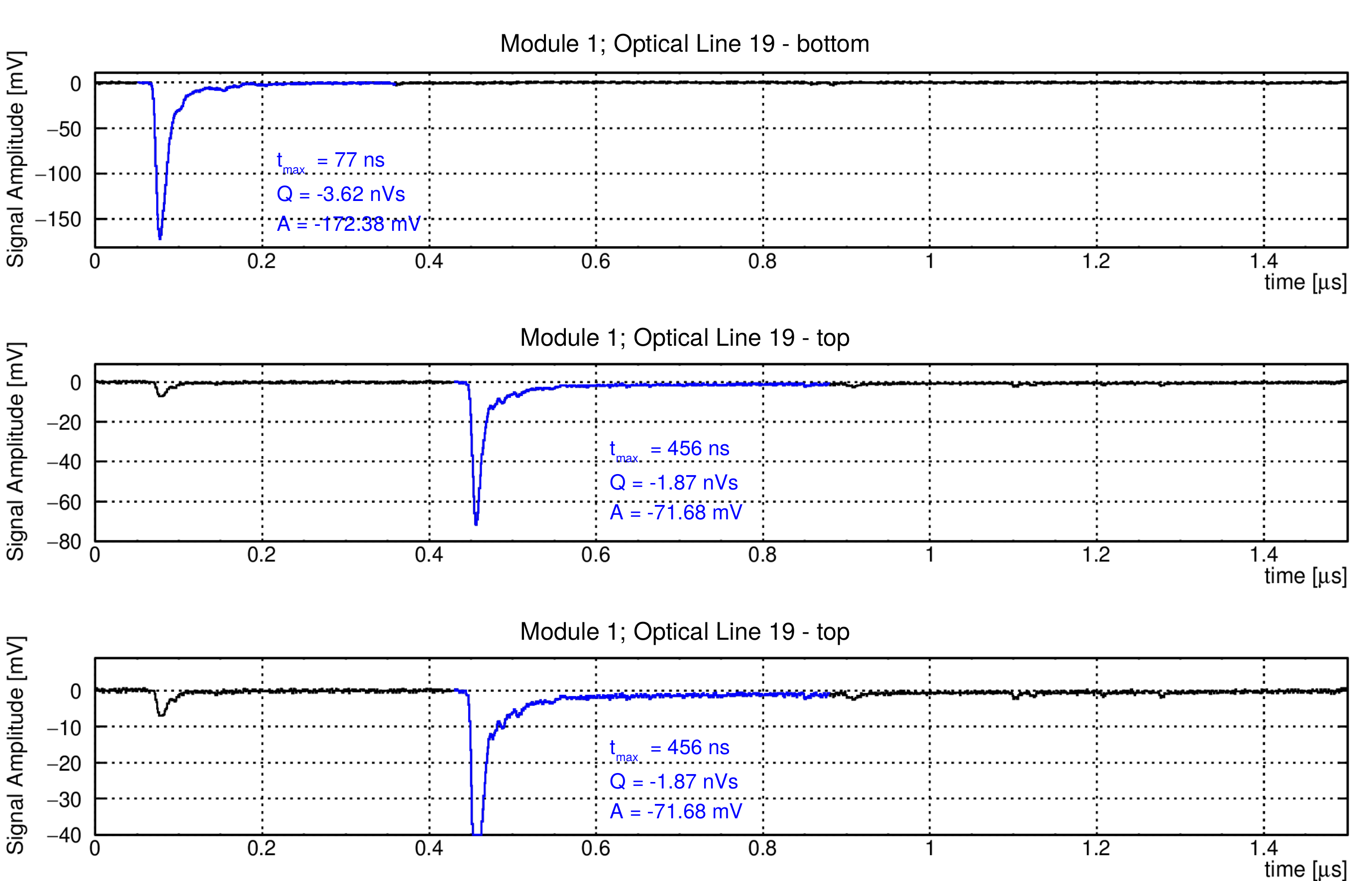}
  \caption{Display of a BiPo event identified as a bulk background event with a signal in coincidence with the prompt signal in the opposite scintillator, from top to bottom: the prompt signal, the delayed signal and a zoom of the delayed signal.}
  \label{fig:coinc-display}
\end{figure}

As discussed previously, a BiPo background event produced by a bulk contamination inside the scintillator volume can be rejected by detecting a signal in coincidence with the prompt signal in the opposite scintillator. An example of event identified as a bulk background event with a coincidence signal is displayed in Figure~\ref{fig:coinc-display}. The lower the energy threshold is to detect a signal, the higher is the background rejection efficiency. However, a possible optical cross-talk between two opposite scintillators also produces a coincidence signal, limiting the bulk background rejection. The level of optical cross-talk is measured in order to define the minimum threshold to be applied on the coincidence signals for the  bulk background rejection.

\begin{figure}[!]
  \centering
  \includegraphics[width=0.45\textwidth]{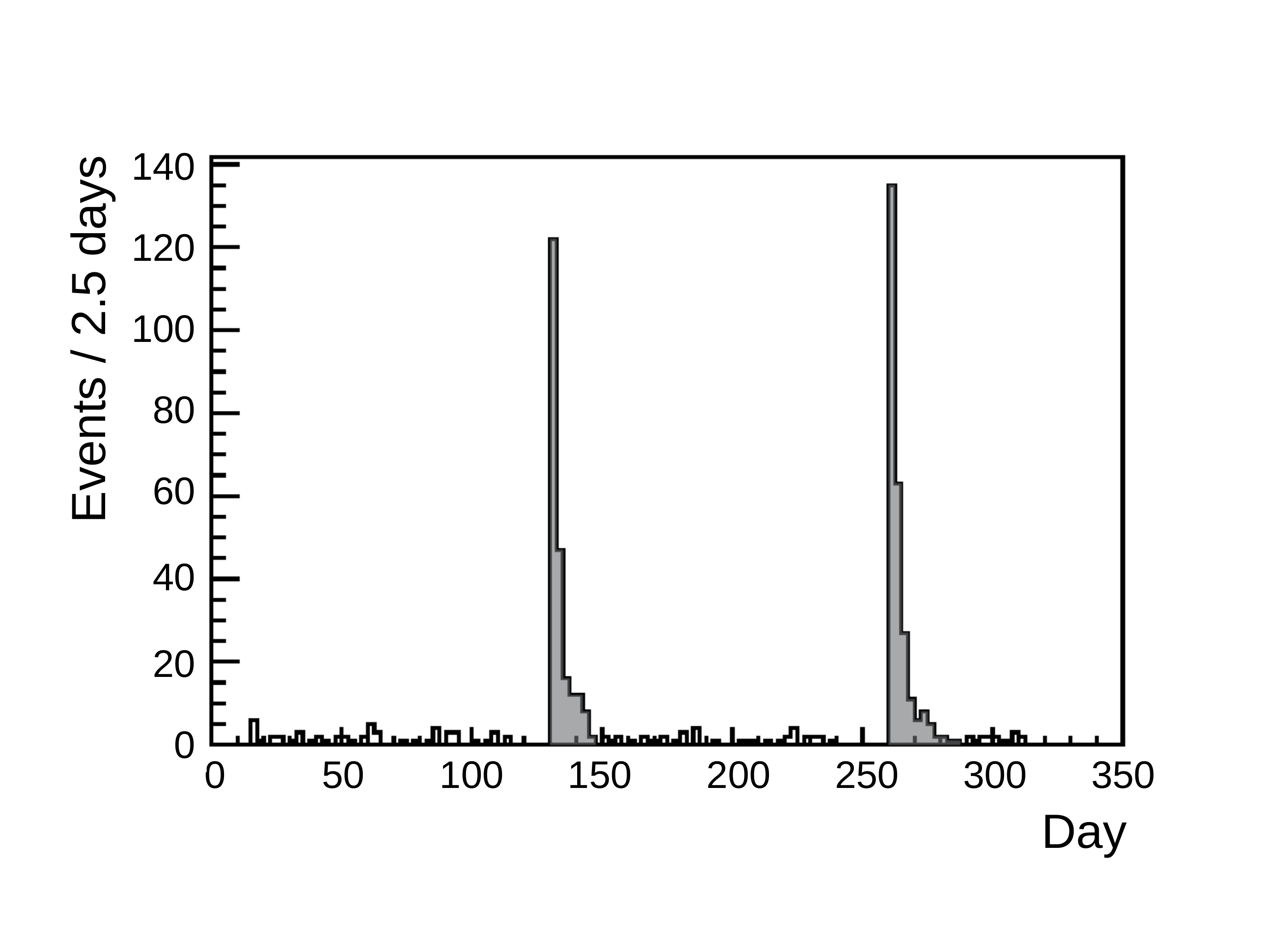}
  \includegraphics[width=0.45\textwidth]{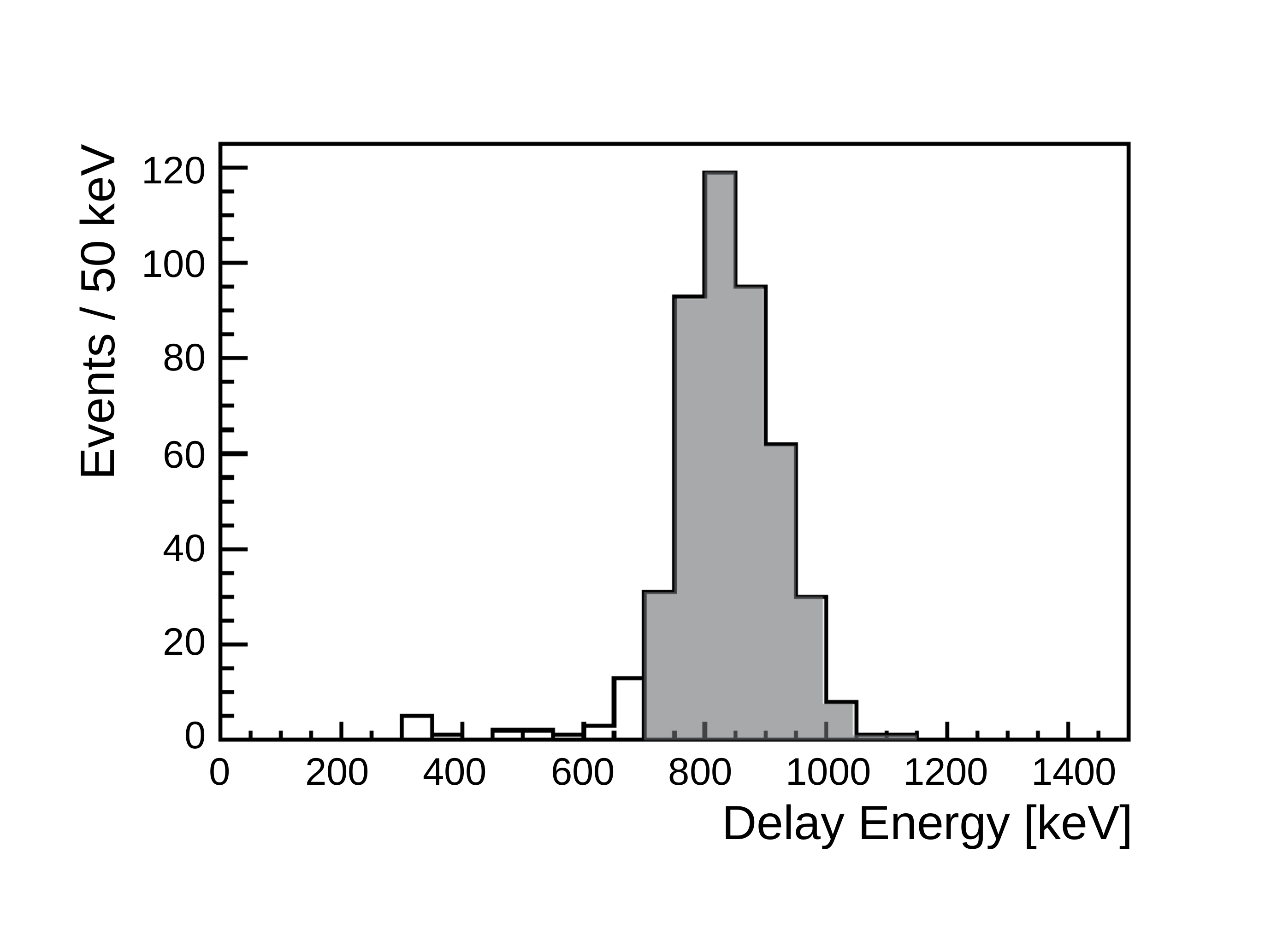}
  \includegraphics[width=0.45\textwidth]{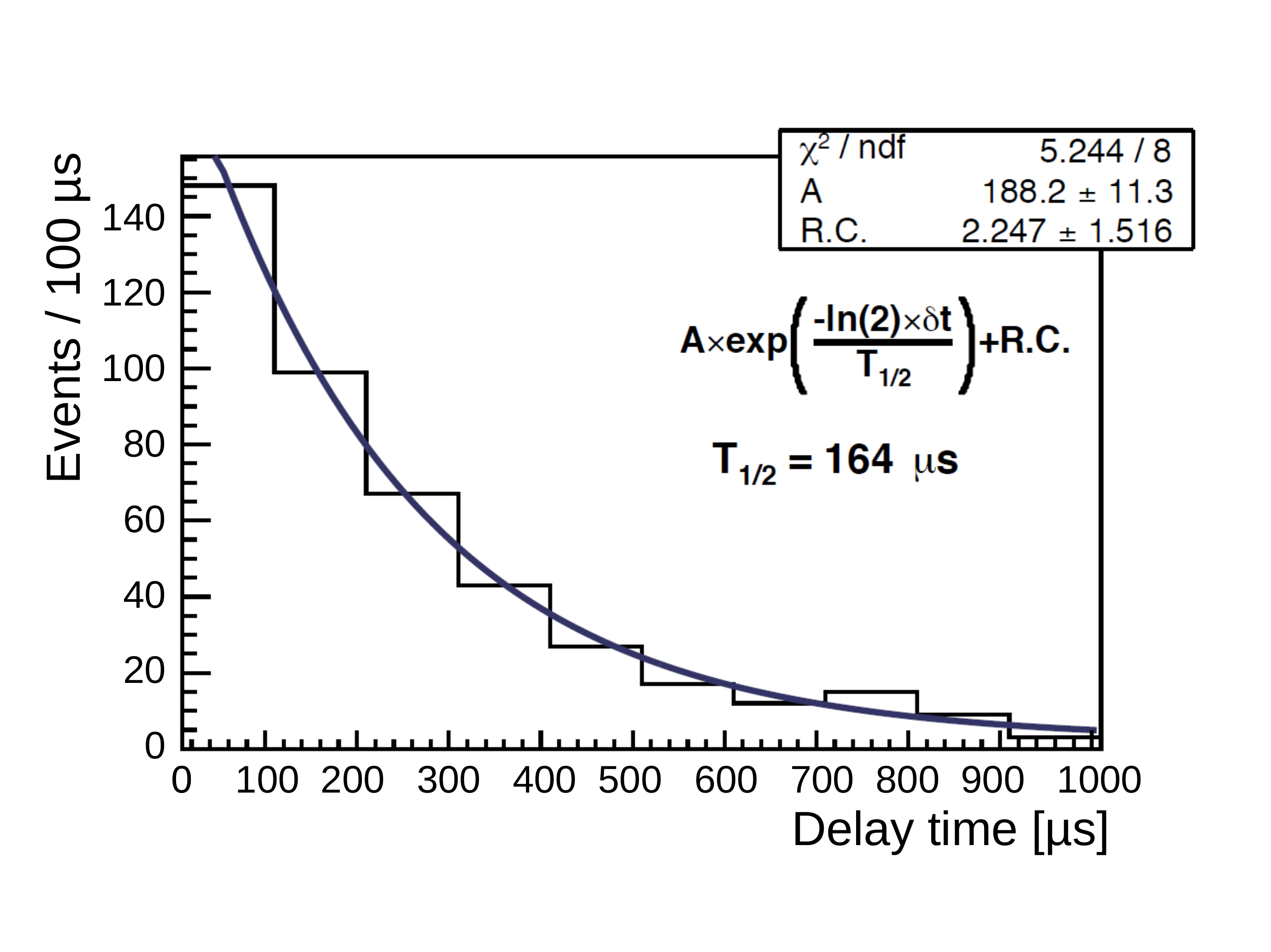}
  \includegraphics[width=0.45\textwidth]{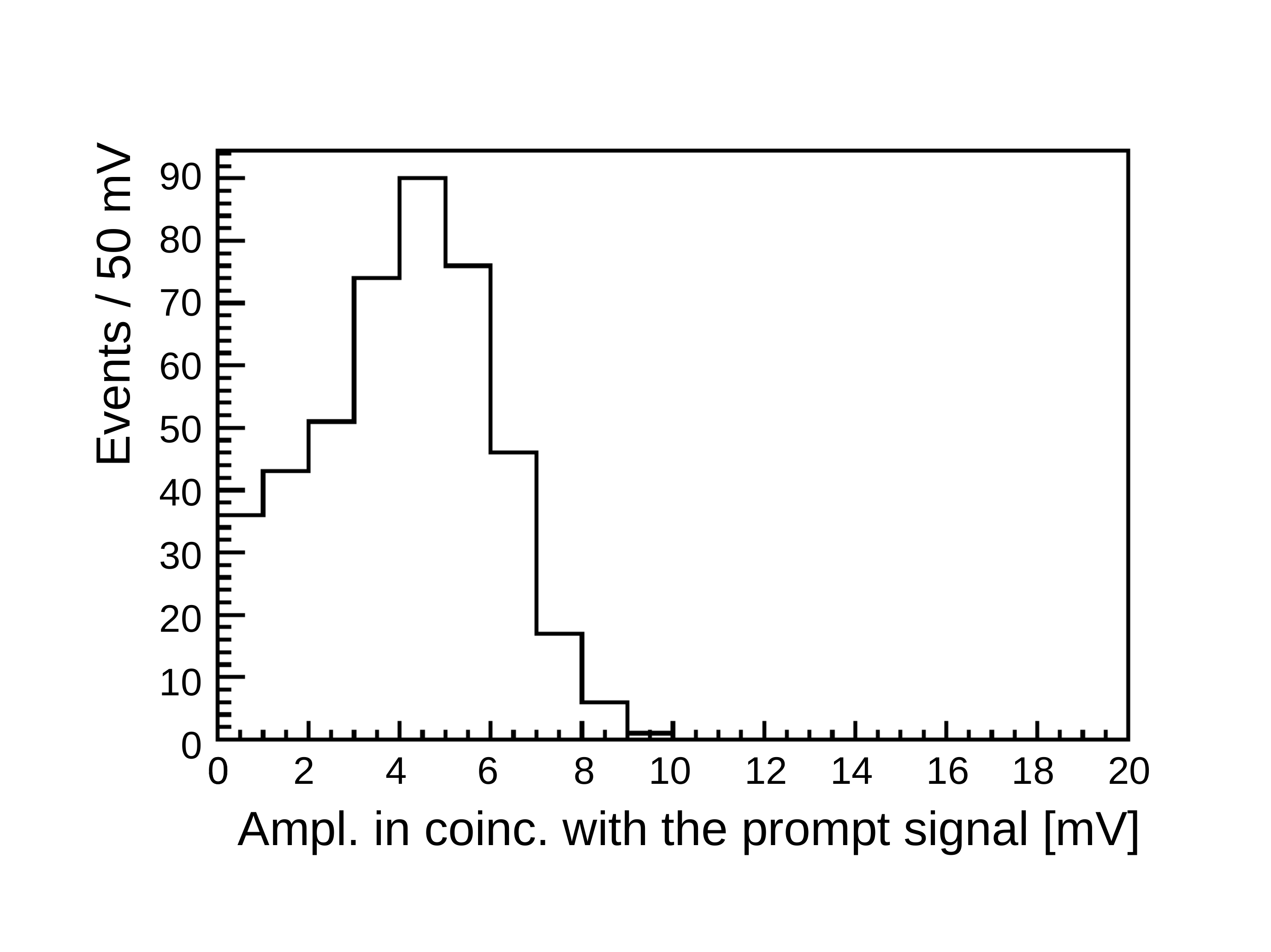}
  \caption{$^{214}$BiPo background events produced by the radon deposition on the surface of the scintillators. Top left: the radon deposition is observed only in the first 15 days after closing the detector. Top right: the energy of the delayed signal is located around 800~keV, as expected for a $^{214}$BiPo event on the surface of the scintillator. Bottom left: the distribution of the delay time between the prompt and the delayed signal is in agreement with $^{214}$Po half-life of 164~$\mathrm{\mu}$s. Bottom right: amplitude of the signal in coincidence with the prompt $\beta$ signal in the opposite scintillator.}
  \label{fig:bulk-bkg-1}
\end{figure}

First, we show that a coincidence signal is observed when a BiPo event is produced by a bismuth ($^{212}$Bi and $^{214}$Bi) bulk contamination, very close to the surface of the scintillator.
We use the background $^{214}$BiPo events produced by the radon deposition on the surface of the scintillators when the detector is opened in a clean room. The radon has a half-life of 3.8 days. It decays to $^{214}$Bi with two successive $\alpha$-decays and a $\beta$-decay. The recoil energy induced by the two successive $\alpha$ decays allows the nucleus to penetrate inside the plastic scintillator. The $^{214}$Bi isotope is no longer on the surface but is slightly inside the scintillator (within a few tens of micrometers). As a consequence, a small signal is expected to be observed in coincidence with the prompt $\beta$ signal for these sets of BiPo events produced by the radon deposition. To test it, the data of the first 15 days of $^{214}$BiPo background measurements (with radon deposition) have been used. 467 $^{214}$BiPo events have been selected. This set of events corresponds to a pure sample of $^{214}$BiPo events as shown in Figure~\ref{fig:bulk-bkg-1}: the energy of the delayed signal is concentrated around 800~keV since the $\alpha$ particle deposits all its energy inside the scintillator, and the distribution of the delay time between the prompt and the delayed signal is in agreement with an exponential decay time with a half-life of 164~$\mathrm{\mu}$s ($^{214}$Po). The amplitude of the signal in coincidence with the prompt signal is also presented in Figure~\ref{fig:bulk-bkg-1}. A clear coincidence is observed, with an average amplitude of about 4~mV. 
It demonstrates that the BiPo-3 detector is able to identify a bismuth bulk contamination ($^{212}$Bi and $^{214}$Bi), very close to the surface. 

\begin{figure}[!]
  \centering
  \includegraphics[scale=0.65]{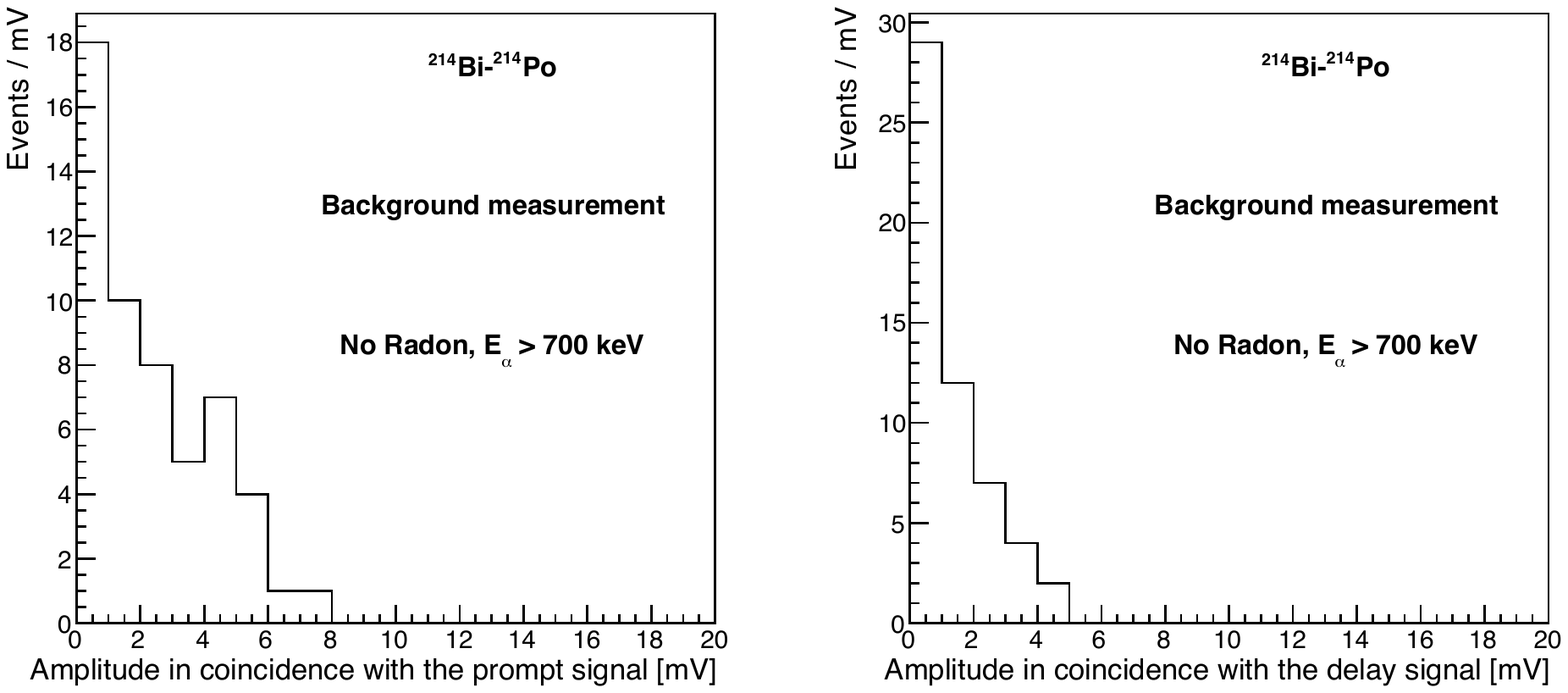}
  \vspace{-9cm}
  \caption{Amplitude of the signal in coincidence with the prompt electron signal (left) and the delayed $\alpha$ signal (right) for the 54 $^{214}$BiPo events selected from the three dedicated background measurements, after rejecting the first 15 days of data (no radon deposition) and after requiring a delayed energy greater than 700~keV (surface background events).}
  \label{fig:bulk-bkg-2}
\end{figure}

\begin{figure}[!]
  \centering
  \includegraphics[scale=0.3]{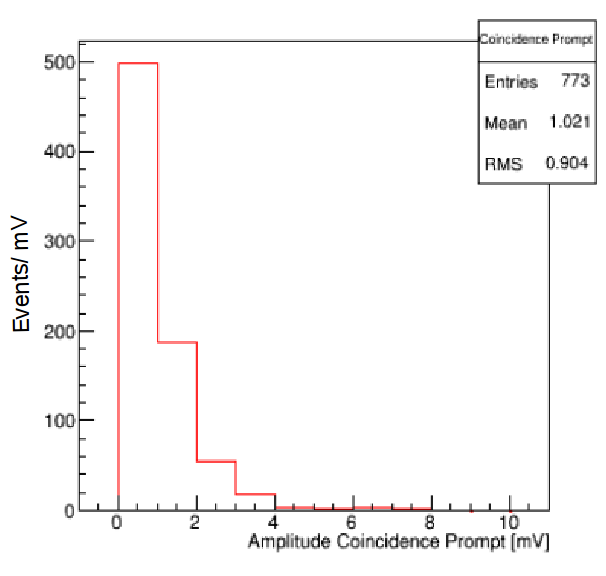}
  \caption{Amplitude of the signal in coincidence with the prompt electron signal for the $^{212}$BiPo events observed in the calibrated aluminium foil. The coincidence amplitude, due to optical cross-talk and noise, is greater than 3~mV for only 4\% of the $^{212}$BiPo events.}
  \label{fig:bulk-bkg-3}
\end{figure}

We then measure the optical cross-talk, using two separate analyses. 
The first analysis is performed with a pure $^{214}$BiPo event sample, produced by the $^{214}$Bi contamination on the surface of the scintillators and selected from the three background measurements. In this case, no signal is expected in coincidence with the delayed $\alpha$ signal.
This sample of events is a sub-sample of the $^{214}$BiPo background events selected in section~\ref{sec:bkg-214}, except that two criteria are modified: the energy of the delayed signal is required to be greater than 700~keV in order to select only the surface background events while rejecting the random coincidences and external radon background, and no criteria are applied on the amplitude in coincidence with the prompt or delayed signal. A total of 54 $^{214}$BiPo events are selected. 
The amplitude distributions of the signals in coincidence either with the prompt $\beta$ signal or with the delayed $\alpha$ signal are presented in Figure~\ref{fig:bulk-bkg-2}. For about 30\% of the  surface background events, a signal above 3~mV is observed in coincidence with the prompt electron signal. 
However, for only 10\%  of the events, a signal above 3~mV is observed in coincidence with the delayed $\alpha$ signal. This coincidence signal corresponds to an optical cross-talk.  
In order to measure more precisely the cross-talk with higher statistics, a second analysis is performed with a pure sample of $^{212}$BiPo events, produced by a calibrated aluminium foil which is used to validate the BiPo measurement, as discussed in section~\ref{sec:alu}. 
A total of 773 $^{212}$BiPo events produced in the aluminium foil are selected. No signal in coincidence with the prompt $\beta$ signal is expected for this set of $^{212}$BiPo events. 
The amplitude distribution of the signal in coincidence with the prompt $\beta$ signal is presented in Figure~\ref{fig:bulk-bkg-3}. 
The coincidence amplitude, due to optical cross-talk and noise, is greater than 3~mV for only 4\% of the $^{212}$BiPo events.
In other words, the detection efficiency of the BiPo-3 detector is reduced by 4\% when the criterion to reject the bulk background is applied (requiring a coincidence amplitude lower than 3~mV).

\section{Validation of the BiPo-3 measurement with a calibrated aluminium foil}
\label{sec:alu}

A 88~$\mathrm{\mu}$m thick calibrated aluminium foil has been measured in the BiPo-3 detector in order to validate the Monte Carlo simulations and to validate the detection efficiency. 
Part of this aluminium foil have been measured by HPGe $\gamma$ spectrometry during 21 days. The measured $^{208}$Tl and $^{214}$Bi activities are:

\begin{eqnarray}
\nonumber \mathcal{A}(^{208}\mathrm{Tl}) & = & 83 \pm 4 \ \mathrm{(stat.)} \pm 15 \ \mathrm{(syst.)} \ \mathrm{mBq/kg}\\
\nonumber \mathcal{A}(^{214}\mathrm{Bi}) & = & 9.2 \pm 3.3 \ \mathrm{(stat.)} \pm 0.9 \ \mathrm{(syst.)} \ \mathrm{mBq/kg}
\end{eqnarray}

The foil was measured successively in the two BiPo-3 modules.
For each module, a unique foil is placed in one half of the module, while two overlaid foils (with a total thickness of 176~$\mathrm{\mu}$m) are placed in the second half, in order to measure separately two different thicknesses. 
The aluminium foils are wrapped in a 4~$\mathrm{\mu}$m thick radiopure polyethylene film in order to prevent  scintillator surface contamination.

\subsection{$^{208}$Tl activity measurement}
\label{sec:alu-tl208}

\begin{figure}[!]
  \centering
  \includegraphics[scale=0.35]{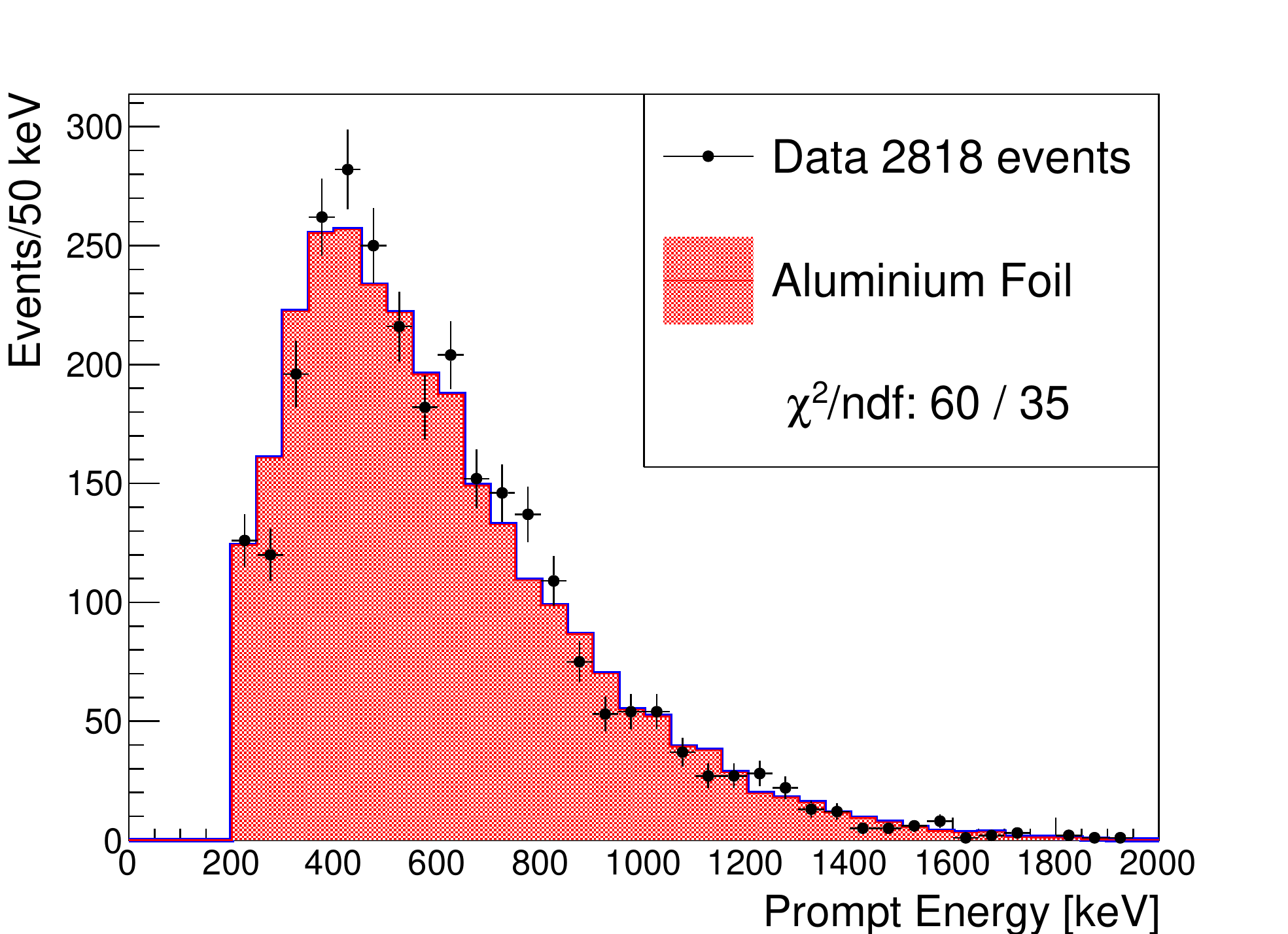}
  \includegraphics[scale=0.35]{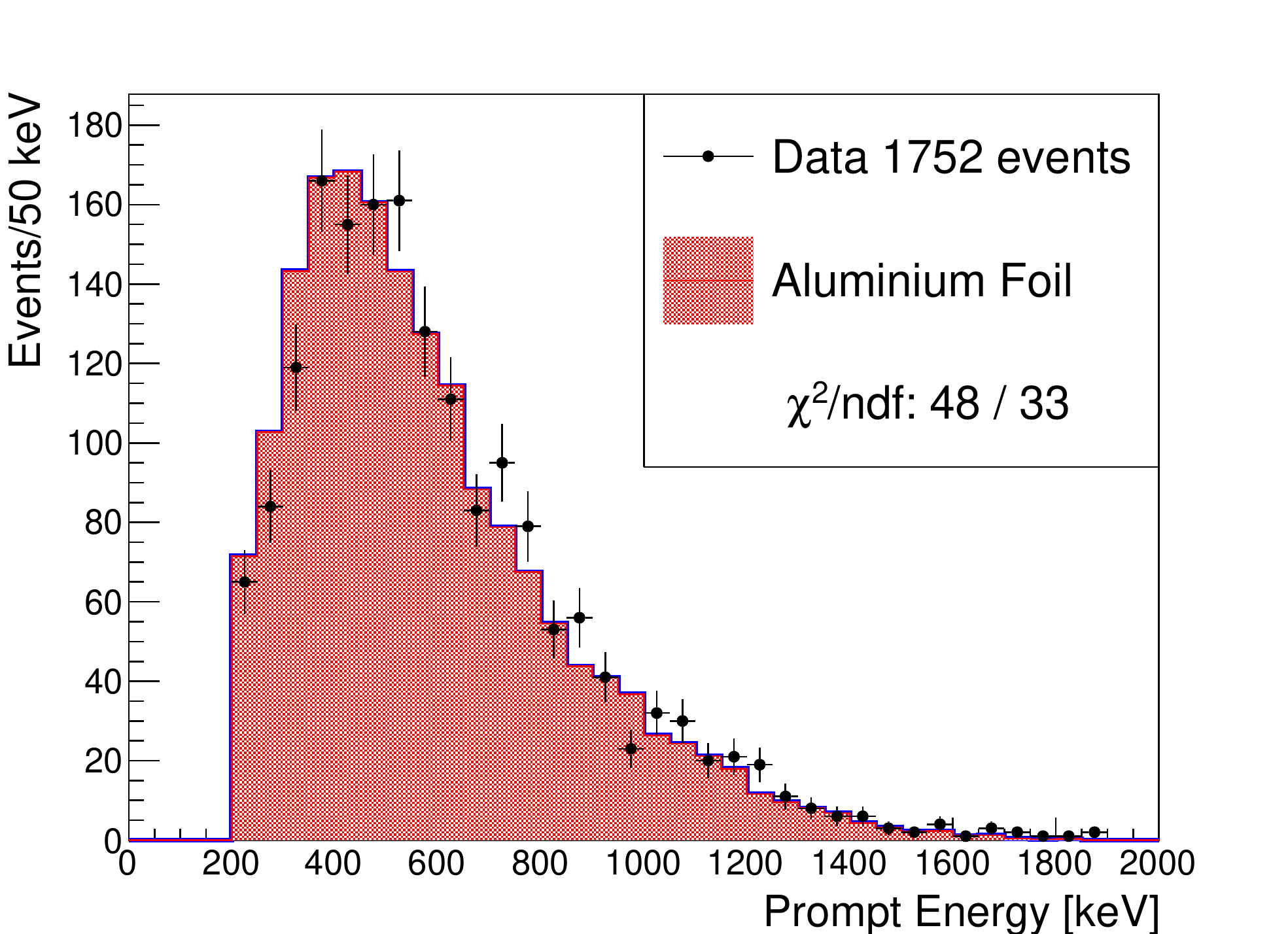}
  \includegraphics[scale=0.35]{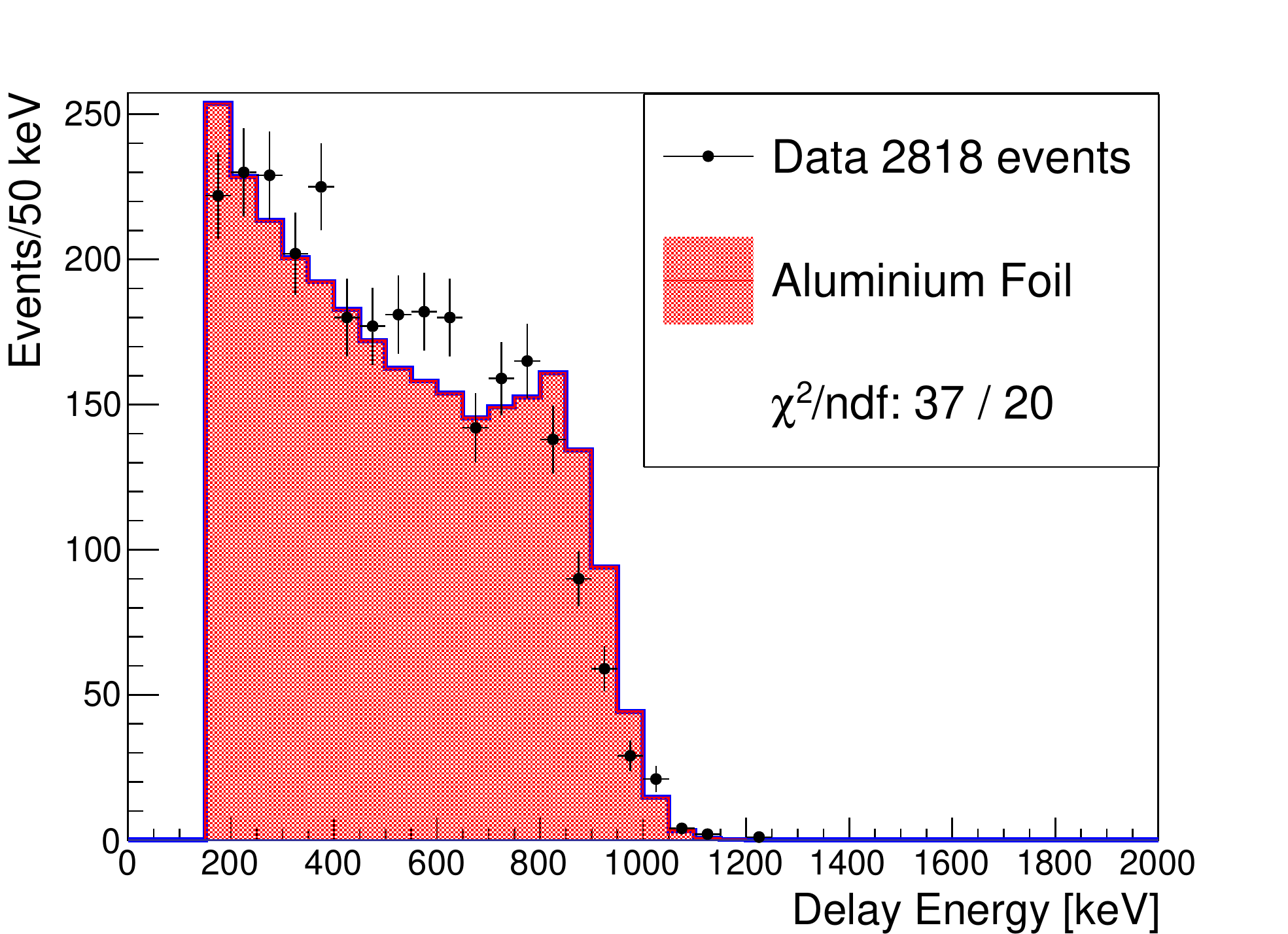}
  \includegraphics[scale=0.35]{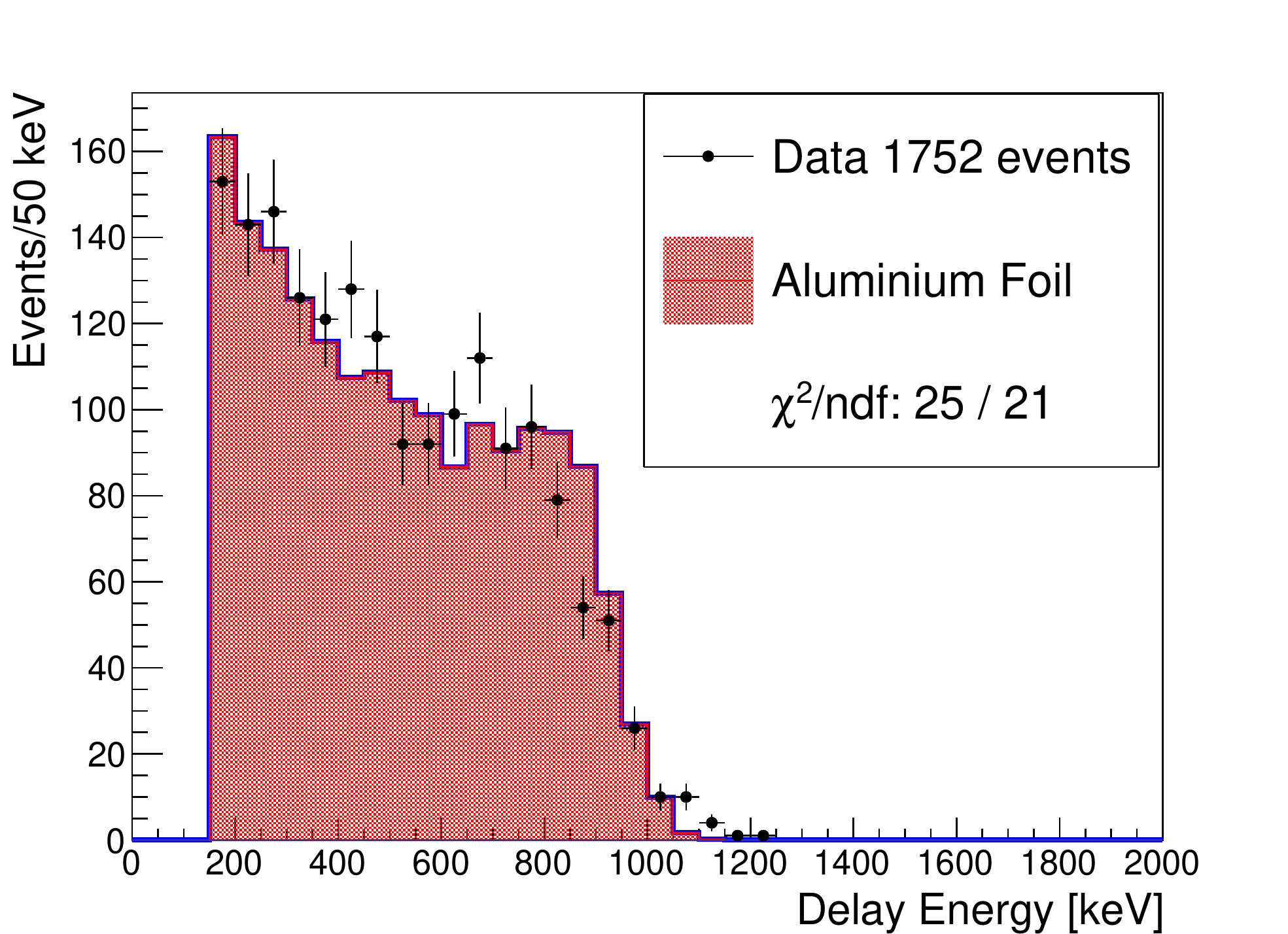}
  \includegraphics[scale=0.35]{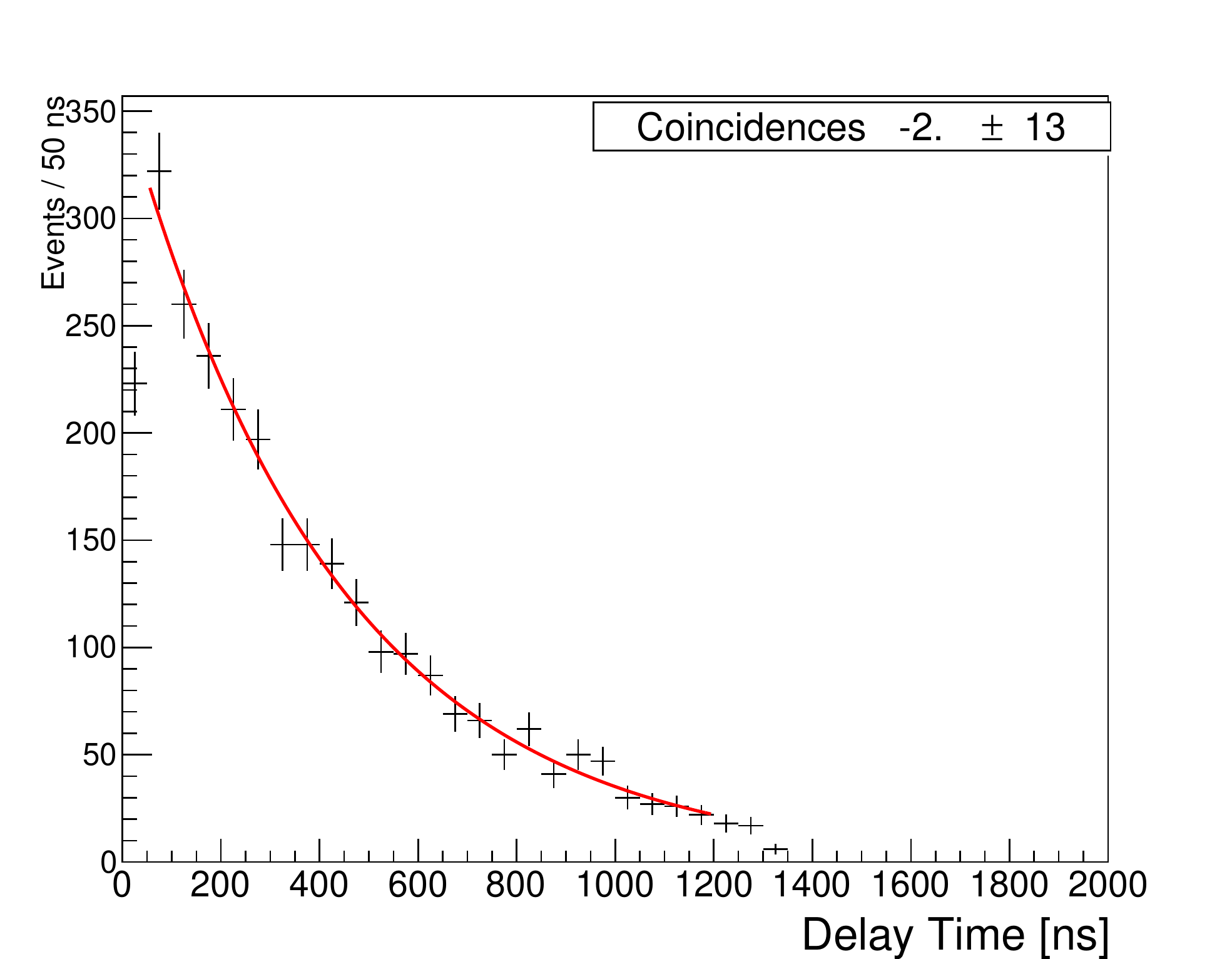}
  \includegraphics[scale=0.35]{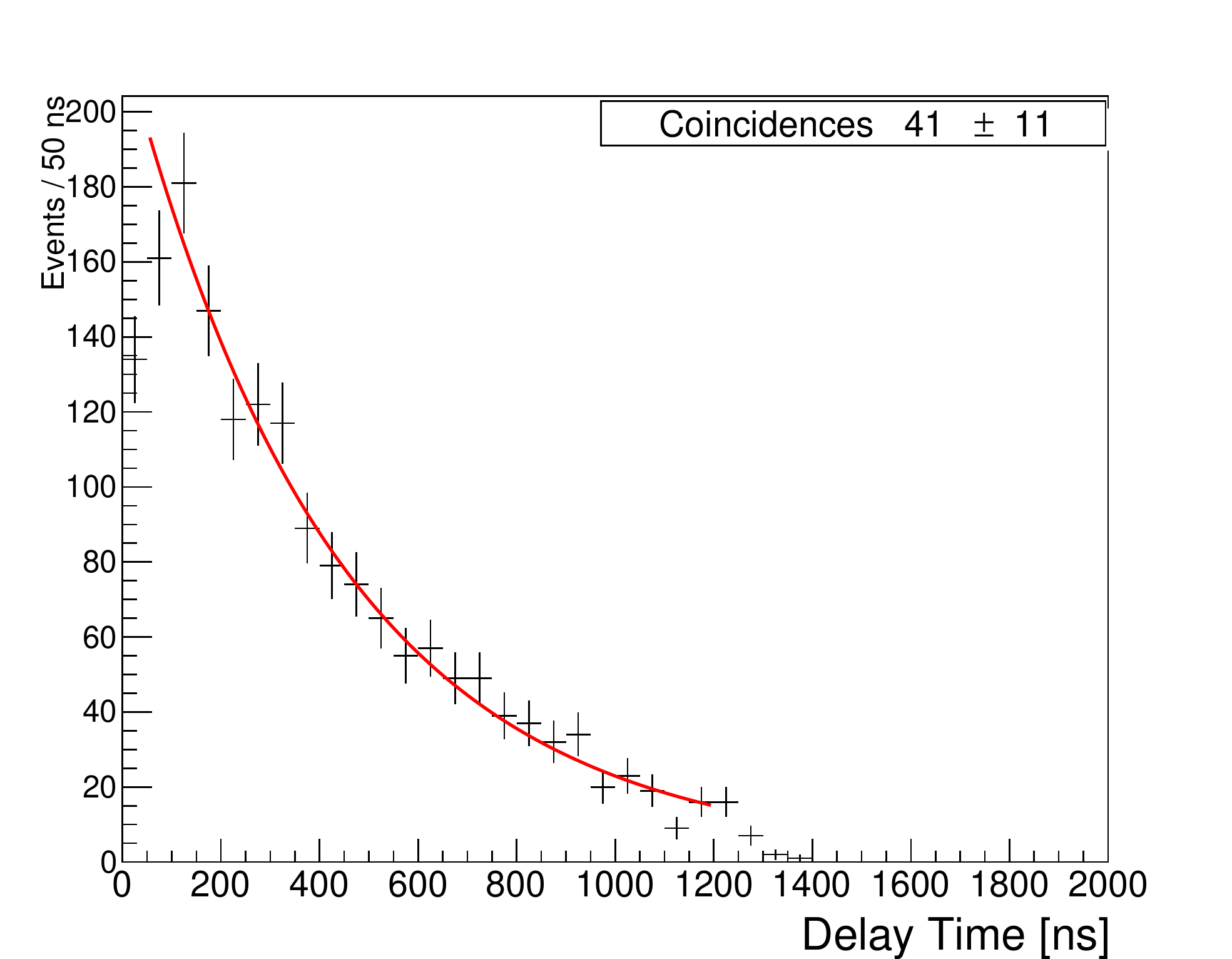}
  \caption{$^{212}$BiPo measurement of the calibrated aluminium foil with Module 1, for a single foil (left plots) and for two overlaid foils (right plots): energy of the prompt electron signal (upper plots), energy of the delayed $\alpha$ signal (middle plots), and delay time (lower plots).}
  \label{fig:alu_212_onefoil}
\end{figure}

The criteria to select the {\it back-to-back} $^{212}$BiPo events are identical to those used for the background measurement (listed in section~\ref{sec:event-selection}).
The duration of measurement, the mass exposure and the number of selected events are given in Table~\ref{tab:alu-212}.
The expected background is negligible for the  $^{212}$BiPo measurement and only $^{212}$Bi contamination inside the aluminium foil is considered. We assume a uniform bulk contamination in the foil.
The detection efficiency, calculated by Monte Carlo simulation, is 5.2$\pm$0.5 \% for a single foil, and 2.3$\pm$0.2 \% for two overlaid foils.
The simulated energy distribution of the delayed $\alpha$ particle is fitted to the observed data, using the likelihood method. Result of the fit is presented in Figure~\ref{fig:alu_212_onefoil}, showing the energy spectra of the prompt and delayed signal. Simulation is in fair agreement with data. The small difference observed for the energy of the $\alpha$ particle comes from the uncertainty on the scintillation quenching factor measurement. The delay time distribution is fitted by an exponential decay with a half-life equal to the $^{212}$Po half-life of 300~ns. 
The observed delay time distribution is in agreement with the expected $^{212}$Po decay time distribution.
The $^{208}$Tl foil activities, measured for one and two foils and successively for the two modules, are summarized in Table~\ref{tab:alu-212}. 

\begin{table}[htb]
\centering
\begin{tabular}{c|c|c|c|c|c}
          & $T_{obs}$ & Mass  &$N_{obs}$ & $\epsilon$ & $\mathcal{A}(^{208}\mathrm{Tl})$ \\
          & (days)   & (g)   &                     & (\%)       & (mBq/kg)                        \\
\hline
\hline
Module 1  &  25.0    &       &                      &            &                                  \\
One foil  &          & 192.6 &  2818                &  5.2 $\pm$ 0.5       &  $74.3 \pm 1.4 \mathrm{(stat.)} \pm 8.9\mathrm{(syst.)}$   \\
Two foils &          & 299.6 &  1752                &  2.3 $\pm$ 0.2     &  $66.2 \pm 1.7 \mathrm{(stat.)} \pm 8.5\mathrm{(syst.)}$   \\
\hline
Module 2  &  40.0    &       &                       &            &                                  \\
One foil  &          & 173.3 &  3919                &  5.2 $\pm$ 0.5       &  $70.8 \pm 1.1 \mathrm{(stat.)} \pm 8.5\mathrm{(syst.)}$   \\
Two foils &          & 330.3   &  3071                &  2.3 $\pm$ 0.2       &  $65.8 \pm 1.2 \mathrm{(stat.)} \pm 7.9\mathrm{(syst.)}$   \\
\end{tabular}
\caption{$^{208}$Tl activities in the aluminium foils measured successively with the two BiPo-3 modules. $T_{obs}$ is the duration of the measurement, $\epsilon$ is the detection efficiency, $N_{obs}$ is the number of observed $^{212}$BiPo events and $\mathcal{A}(^{208}\mathrm{Tl})$ is the corresponding $^{208}$Tl  activity of the foil.}
\label{tab:alu-212}
\end{table}

The different contributions to the systematic uncertainty on the reconstructed activity are estimated as follows. Main contribution comes from the associated uncertainty to the detection efficiency. As it has been discussed in section \ref{sec:detection-efficiency}, this comprises the systematics associated to the uncertainties on the energy calibration, the $\alpha$ particles quenching factor, the registered signals with long delay, the light collection efficency, the noise rejection and the thickness of the sample. When there is a sample between the scintillators, the computed detection efficiency has an associated systematic of 10.2 \%. Moreover, in this case the activity measurement has been carried out for several values of the  energy threshold for the prompt signal (from 200 to 600~keV) and for the delayed signal (from 100 to 600~keV) in order to estimate the contribution of the simulated energy spectrum. An additional systematic uncertainty of 6\% has been estimated, which is correlated to that associated to the energy calibration. Finally, a 6\% systematic uncertainty between the two BiPo-3 modules has been estimated from the difference between the different activity measurements. All these components add up to a total systematic uncertainty of 12\%. 

The final result of the measured $^{208}$Tl activity, combining the two foil geometries (single and two overlaid foils) and the two BiPo-3 modules, is:

\begin{eqnarray}
\nonumber \mathcal{A}(^{208}\mathrm{Tl}) & = & 70.5 \pm 0.7 \ \mathrm{(stat.)} \pm 8.5 \ \mathrm{(syst.)} \ \mathrm{mBq/kg}
\end{eqnarray}

The $^{208}$Tl activity measurement is in agreement, within uncertainties, with the activity reported by the HPGe measurement.

\begin{figure}[!]
  \centering
  \includegraphics[scale=0.4]{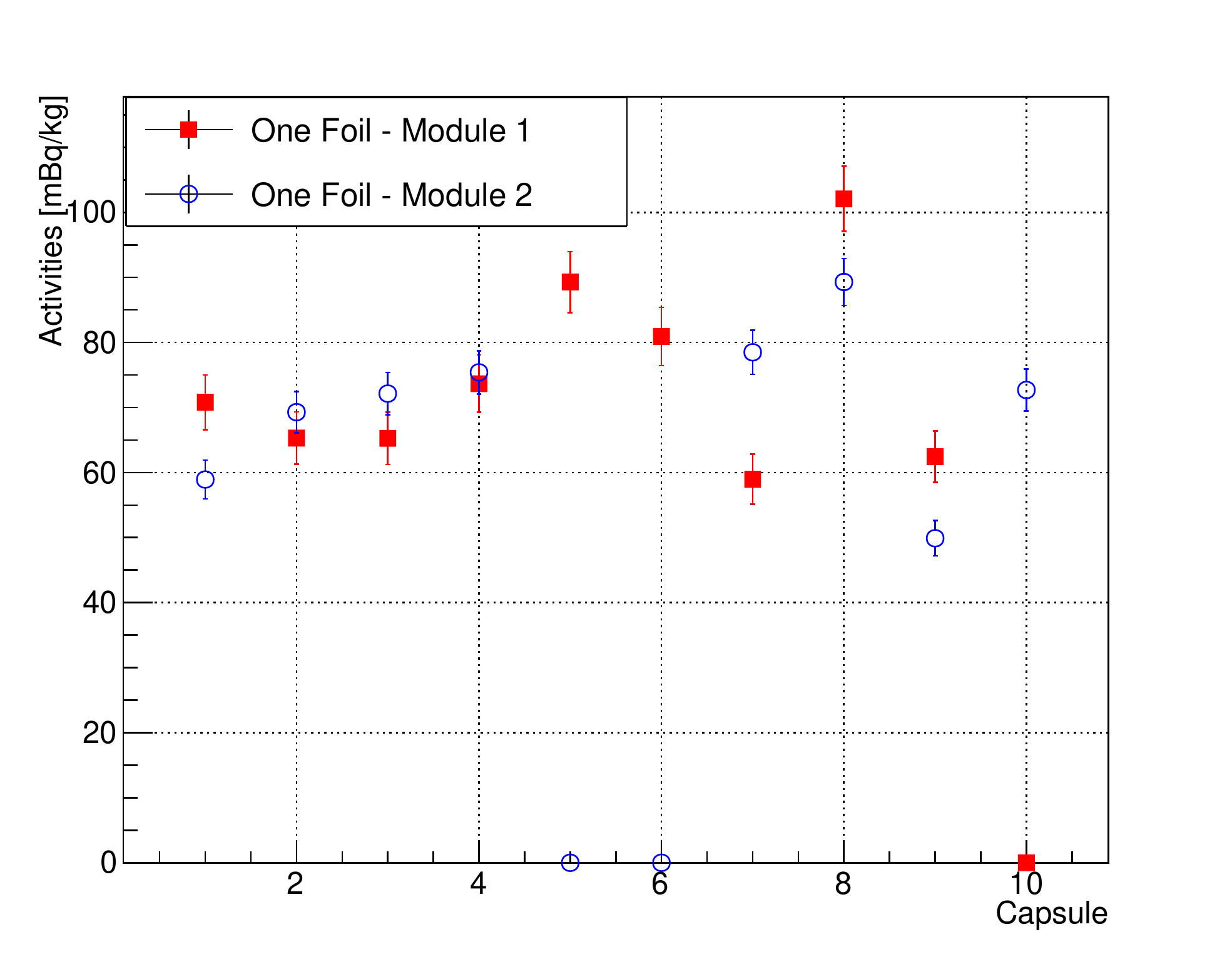}
  \caption{Comparison of the $^{208}$Tl activity measured on individual pairs (capsules) of optical sub-modules, in the two BiPo-3 modules, for the one-foil sample. Modules showing zero activity are due to different hardware problems during the data taking and have not been considered for the analysis.}
  \label{fig:alu_212_submodule}
\end{figure}

The sample measured by the HPGe corresponds to one of the ends of the full foil measured with BiPo and it corresponds to about 1/10 of the full foil. 
In order to compare exactly the same sample and to test the homogeneity of the result, activity measurements have been carried out on individual pairs of optical sub-modules. Figure~\ref{fig:alu_212_submodule} displays the comparison of the activities in the two modules for the one-foil sample. The activities obtained with capsules 1 or 10, at the end of the foil, are well within the combined result for the full foil and are in agreement with the HPGe measurement.
Inhomogeneities in the measured activities are observed in different pairs of sub-modules. Same variations appear in the two BiPo-3 modules for the same locations on the foil and similar result is obtained with the measurement of the two-foils sample. It seems that non homogeneous $^{232}$Th contamination exists in the aluminium foil.

\subsection{$^{214}$Bi activity measurement}
\label{sec:foil-214}

The criteria to select the {\it back-to-back} $^{214}$BiPo events are identical to those used for the background measurement (listed in section~\ref{sec:event-selection}), except for the energy thresholds. 
The random coincidence background becomes relatively high for the $^{214}$BiPo measurement because of the activity of the aluminium foil itself. In contrast, the surface background is negligible. In order to reduce the random coincidences, the energy threshold of the prompt signal is increased to 300~keV, while the energy threshold of the delayed signal is set to 150~keV in order to study the detection efficiency for $\alpha$ particles at low energy.  
With the two overlaid foils sample, the rate of random coincidences is too high.
Therefore, we present here the results for the single foil sample only. 

The duration of measurement, the mass exposure and the number of selected events are given in Table~\ref{tab:alu-214}.
The detection efficiency, calculated by Monte Carlo simulation, is 2.7$\pm$0.3\%.

\begin{table}[htb]
\centering
\begin{tabular}{c|c|c|c|c|c|c|c}
 & $T_{obs}$ & Mass & $\epsilon$ &$N_{obs}$  & \multicolumn{2}{c}{$N_{fit}$} & $\mathcal{A}(^{214}\mathrm{Bi})$ \\
 & (days)   & (g)  & (\%)       &         & Foil    & RC        & (mBq/kg) \\
\hline
\hline
Mod. 1 &  11.6  & 192.6 &  2.7 $\pm$ 0.3 &  250 & 62.0 & 185.6  &  $11.9 \pm 2.8 \mathrm{(stat.)} \pm 1.4 \mathrm{(syst.)}$  \\
\hline
Mod. 2 &  25.8  & 186.7   &  2.7 $\pm$ 0.3 &  477 & 95.5 & 354.0  &  $8.5 \pm 1.8 \mathrm{(stat.)} \pm 1.0 \mathrm{(syst.)}$  \\
\end{tabular}
\caption{$^{214}$Bi activities in the aluminium foil measured successively with the two BiPo-3 modules. $T_{obs}$ is the duration of the measurement, $\epsilon$ is the detection efficiency, $N_{obs}$ is the number of observed $^{214}$BiPo events, $N_{fit}$ is the number of fitted events to the delayed energy spectrum for the $^{214}$Bi foil contamination and for the random coincidences (RC) and $\mathcal{A}(^{214}\mathrm{Bi})$ is the corresponding activity of the foil in $^{214}$Bi.}
\label{tab:alu-214}
\end{table}

\begin{figure}[!]
  \centering
  \includegraphics[scale=0.37]{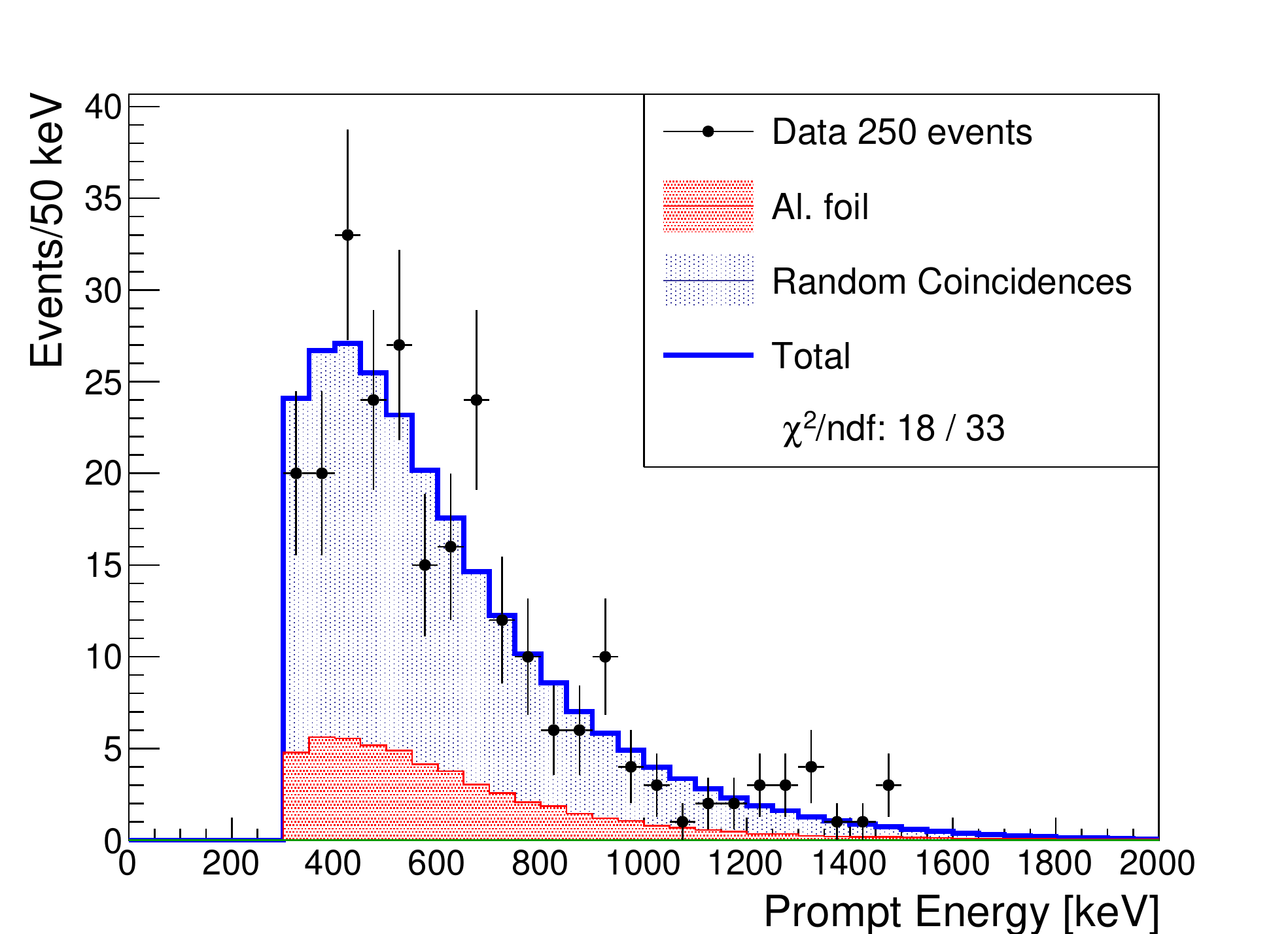}
  \includegraphics[scale=0.37]{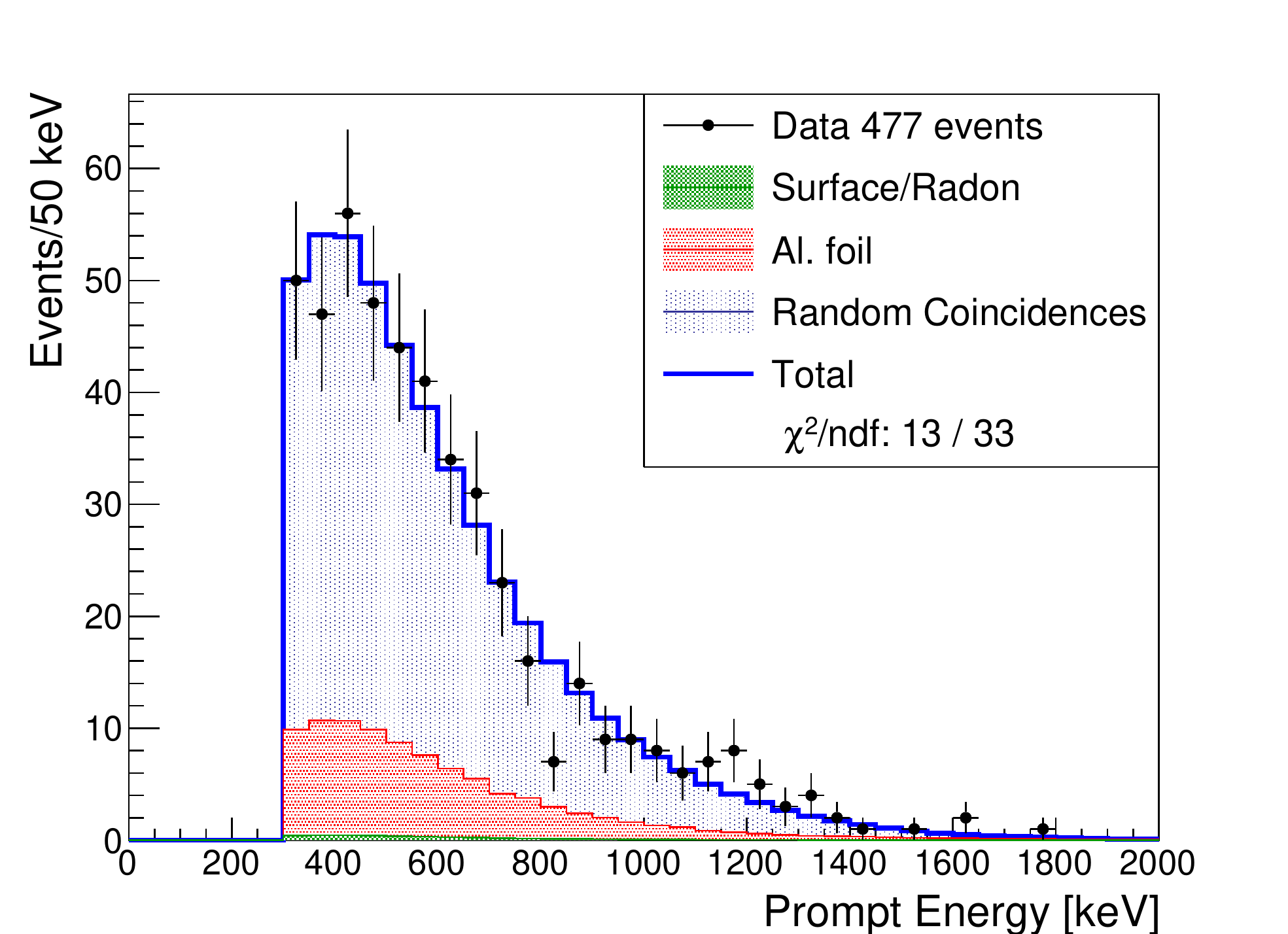}
  \includegraphics[scale=0.37]{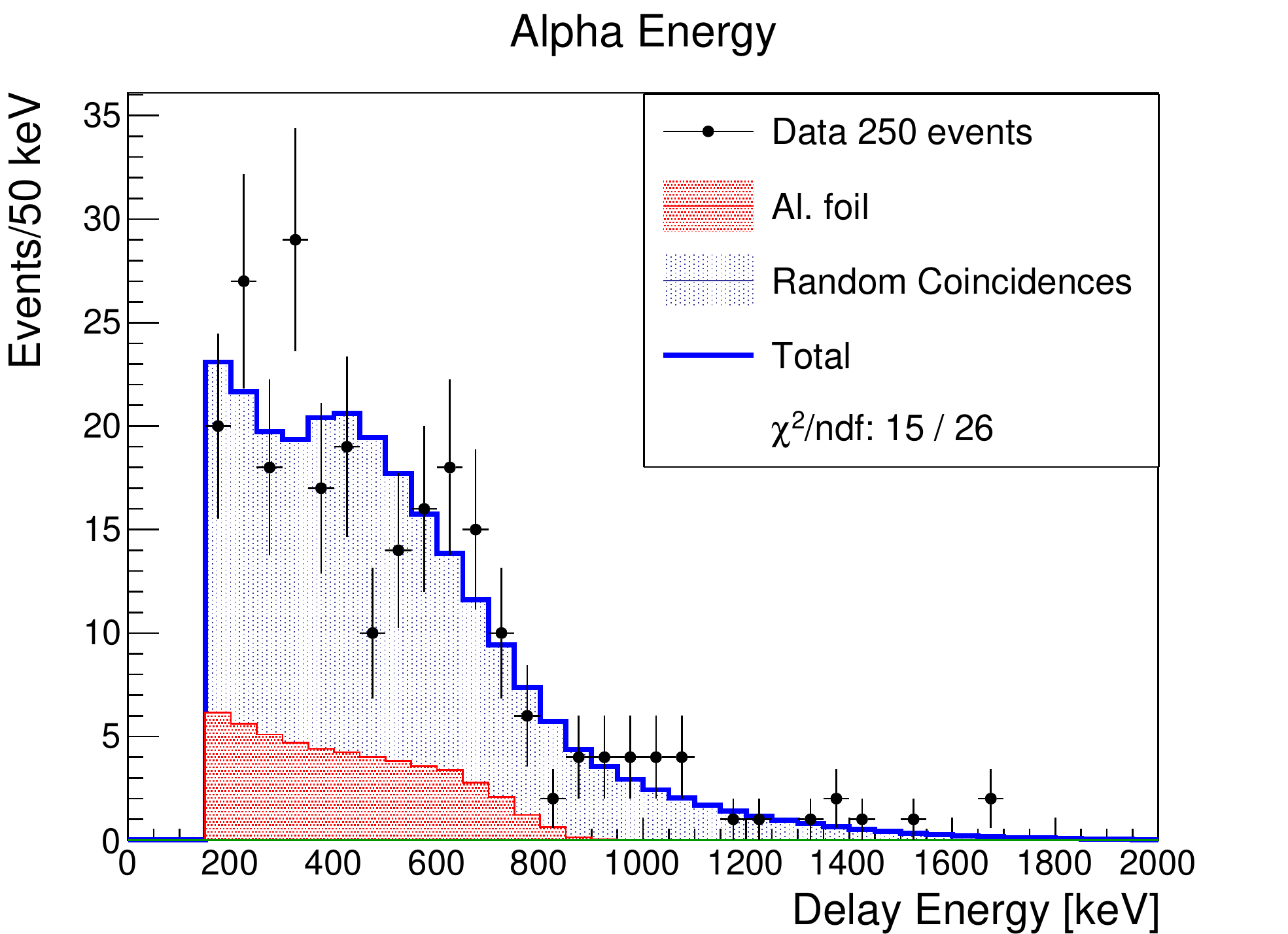}
  \includegraphics[scale=0.37]{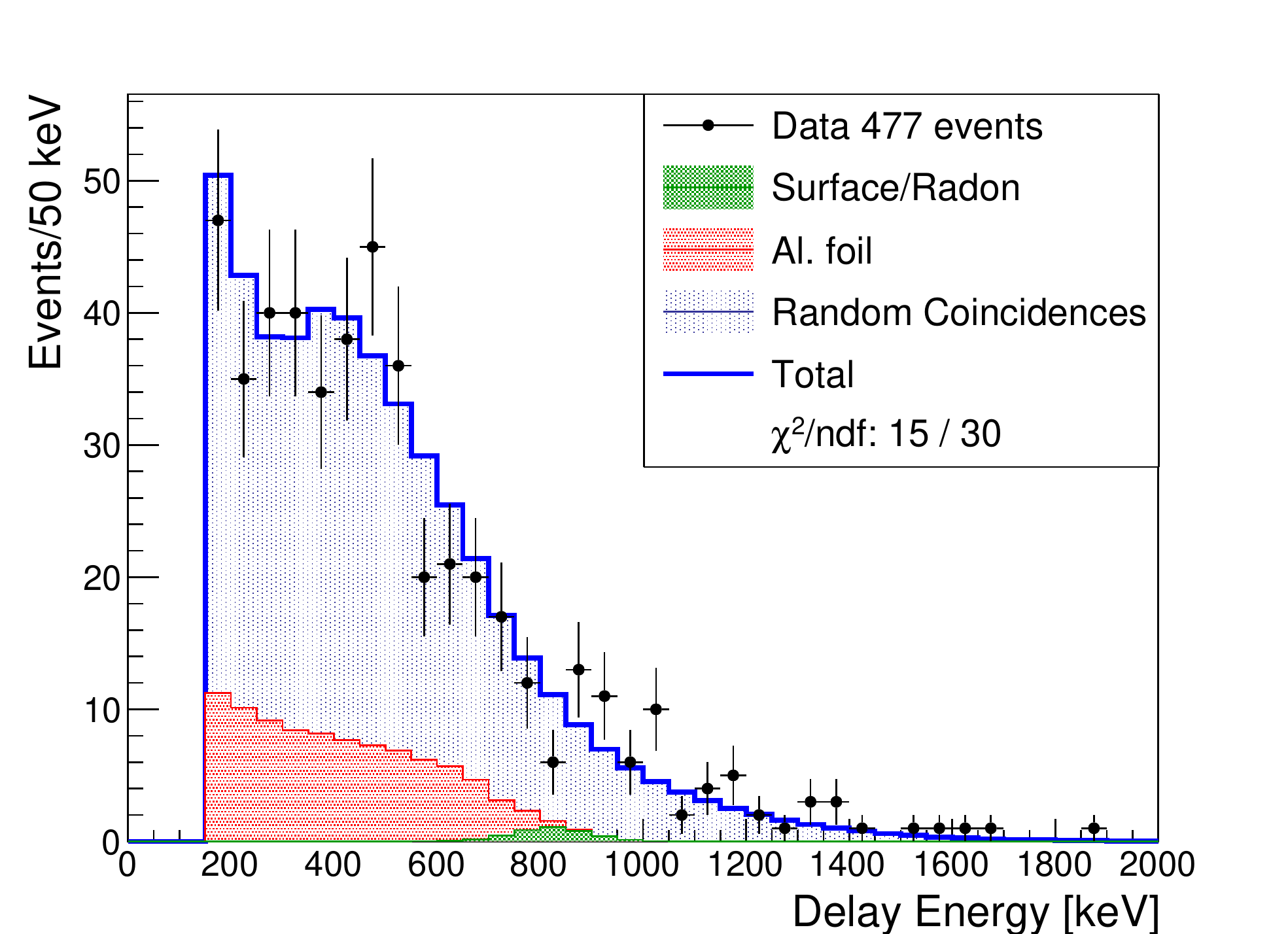}
  \includegraphics[scale=0.37]{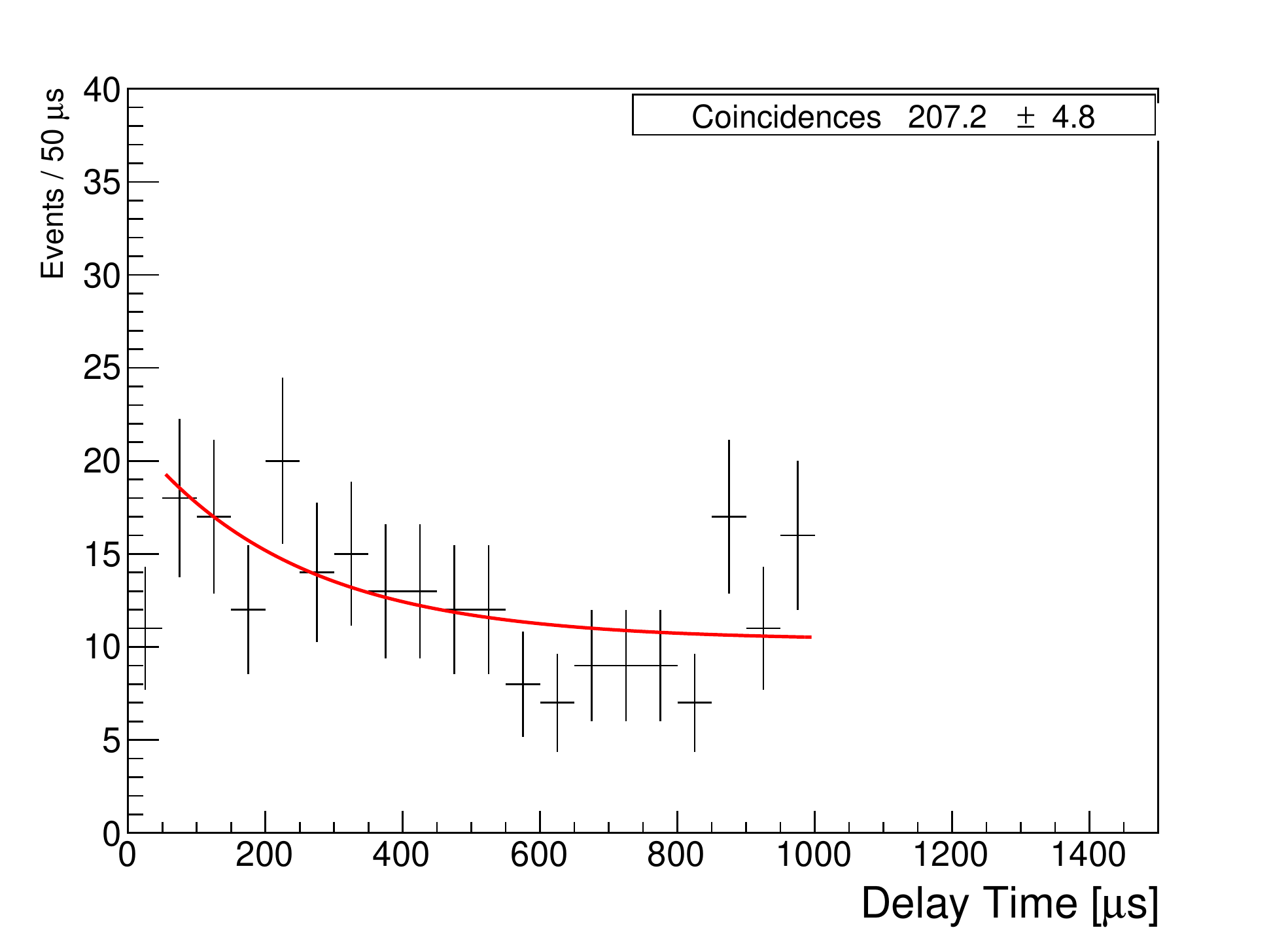}
  \includegraphics[scale=0.37]{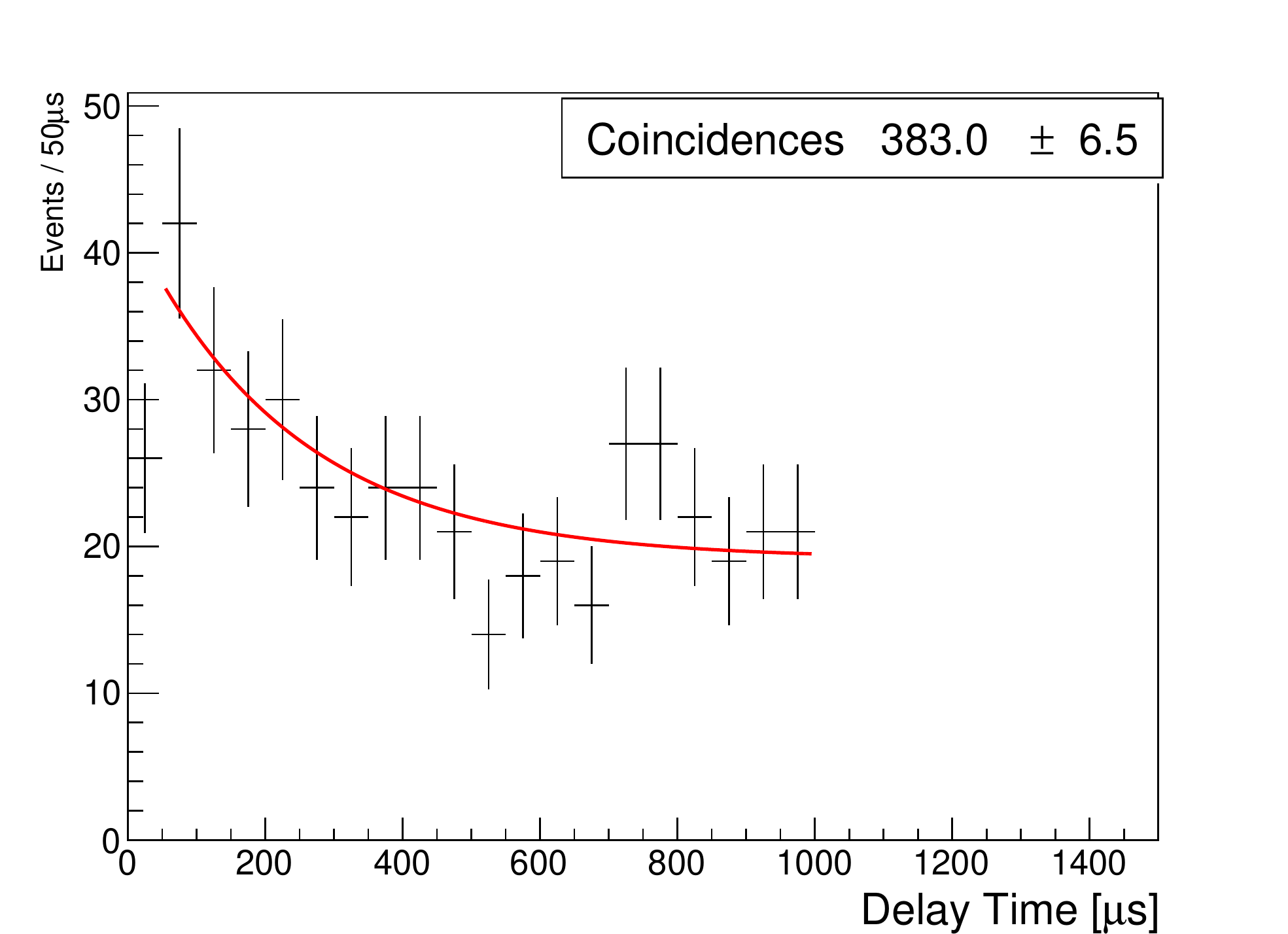}
  \caption{$^{214}$BiPo measurement of the calibrated aluminium foil, for a single foil geometry, with Module 1 (left plots) and  Module 2 (right plots): energy of the prompt electron signal (upper plots), energy of the delayed $\alpha$ signal (middle plots), and delay time (lower plots).}
  \label{fig:alu-214-onefoil}
\end{figure}

The energy spectrum of the random coincidence background is measured independently by selecting events with a signal detected in one scintillator and no signal detected in coincidence in the opposite scintillator. 
The random coincidence energy spectrum, and the simulated energy spectrum  of the delayed $\alpha$ particle from $^{214}$Bi contamination inside the aluminium foil, are simultaneously fitted to the observed data using the likelihood method.  
Results of the fit and of the measured $^{214}$Bi activity of the foil are given in Figure~\ref{fig:alu-214-onefoil} and Table~\ref{tab:alu-214}. The $^{214}$Bi activities measured independently with the two modules are in agreement with each other, and are in agreement with the activity reported by the HPGe measurement. 
The delay time distribution is fitted separately by a constant plus an exponential decay with a half-life equal to the $^{214}$Po half-life of 164~$\mathrm{\mu}$s. 
The fitted constant value corresponds to a number of random coincidences of $205 \pm 21$ for Module 1 and $379 \pm 29$ for Module 2. They are in agreement with the number of random coincidences fitted to the delayed energy spectrum (see Table~\ref{tab:alu-214}). 

\begin{figure}[!]
  \centering
  \includegraphics[scale=0.37]{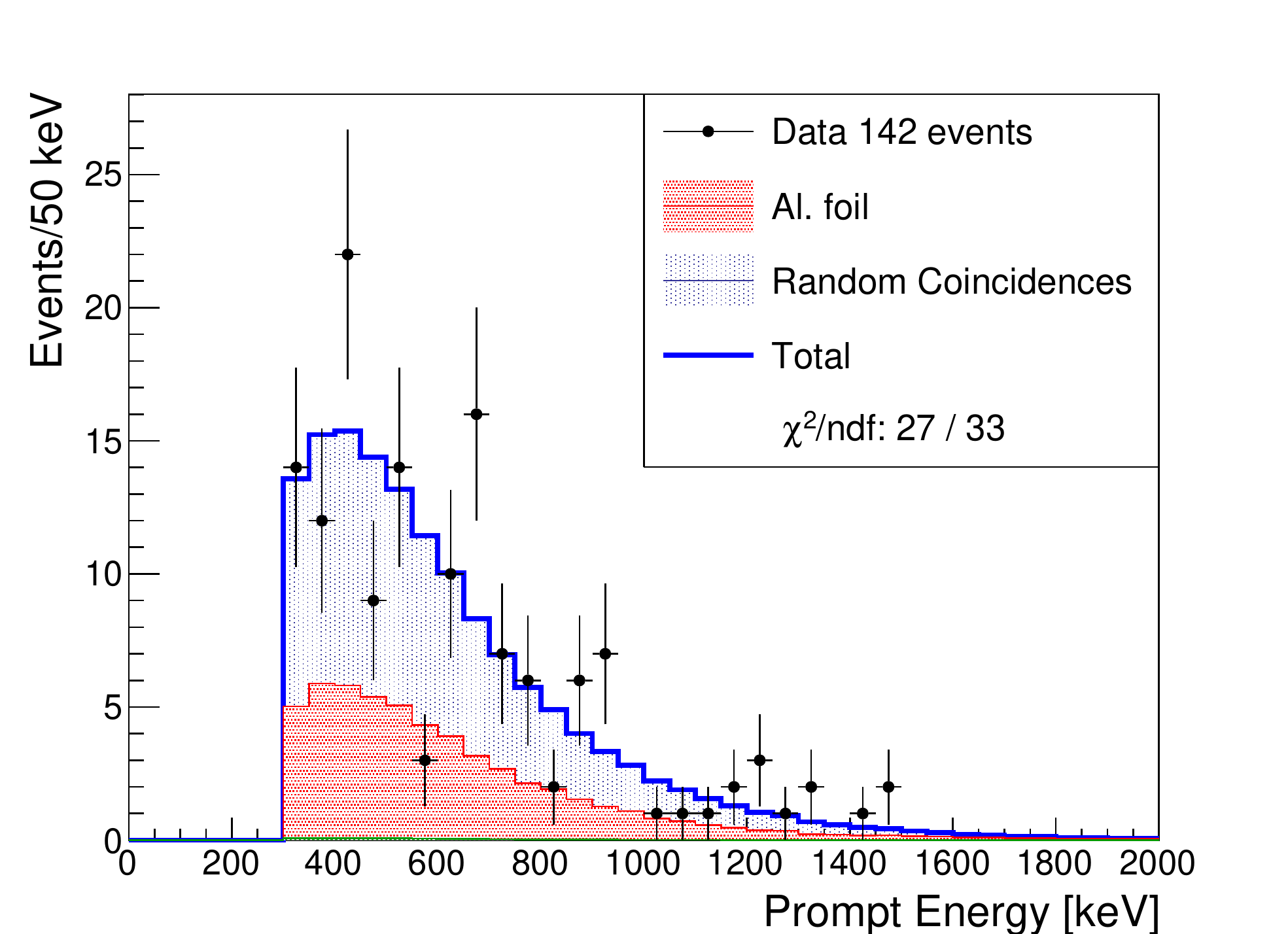}
  \includegraphics[scale=0.37]{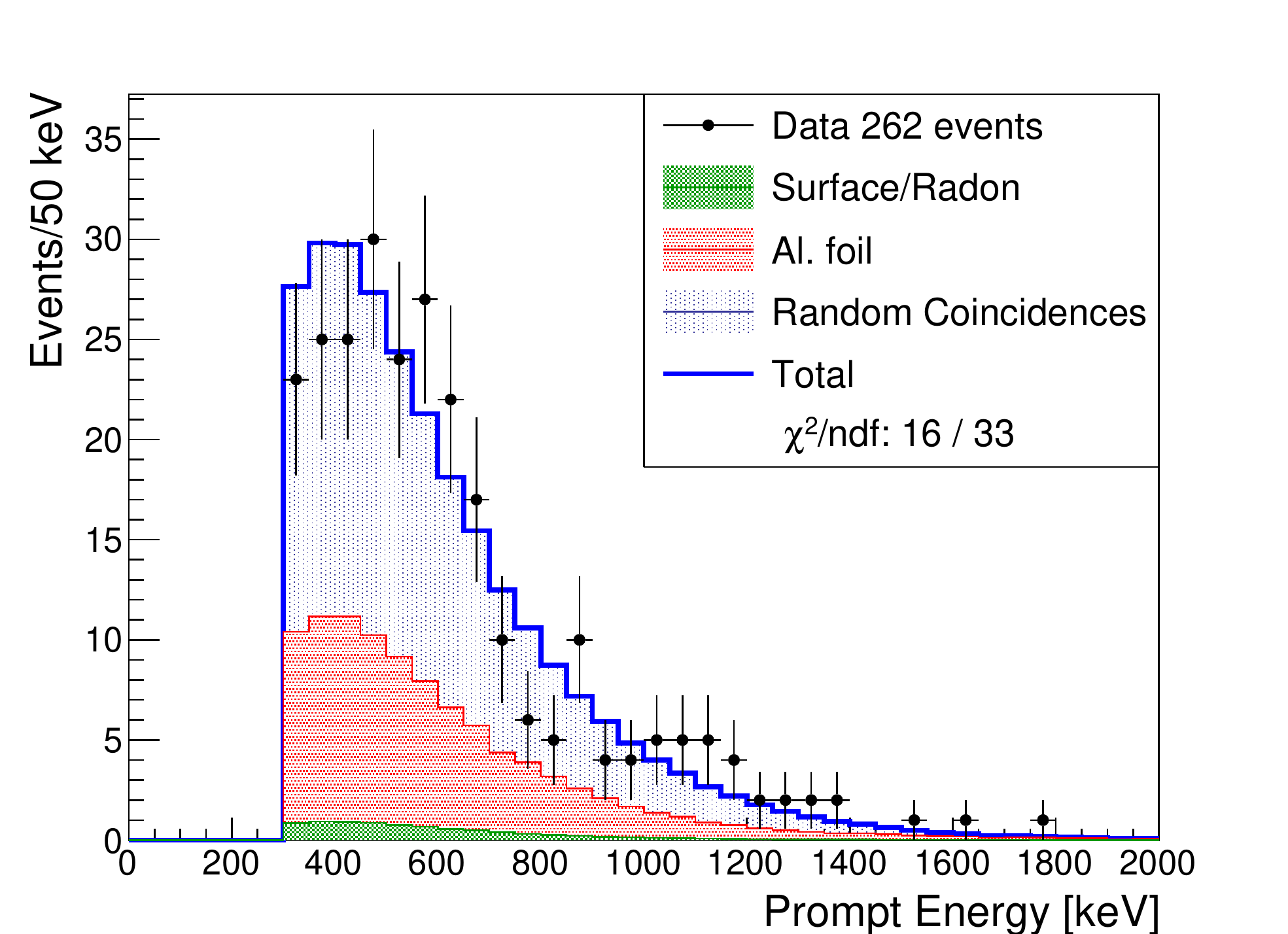}
  \includegraphics[scale=0.37]{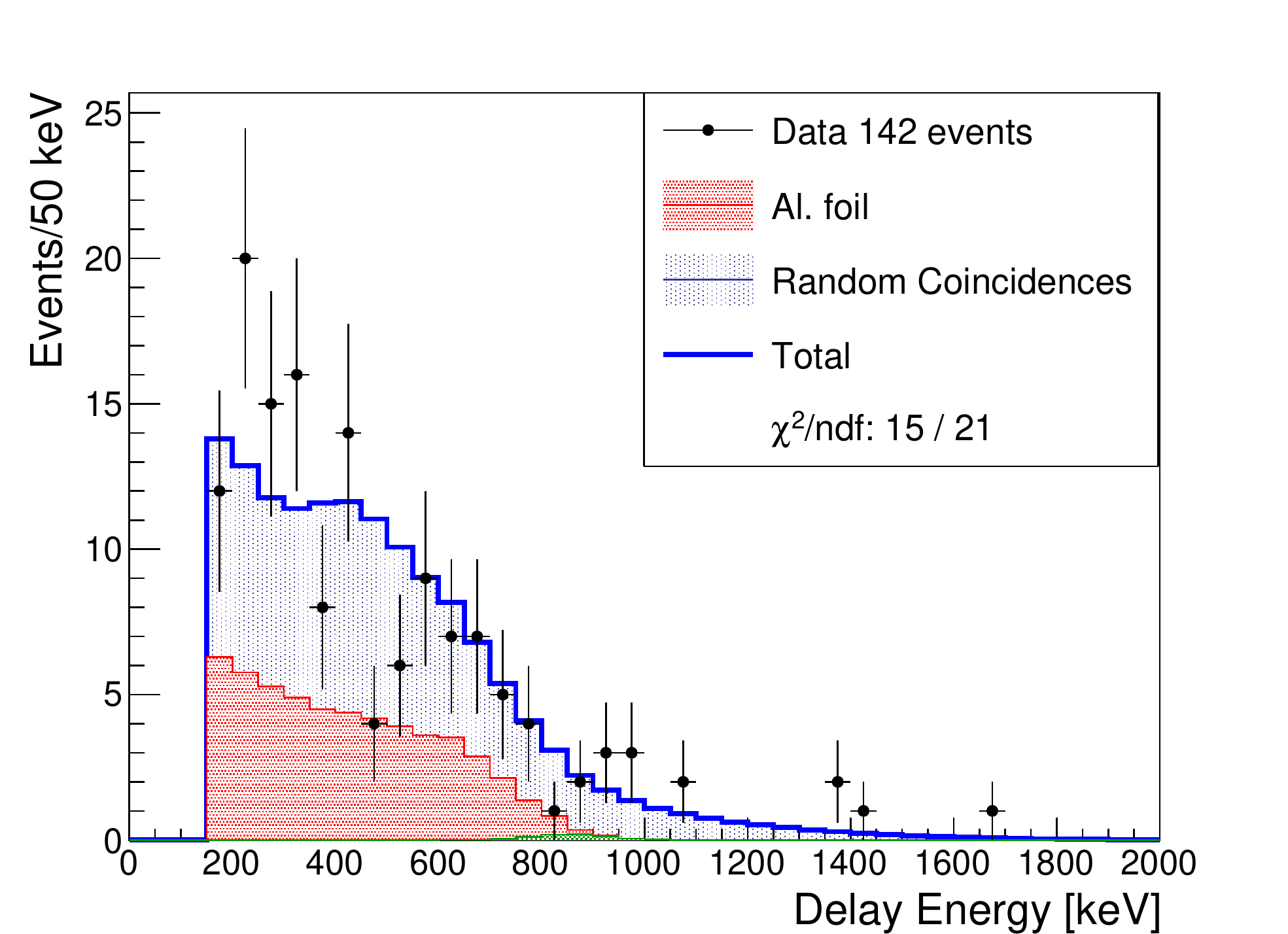}
  \includegraphics[scale=0.37]{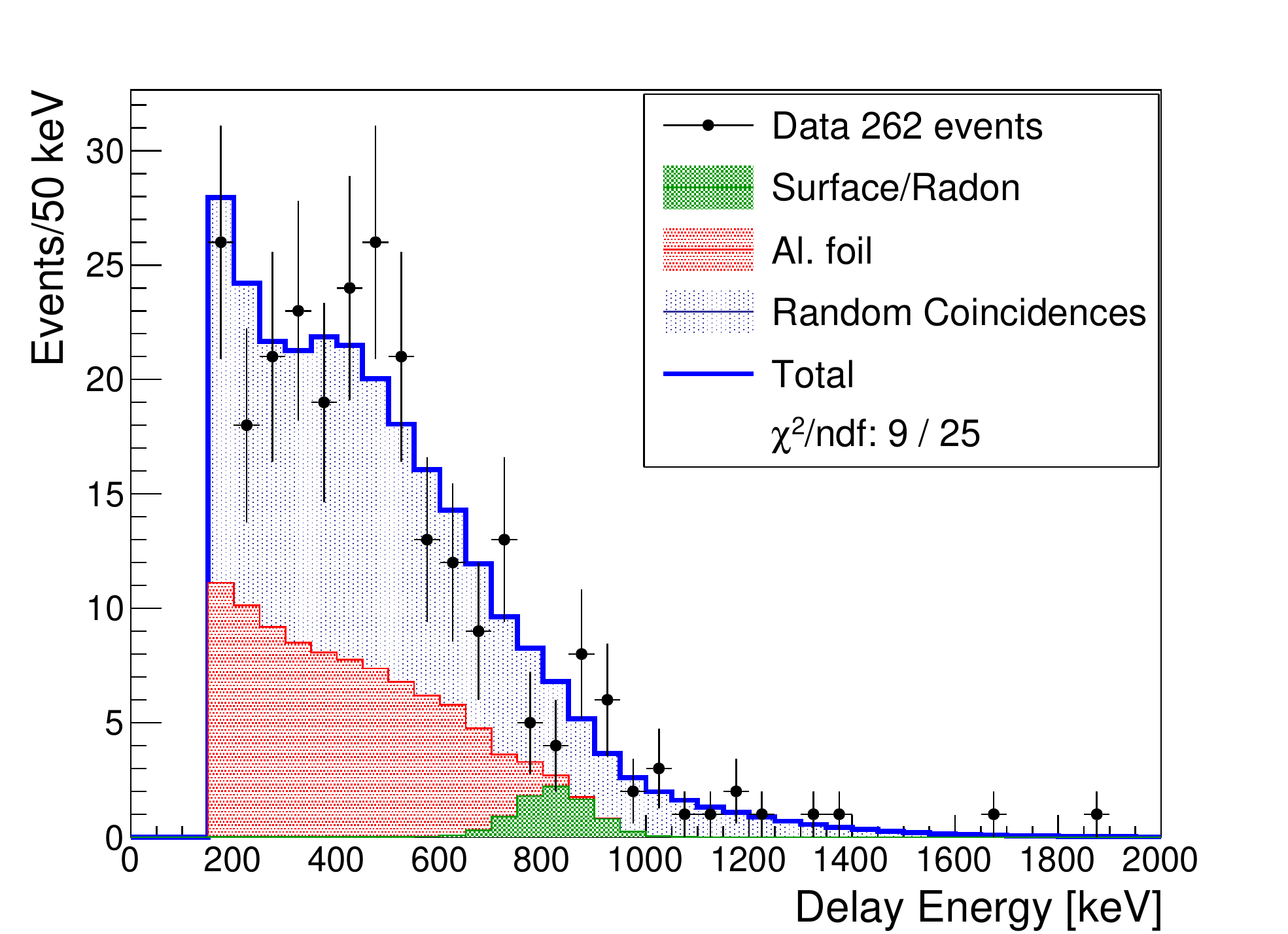}
  \includegraphics[scale=0.37]{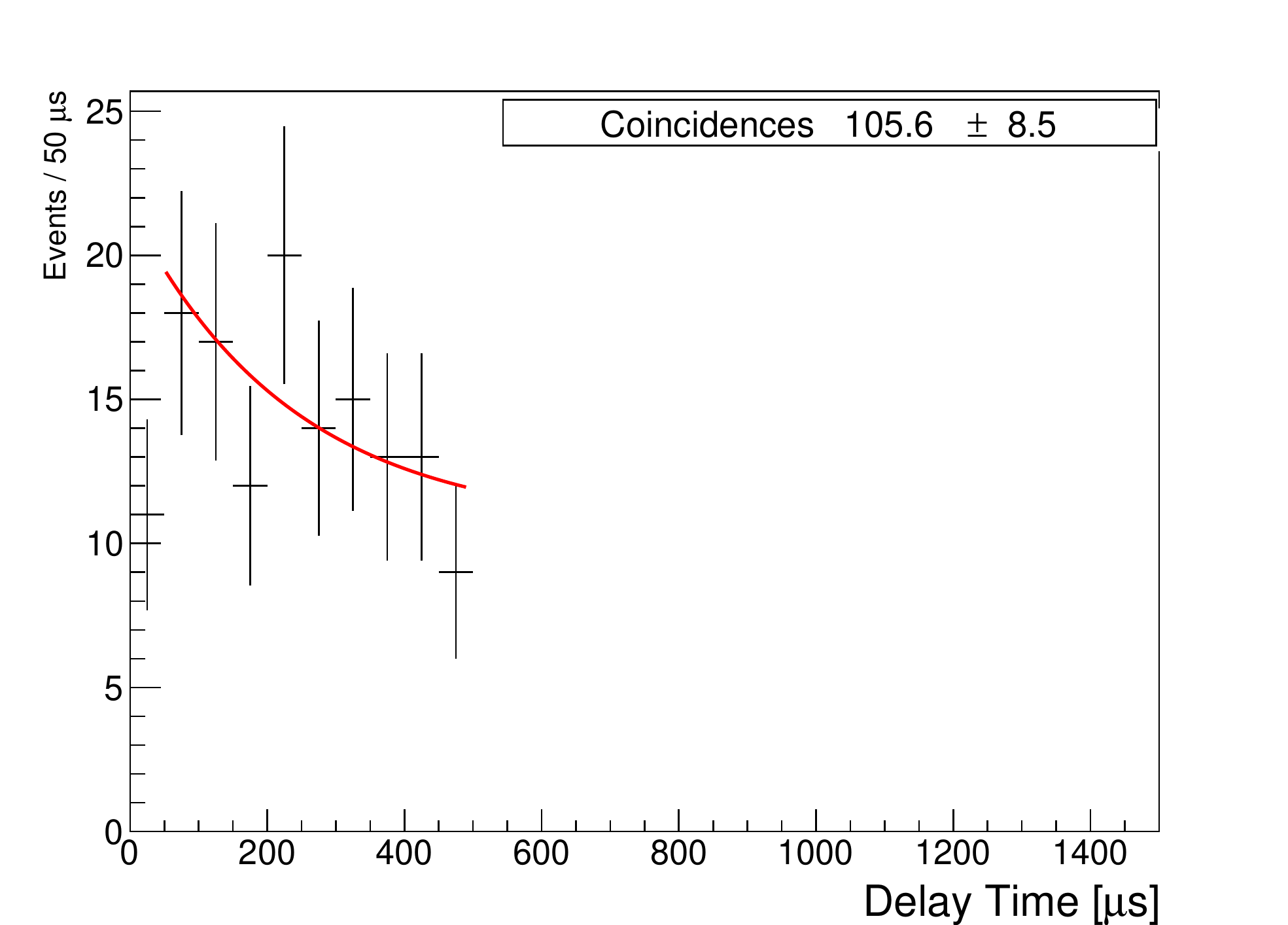}
  \includegraphics[scale=0.37]{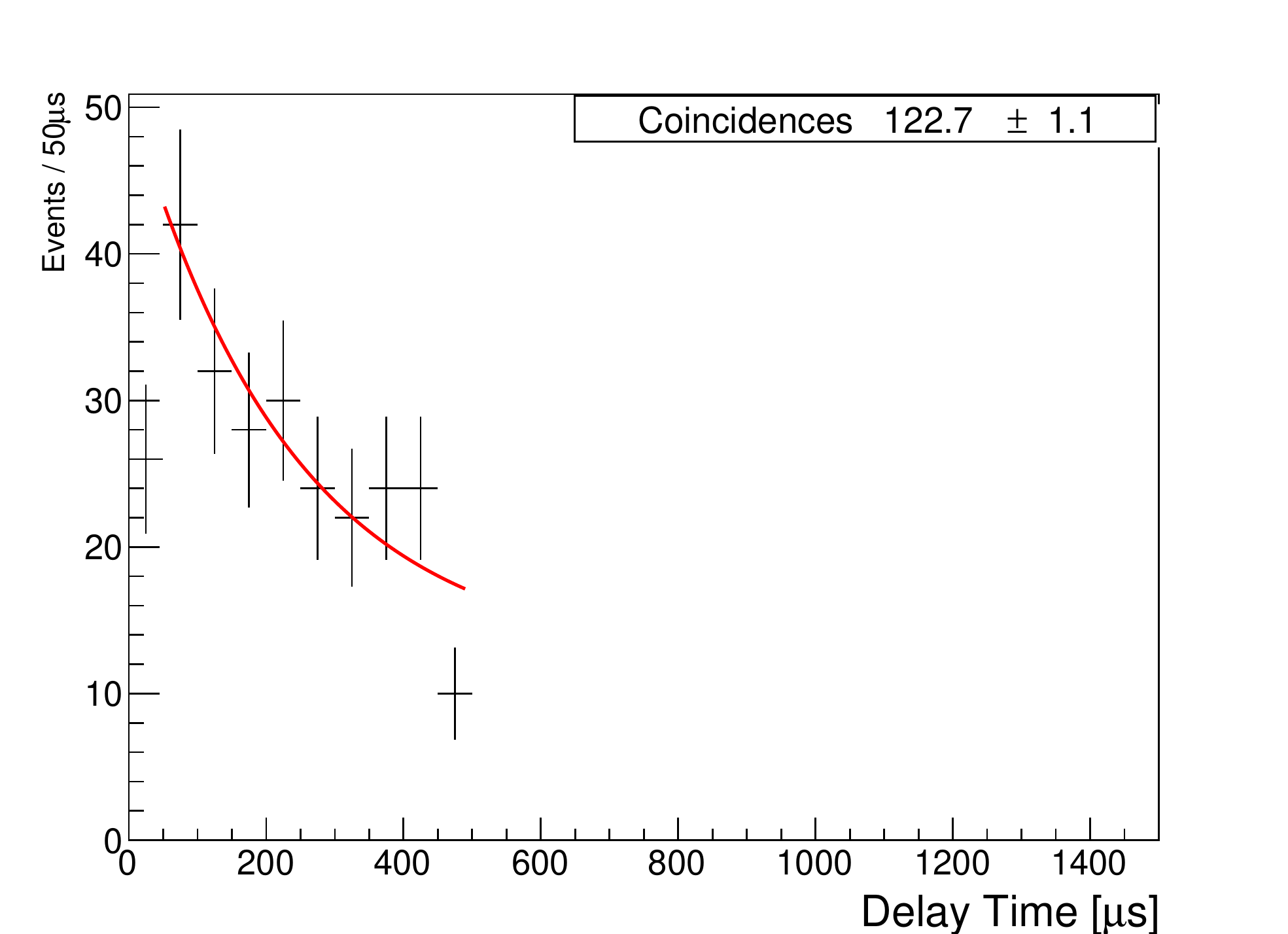}
  \caption{Same as Figure~\ref{fig:alu-214-onefoil} with the additional criteria of a delay time lower than three times the $^{214}$Po half-life of 164~$\mathrm{\mu}$s.}
  \label{fig:alu-214-onefoil-dtcut}
\end{figure}

The same analysis is also performed by adding the requirement of a delay time lower than three times the $^{214}$Po half-life of 164~$\mathrm{\mu}$s, in order to reduce the random coincidences background. The results of the fit of the delayed energy spectrum  are given in Figure~\ref{fig:alu-214-onefoil-dtcut} and Table~\ref{tab:alu-214-dtcut}. The signal-over-background ratio is increased by almost 60\% while the measured activity is slightly reduced by about 10\% but in agreement within uncertainty with the previous measurement. 

\begin{table}[htb]
\centering
\begin{tabular}{c|c|c|c|c|c|c|c}
 & $T_{obs}$ & Mass & $\epsilon$ &$N_{obs}$  & \multicolumn{2}{c}{$N_{fit}$} & $\mathcal{A}(^{214}\mathrm{Bi})$ \\
 & (days)   & (g)  & (\%)       &         & Foil    & RC        & (mBq/kg) \\
\hline
\hline
Mod. 1  &  11.6  & 192.6 &  2.4 $\pm$ 0.2 &  142 & 49.4 & 92.8   &  $10.7 \pm 2.4 \mathrm{(stat.)} \pm 1.3 \mathrm{(syst.)}$  \\
\hline
Mod. 2  &  25.8  & 186.7 &  2.4 $\pm$ 0.2 &  262 & 73.9 & 177.0  &  $7.5 \pm 1.5 \mathrm{(stat.)} \pm 0.9 \mathrm{(syst.)}$   \\
\end{tabular}
\caption{Same as Table~\ref{tab:alu-214} with the additional criteria  of a delay time lower than three times the $^{214}$Po half-life of 164~$\mathrm{\mu}$s.}
\label{tab:alu-214-dtcut}
\end{table}

\section{Measurement of the first SuperNEMO $^{82}$Se double $\beta$-decay source foils}
\label{sec:sn-foil-measurement}

Several samples have been measured in BiPo-3: most of them are samples used for the SuperNEMO double $\beta$-decay source foils production, but also others used in other double $\beta$-decay or rare events experiments like CUORE~\cite{cuore}, LUMINEU~\cite{lumineu} and Micromegas TPC~\cite{micromegas-DBD, micromegas-DM}. 
We present here the results of measurements of the $^{208}$Tl ($^{212}$BiPo) and $^{214}$Bi ($^{214}$BiPo) activities of the first SuperNEMO double $\beta$-decay source foils. The results of measurement of the other samples will be presented in dedicated articles.

The SuperNEMO foils are in the form of strips, 270~cm long and 13.5~cm wide.  
To produce enriched $^{82}$Se foils for the SuperNEMO experiment, thin and chemically purified $^{82}$Se powder is mixed  with PVA glue and then deposited between Mylar foils.  
The area density of $^{82}$Se powder and PVA mixture is 41~mg/cm$^2$ (the area density of the $^{82}$Se powder only is 37~mg/cm$^2$) for the first foils reported here. 
The Mylar foil is 12~$\mathrm{\mu}$m thick and has been irradiated at JINR Dubna (Russia) with an ion beam and then etched in a sodium hydroxide solution. This produces a large number of microscopic holes in order to ensure a good bond and to allow water evaporation during the drying of PVA. 

\subsection{Analysis method}

The analysis method for the $^{212}$Bi and $^{214}$Bi contamination measurements inside the samples is similar to the method used for the background and the aluminium foils measurements. 

The criteria to select the {\it back-to-back} BiPo events are identical to those used for the background measurement (listed in section~\ref{sec:event-selection}).
For the $^{214}$Bi measurement, the delay time between the prompt and the delayed signal is required to be lower than 492~$\mathrm{\mu}$s (three times the $^{214}$Po half-life). This reduces the random coincidences background, as explained in the measurement of the calibrated aluminium foil (section~\ref{sec:foil-214}).

We search for an excess of BiPo events above the background expectation in the delayed energy spectrum. 
The background components are random coincidences and  bismuth ($^{212}$Bi and $^{214}$Bi) contamination on the scintillator surface. 
For the $^{82}$Se foils, the bismuth contamination inside the irradiated Mylar is added as an extra component of background. 
The number of random coincidences is fixed to the expected value calculated from the single rates of scintillators using the same set of data (as explained in section~\ref{sec:bkg-analysis-method}). 
The surface background and the irradiated Mylar counting rates are constrained by the quoted values from the dedicated measurements, within the 90\%~C.L. uncertainty interval (see the combined values in Tables~\ref{tab:bkg-bipo212} and \ref{tab:bkg-bipo214} for the surface background, and Tables~\ref{tab:foil-components-212} and \ref{tab:foil-components-214} for the irradiated Mylar).

The delayed energy spectra of the background components are then simultaneously fitted to the observed data, using the likelihood method.
The energy spectrum of each component is calculated by Monte Carlo, except for the random coincidence background for which the energy spectrum is measured using the single counting events.

\subsection{Measurement of the raw materials}
\label{sec:results-raw-materials}

Before producing the $^{82}$Se foils, the raw materials have been first measured separately with the BiPo-3 detector (PVA, Mylar before and after irradiation). Dedicated PVA pads ($300 \times 300$~mm$^2$ and 195~$\mu$m thick) have been produced for the PVA measurement. The PVA and the irradiated Mylar have been measured in Module 1. The raw Mylar (before irradiation) has been measured in Module 2. 
The results of these measurements are presented in Tables~\ref{tab:foil-components-212} and \ref{tab:foil-components-214}, respectively for the $^{212}$Bi and $^{214}$Bi activities of the bulk. 
Results are given with two different criteria to select the BiPo events. In the first case, a single energy threshold is applied to the delayed signal. In the second case, an upper limit of 700~keV (600~keV) for $^{212}$Bi ($^{214}$Bi) is required for the energy of the delayed signal, thus rejecting the surface background. 

\begin{table}[htb]
\centering
\begin{tabular}{l|c|c|c|c|c|c|c}
Sample & $T_{obs}$ & Mass    & $S$      & $N_{bkg}$  & $N_{obs}$ & $\epsilon$ & $\mathcal{A}(^{208}\mathrm{Tl})$  \\
       &  (days)  & (g)  & (m$^2$)  &           &          & (\%)       & ($\mathrm{\mu}$Bq/kg)           \\
       &          &      &          &           &          &            & 90\% C.L.                       \\
\hline
\hline
Raw Mylar                   & 76.5 & 108.1  & 1.62 &             &     &       &         \\
$E(\alpha)>150$~keV     &       &       &      & 3.0$\pm$0.5 &  5  &  7.6 $\pm$ 0.8  & $< 65$   \\
$150<E(\alpha)<700$~keV &       &       &      & 0.5$\pm$0.2 &  1  &  5.5 $\pm$ 0.6  & $< 49$   \\

\hline
Irrad. Mylar            & 44.4  &  200  & 3.06 &              &     &       &              \\
$E(\alpha)>150$~keV     &       &       &      & 3.4$\pm$0.6  & 16  &  9.4 $\pm$ 1.0  & $98^{+62}_{-47}$   \\
$150<E(\alpha)<700$~keV &       &       &      & 0.5$\pm$0.2  & 10  &  7.7 $\pm$ 0.8  & $90^{+63}_{-42}$   \\
\hline
PVA                     & 137.2 &  210  & 1.8  &              &     &      &         \\
$E(\alpha)>150$~keV     &       &       &      & 5.5$\pm$1.0  & 3   &  3.9 $\pm$ 0.4 & $< 15$  \\
$150<E(\alpha)<700$~keV &       &       &      & 0.9$\pm$0.3  & 0   &  3.0 $\pm$ 0.3 & $< 12$  \\
\end{tabular}
\caption{Results of  $^{208}$Tl activity measurements of raw Mylar, irradiated Mylar and PVA. $T_{obs}$ is the duration of the measurement; $S$ is the active scintillator surface area; $N_{bkg}$ is the expected background from the fit of the experimental data (surface background + coincidences); $N_{obs}$ is the number of observed $^{212}$BiPo events; $\epsilon$ is the detection efficiency.}
\label{tab:foil-components-212}
\end{table}

\begin{table}[htb]
\centering
\begin{tabular}{l|c|c|c|c|c|c|c}
Sample                 & $T_{obs}$ & Mass  & $S$      & $N_{bkg}$  & $N_{obs}$ & $\epsilon$ & $\mathcal{A}(^{214}\mathrm{Bi})$  \\
                      &  (days)  & (g)  & (m$^2$)  &           &          & (\%)       & ($\mathrm{\mu}$Bq/kg)           \\
                      &          &      &          &           &          &            & 90\% C.L.                       \\
\hline
\hline
Raw Mylar                    &  31.1 &  108 &  1.62 &     &      &     &         \\
$E(\alpha)>300$~keV     &      &      &        &  2.8$\pm$0.8  &  2 &  3.40 $\pm$ 0.35 & $< 305$  \\
$300<E(\alpha)<600$~keV &     &       &       &  2$\pm$0.7  &  0   & 2.17 $\pm$ 0.22 & $< 195$  \\
\hline
Irrad. Mylar                   & 31.3 &  190   &  2.88   &                  &          &     &         \\
$E(\alpha)>300$~keV     &   &      &            &   7.0$\pm$2.1       &  12       &  3.48 $\pm$ 0.35 & $-$  \\
$300<E(\alpha)<600$~keV &       &         &                &  5.2$\pm$2.3     &   7   & 2.15 $\pm$ 0.22          & $<688 $  \\
\hline
PVA                             &  124.6 &  230 &  1.8 &     &     &     &         \\
$E(\alpha)>300$~keV     &     &     &     &  41.3$\pm$12.4 &  76 &  1.45 $\pm$ 0.15 & $-$  \\
$300<E(\alpha)<600$~keV  &        &      &      &  8.6$\pm$3.0 &  13 &  0.92 $\pm$ 0.09 & $< 505$  \\
\end{tabular}
\caption{Results of  $^{214}$Bi activity measurements of raw Mylar, irradiated Mylar and PVA. $T_{obs}$ is the duration of the measurement; $S$ is the active scintillator surface area; $N_{bkg}$ is the number of expected background from the fit of the experimental data (surface background + coincidences); $N_{obs}$ is the number of observed $^{214}$BiPo events; $\epsilon$ is the detection efficiency.}
\label{tab:foil-components-214}
\end{table}

The Mylar before irradiation and the PVA are very pure in $^{208}$Tl. For the PVA samples, using the statistical analysis approach described in~\cite{feldman-cousins}, an upper limit of $\mathcal{A}(^{208}\mathrm{Tl}) < 12$~$\mu$Bq/kg (90\%~C.L.) is obtained with an exposure of 28.8~kg$\times$days. A $^{208}$Tl contamination is observed inside the Mylar after irradiation at a level of $\mathcal{A}(^{208}\mathrm{Tl}) = 90^{+63}_{-42} \ \mu$Bq/kg (90\%~C.L.). This contamination is taken into account in the measurement of the enriched  $^{82}$Se foils. It is worth mentioning that for these results, as well as for all the other measured samples, a contamintaion value is given if the lower limit of the obtained interval is higher than zero. Otherwise, the upper limit of the interval is quoted as the corresponding upper limit of the measurement.

\begin{figure}[!]
  \centering
  \includegraphics[scale=0.4]{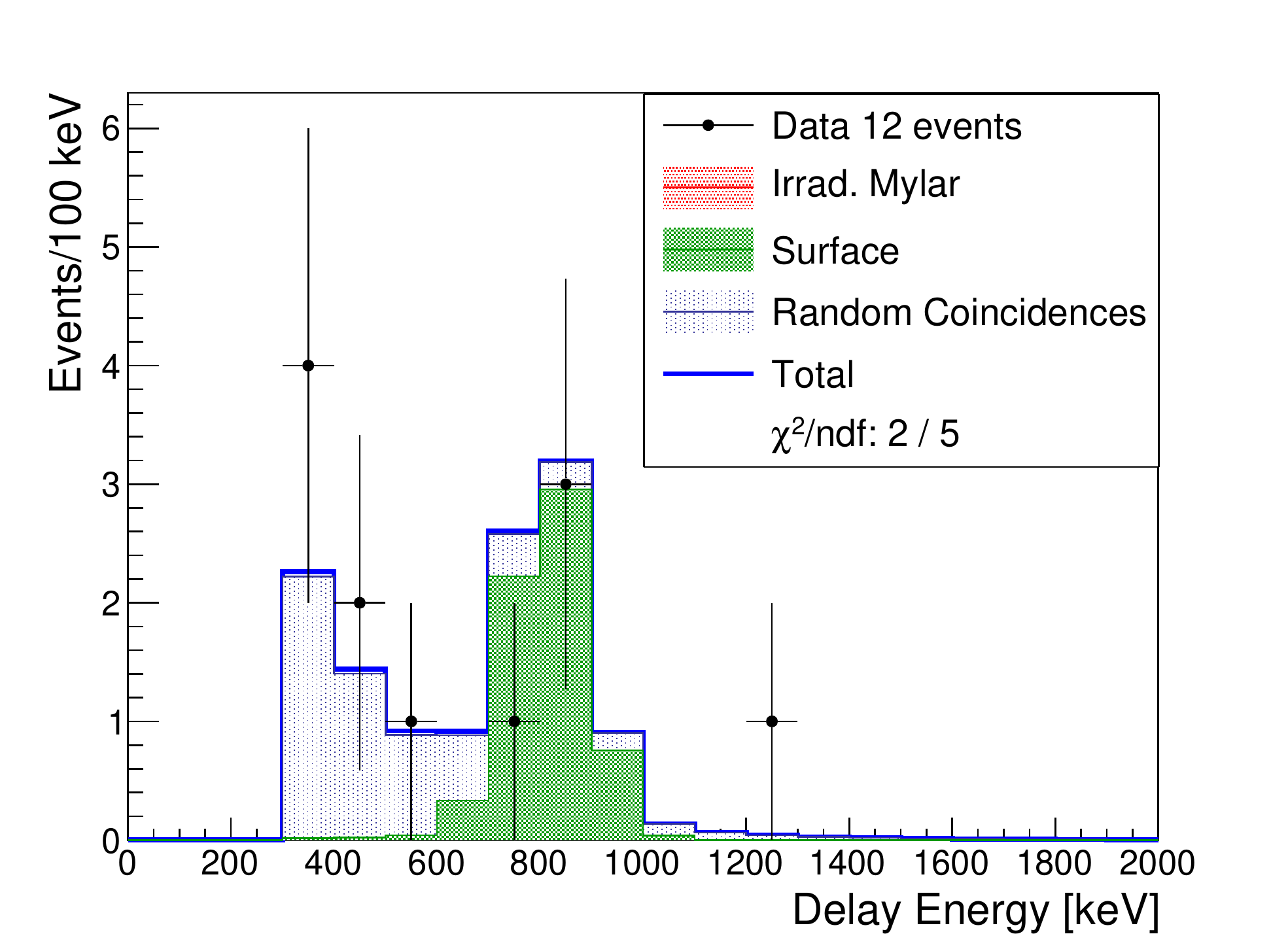}
  \caption{Distributions of the delayed energy for the $^{214}$BiPo measurement of irradiated Mylar. 
  The excess of observed events with a delayed energy between 600 and 900~keV corresponds to a $^{214}$Bi contamination on the surface of the sample (green histogram). For delayed energy lower than 600~keV, no significant excess of events is observed at 90\%~C.L., and data are in agreement to the  expected background from the random coincidences (blue histogram).}
  \label{fig:irrad-mylar-edelay}
\end{figure}

For the $^{214}$BiPo measurements, no significant excess of events is observed  above the background expectation in the energy range corresponding to bulk events (delayed energy $<$ 600 keV),  and upper limits in $^{214}$Bi activity are set. 
However, a surface contamination in  $^{214}$Bi is observed on the PVA pads and the irradiated Mylar. It corresponds to an excess of events with a delayed energy between 600 and 900~keV, in agreement with a surface contamination, as shown in Figure~\ref{fig:irrad-mylar-edelay} for the irradiated Mylar measurement. 
By adding this background contribution to the fit, a surface contamination in  $^{214}$Bi is estimated to be $3.7^{+6.8}_{-3.5} \ \mathrm{\mu Bq/m}^2$ for the irradiated Mylar, and $31\pm8 \ \mathrm{\mu Bq/m}^2$ for the PVA pads (90\% C.L. intervals). 

\subsection{Measurement of the enriched $^{82}$Se foils}

\begin{figure}[h]
  \centering
  \includegraphics[scale=0.5]{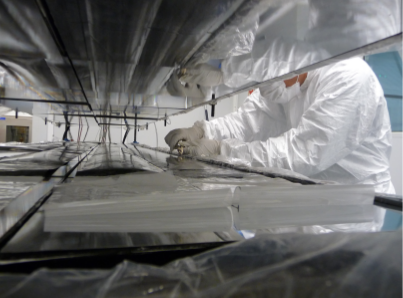}
  \caption{The first BiPo-3 module open in the clean room in LSC, during the installation of the SuperNEMO enriched $^{82}$Se foils.}
  \label{fig:photo-se-foils}
\end{figure}

Two first SuperNEMO $^{82}$Se foils have been measured from August 2014 to December 2014, using half of the available surface area of Module 1. The second half has been kept empty and has been used to control the background, as discussed in section~\ref{sec:bkg}.
In December 2014, two additional foils have been installed besides the two first foils in the second half of the module. Then the four foils have been measured from December 2014 to June 2015. 
A picture of the installation of the  SuperNEMO $^{82}$Se foils inside the BiPo-3 detector is shown in Figure~\ref{fig:photo-se-foils}.
Combining the two sets of measurements, the total duration of measurement is 262~days for the $^{212}$BiPo measurement (after rejecting the first three days to suppress the background induced by the thoron deposition) and 241.1~days for the $^{214}$BiPo measurement (after rejecting the first fifteen days to suppress the background induced by the radon deposition), the total effective mass of the $^{82}$Se+PVA mixture is 359~g (352~g), and the effective scintillators surface area is 2.13~m$^2$ (1.97~m$^2$) for the $^{212}$BiPo ($^{214}$BiPo) measurement. 

\begin{figure}[h]
  \centering
  \includegraphics[scale=0.37]{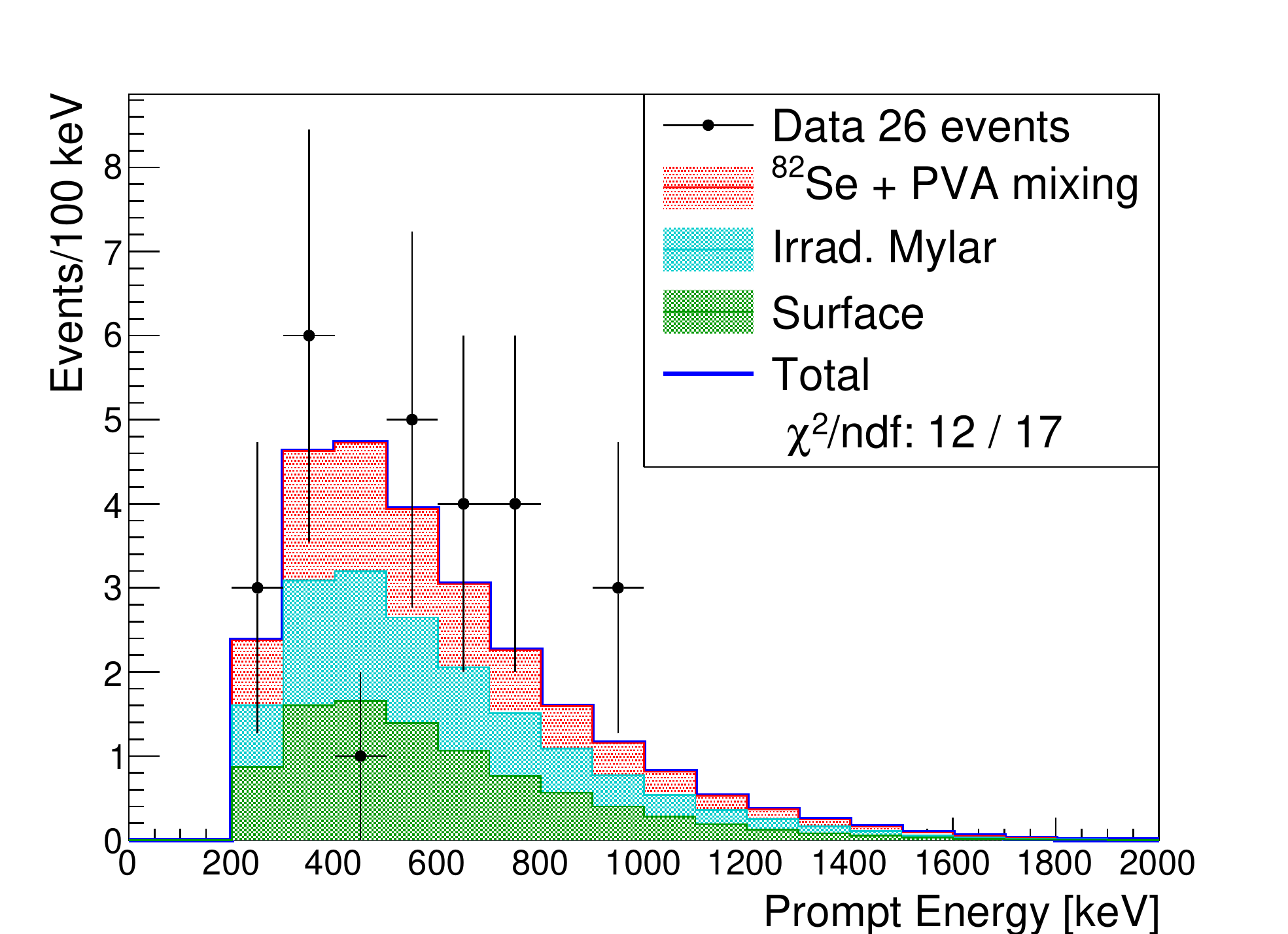}
  \includegraphics[scale=0.37]{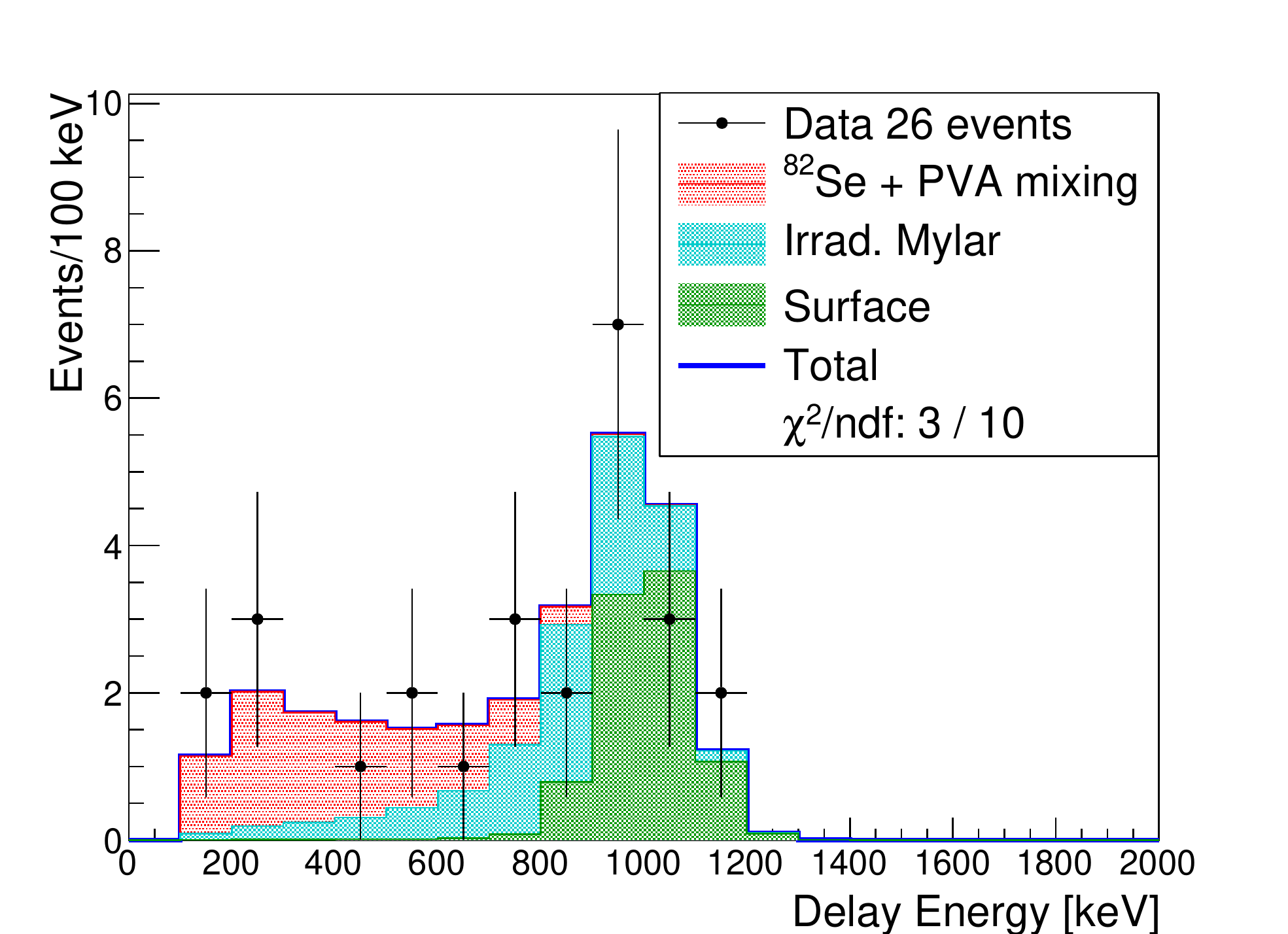}
  \includegraphics[scale=0.37]{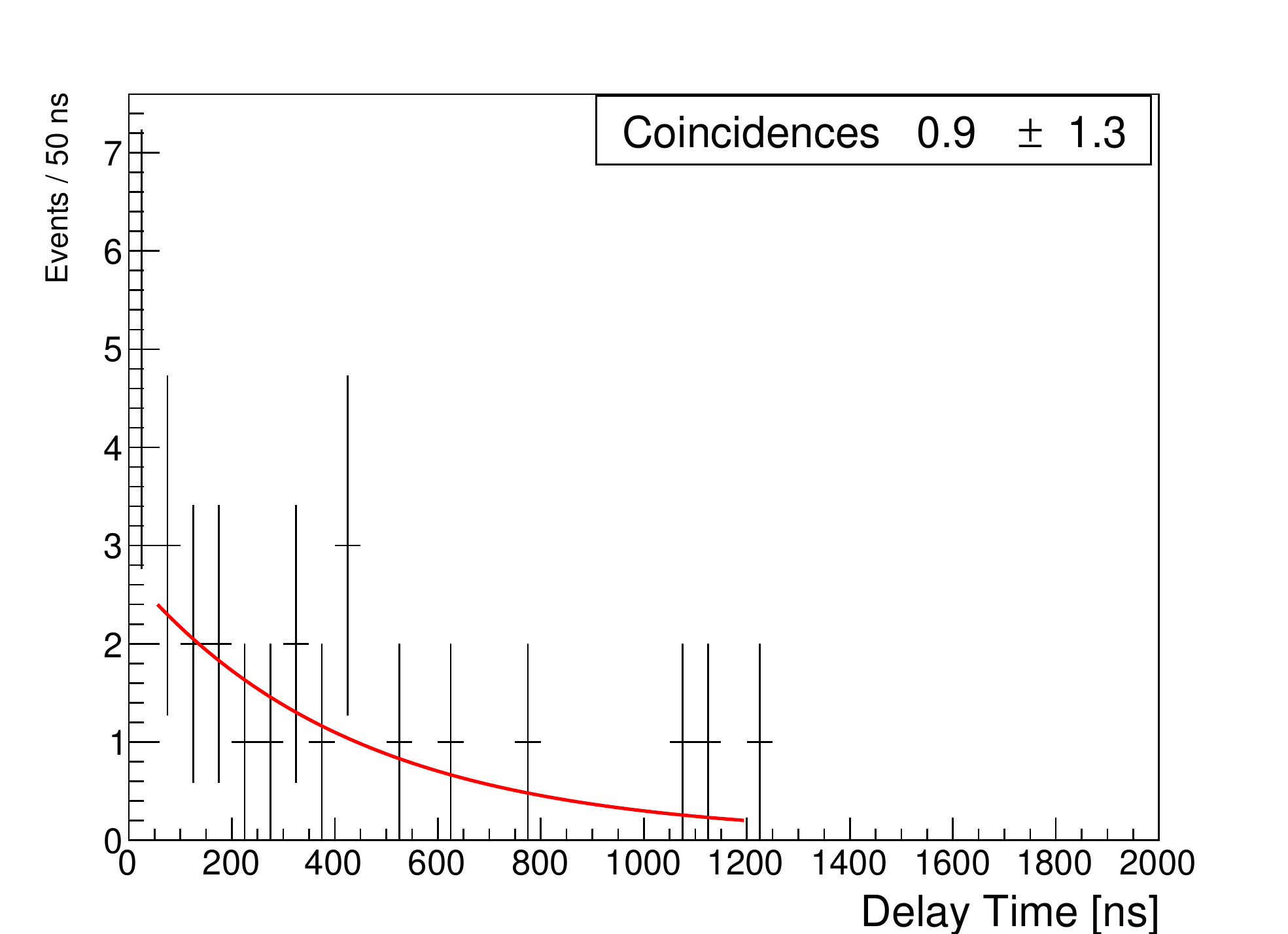}
  \caption{Distributions of the prompt and delayed energy and distribution of the delay time, for the $^{212}$BiPo measurement of the four first enriched $^{82}$Se SuperNEMO foils, with 262 days of data collection, an effective mass of $^{82}$Se+PVA mixture of 359~g and a surface area of measurement of 2.13~m$^2$. The data are compared to the expected background from the $^{212}$Bi contamination on the surface of the scintillators (green histogram) and in the irradiated Mylar (light blue histogram). The excess of observed events with a low delayed energy corresponds to a $^{212}$Bi contamination inside the Se+PVA mixture (red histogram) of 17.8$^{+16.4}_{-11.5} ~\mathrm{\mu Bq/kg} \ (90\% \mathrm{C.L.})$. The delay time distribution is fitted a posteriori by an exponential decay with an half-life set to the value of the $^{212}$Po decay half-life (300~ns) plus a constant value for a flat random coincidence distribution decay.}
  \label{fig:se-foil-212}
\end{figure}

\begin{figure}[h]
  \centering
  \includegraphics[scale=0.37]{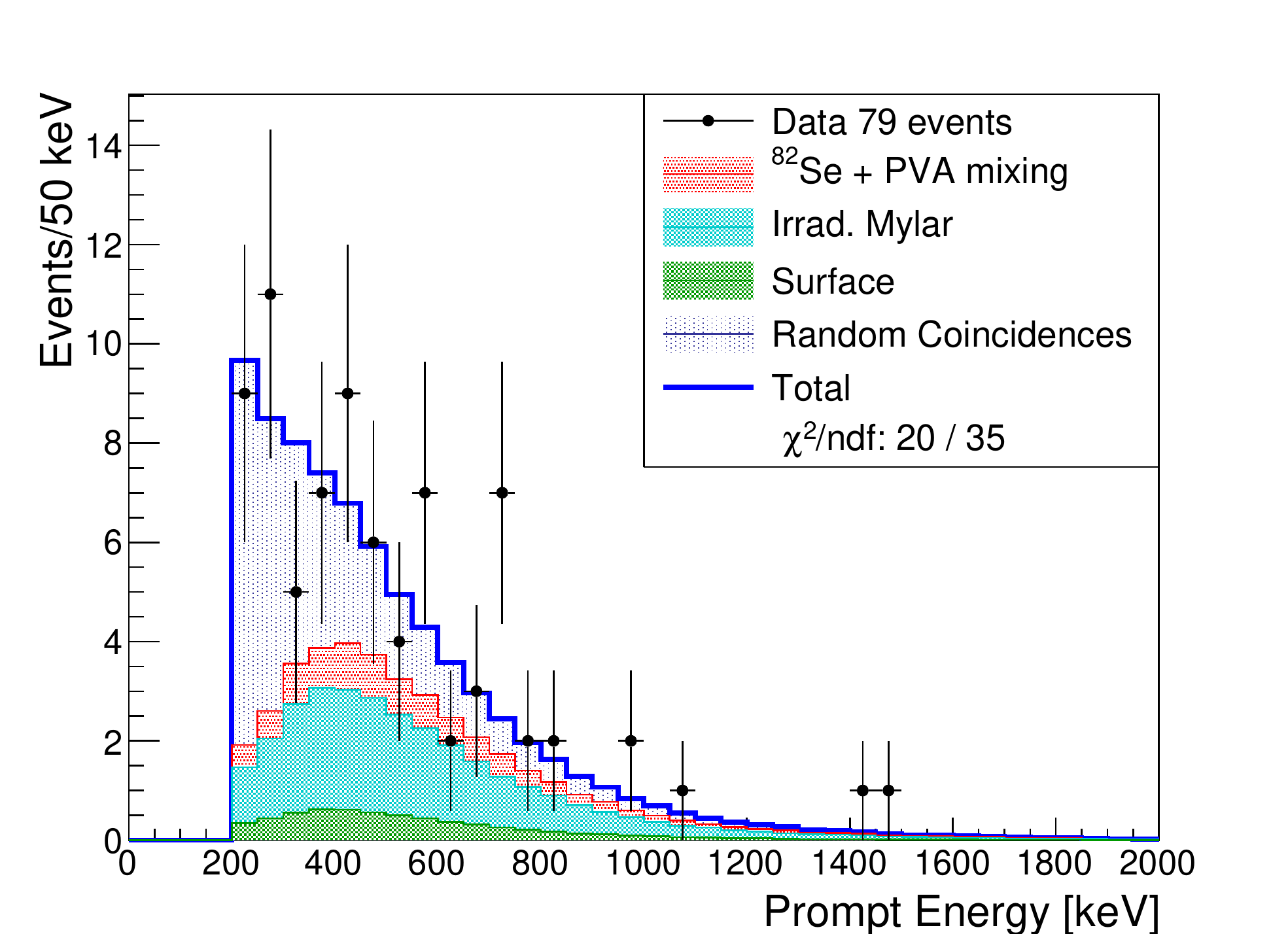}
  \includegraphics[scale=0.37]{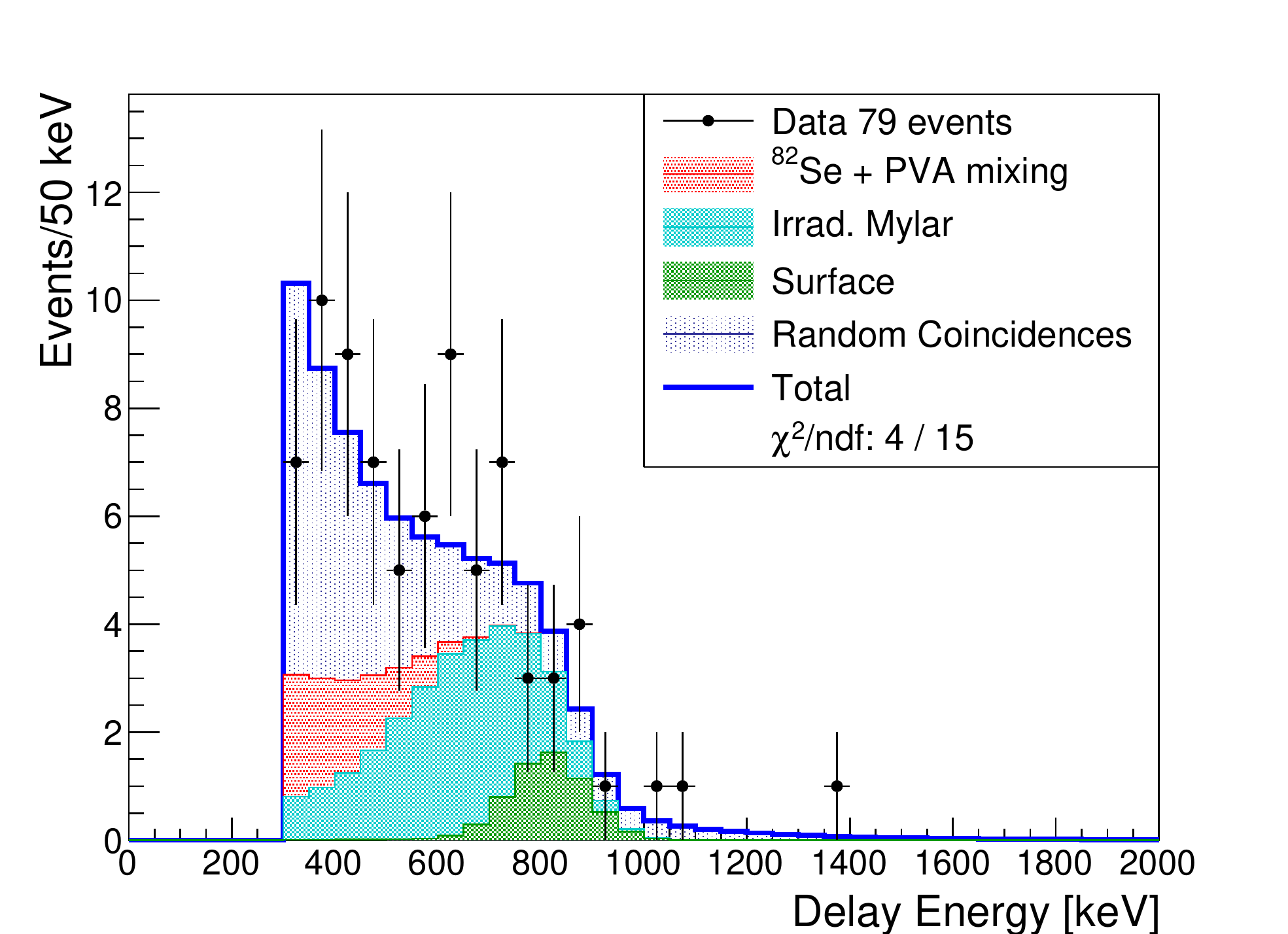}
  \includegraphics[scale=0.37]{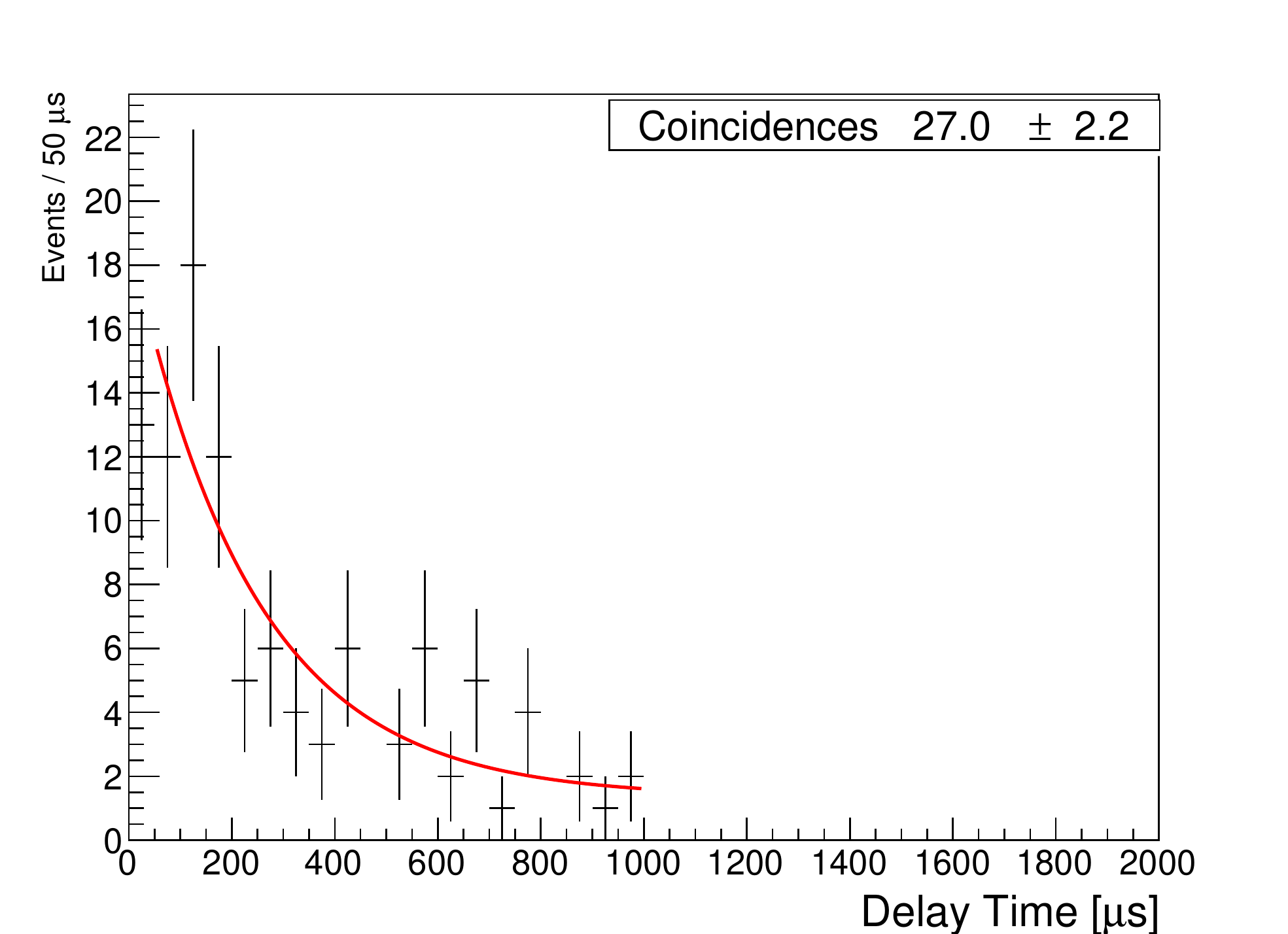}
  \caption{Distributions of the prompt and delayed energy and distribution of the delay time, for the $^{214}$BiPo measurement of the four first enriched $^{82}$Se SuperNEMO foils, with 241.1~days of data collection, an effective mass of $^{82}$Se+PVA mixture of 352~g and a surface area of measurement of 1.97~m$^2$. The delay time is presented before requiring a delay time lower than 492~$\mu$s (three times the $^{214}$Po half-life of 164~$\mathrm{\mu}$s). 
  The data are compared to the expected background from the random coincidences (dark blue) and from the fitted $^{214}$Bi contamination on the surface of the scintillators (green histogram) and in the irradiated Mylar (light blue histogram). 
  No significant excess of data events is observed at 90\% C.L. It corresponds to an upper limit of the $^{214}$Bi activity of $  \mathcal{A}(^{214}\mathrm{Bi}) < 300 ~\mathrm{\mu Bq/kg} \ 90\% \mathrm{C.L.}$ (red histogram). The delay time distribution is fitted a posteriori by an exponential decay with an half-life set to the value of the $^{214}$Po decay half-life plus a constant value for a flat random coincidence distribution decay.}
  \label{fig:se-foil-214}
\end{figure}

The energy spectra of the prompt and delayed signals are presented in Figures~\ref{fig:se-foil-212} and \ref{fig:se-foil-214} for the $^{212}$BiPo and $^{214}$BiPo measurements, respectively. The data are compared to the result of the fit. 
The numbers of fitted events from each background component and from bismuth ($^{212}$Bi and $^{214}$Bi) contamination inside the $^{82}$Se+PVA mixture are summarized in Tables~\ref{tab:se-foil-fit-212} and \ref{tab:se-foil-fit-214} for the $^{212}$BiPo and $^{214}$BiPo measurements, respectively. 
In order to reject the surface background and to reduce the background contribution from the bismuth contamination inside the irradiated Mylar, an upper limit on the delayed energy is added (700~keV for $^{212}$BiPo and 600~keV for $^{214}$BiPo), allowing to increase the signal over background ratio. 

\begin{table}[h]
\centering
\begin{tabular}{l||c|c||c|c}
 $^{212}$BiPo &  \multicolumn{2}{c}{$E_{\alpha}>150 \ \mathrm{keV}$} &  \multicolumn{2}{c}{$150 < E_{\alpha} < 700 \ \mathrm{keV}$}  \\
             & Expected          & Fitted  &  Expected         & Fitted \\
\hline
Surf. Bkg.   & $9.3 \pm 1.7$     & $9.1 \pm 1.4$     & $0.085 \pm 0.015$ & $0.08  \pm 0.01$ \\
Irrad. Mylar & $13.1 \pm 7.5$    & $8.4  \pm 3.5$     & $2.9 \pm 1.6$     & $1.9  \pm 0.8$   \\
Rand. Coinc. & 0.27              & 0.27   & 0.23              & 0.23 \\ 
$^{82}$Se+PVA &                   & $7.1  \pm 2.8$    &                   & $6.8  \pm 2.7$   \\
Total Fit    &                   & 24.9 $\pm$ 4.7   &                   & 9.0 $\pm$ 2.8 \\
\hline
Total Data   &                   & 26   &                   & 9    \\
\end{tabular}
\caption{Number of expected background events (surface background, irradiated Mylar and random coincidences), number of fitted events (background events and signal events from $^{212}$BiPo decays inside the $^{82}$Se+PVA mixture) and number of observed events, after 262~days of measurement of the first SuperNEMO $^{82}$Se source foils, with an effective mass of $^{82}$Se+PVA mixture of 359 g and an effective scintillators surface area of 2.13~m$^2$. The expected background events are calculated by using the measured activities which have been reported in this paper. 
For the fit, the number of random coincidences is fixed to the expected value while the numbers of surface background events and irradiated Mylar events are constrained by the expected numbers within the 1$\sigma$ uncertainty interval. Number of expected and fitted events are also given in the delayed energy range where the signal events from $^{212}$BiPo decays inside the $^{82}$Se+PVA mixture are expected ($150 < E_{\alpha} < 700 \ \mathrm{keV}$).}
\label{tab:se-foil-fit-212}
\end{table}

\begin{table}[h]
\centering
\begin{tabular}{l||c|c||c|c}
$^{214}$BiPo &  \multicolumn{2}{c}{$E_{\alpha}>300 \ \mathrm{keV}$} &  \multicolumn{2}{c}{$300 < E_{\alpha} < 600 \ \mathrm{keV}$}  \\
             & Expected           & Fitted  &  Expected         & Fitted \\
\hline
Surf. Bkg.   & $3.6 \pm 2.4 $     & $6.2  \pm 4.3$    & $0.05 \pm 0.03$   & $0.08 \pm 0.06$  \\
Irrad. Mylar & $<$25              & $25.0  \pm 4.9$  & $<9.7$             & $9.7 \pm 1.9$  \\
Rand. Coinc. & $29.7 \pm 5.4$     &$35.1 \pm 9.4$   & $22.0 \pm 4.7$    &$ 26.1 \pm 6.9$ \\
$^{82}$Se+PVA &                    & $9.2 \pm 6.8$    &                   & $8.9 \pm 6.6$  \\
Total Fit    &                    & 75.3 $\pm$ 13.3   &                   & 44.7 $\pm$ 9.7 \\
\hline 
Total Data   &                    & 79      &                   & 44    \\
\end{tabular}
\caption{Number of expected background events (surface background, irradiated Mylar and random coincidences), number of fitted events (background events and signal events from $^{214}$BiPo decays inside the $^{82}$Se+PVA mixture) and number of observed events, after 241.1~days of measurement of the first SuperNEMO $^{82}$Se source foils, with an effective mass of $^{82}$Se+PVA mixture of 352~g and an effective scintillators surface area of 1.97~m$^2$.
The expected background events are calculated by using the measured activities which have been reported in this paper. 
For the fit, the number of random coincidences is fixed to the expected value while the numbers of surface background events and irradiated Mylar events are constrained by the expected numbers within the 1$\sigma$ uncertainty interval. Number of expected and fitted events are also given in the delayed energy range where the signal events from $^{214}$BiPo decays inside the $^{82}$Se+PVA mixture are expected ($300 < E_{\alpha} < 600 \ \mathrm{keV}$).}
\label{tab:se-foil-fit-214}
\end{table}

For the $^{212}$BiPo measurement (Table~\ref{tab:se-foil-fit-212}), the fitted activities of the surface background and of the irradiated Mylar background are contained within the uncertainty interval (90\% C.L.) of the expected values, indicating the reliability of the fit. With a delayed energy lower than 700~keV, 9 $^{212}$BiPo events are observed and 2.2 background events are expected from the fit. The excess of observed events above the fitted background is in agreement with a $^{212}$Bi contamination inside the $^{82}$Se+PVA mixture. 
Taking into account the detection efficiency of 2.65$\pm$0.27 \% (calculated by simulating $^{212}$BiPo events distributed uniformly inside the $^{82}$Se+PVA mixture), this corresponds to a $^{208}$Tl activity of the $^{82}$Se+PVA mixture 
of:
$$ \mathcal{A}(^{208}\mathrm{Tl}) = 17.8^{+16.4}_{-11.5} \ \mathrm{\mu Bq/kg} \  \mathrm{(90\% \ C.L.)} $$
Adding the activity measured in the irradiated Mylar (Table~\ref{tab:foil-components-212}), weighted by the Mylar mass, the global $^{208}$Tl activity of the four first enriched $^{82}$Se foils of SuperNEMO is:
$$ \mathcal{A}(^{208}\mathrm{Tl}) = 24^{+20}_{-15} \ \mathrm{\mu Bq/kg} \  \mathrm{(90\% \ C.L.)} $$
This activity is larger than the SuperNEMO required activity, by a factor of about 5 to 20. 
Although this is acceptable for the first SuperNEMO module (named demonstrator), other $^{82}$Se foils are under development with different purification methods and production techniques in order to reach the required radiopurity. 

For the $^{214}$BiPo measurement (Table~\ref{tab:se-foil-fit-214}), 
with a delayed energy lower than 600~keV, 44 $^{214}$BiPo events are observed and 35.9 background events are expected from the fit. 
The excess of observed events above the fitted background corresponds to 1.35$\sigma$ excess.
Considering it as a background fluctuation, and taking into account the detection efficiency of 0.66 $\pm$ 0.07\% (calculated by simulating $^{214}$BiPo events distributed uniformly inside the $^{82}$Se+PVA mixture), an upper limit at 90\%~C.L. is set to the $^{214}$Bi contamination of the $^{82}$Se+PVA mixture:
$$ \mathcal{A}(^{214}\mathrm{Bi}) < 300 \ \mathrm{\mu Bq/kg} \  \mathrm{(90\% \ C.L.)} $$
The sensitivity to $^{214}$BiPo decays is reduced by the uncertainty on the irradiated Mylar measurement.
We note that the fitted activity of the irradiated Mylar is at its maximum allowed value, constrained by the upper limit (90\% C.L.) from the dedicated measurements (see Table~\ref{tab:foil-components-214}).
A new measurement of the irradiated Mylar with higher exposure is ongoing in order to improve the sensitivity of the $^{214}$BiPo measurement. Random coincidences are also a limitation for the $^{214}$BiPo measurement. The reduction of coincidences is foreseen by improving the shield tightness and by applying a pulse shape discrimination, as developed with the BiPo-1 prototype~\cite{bipo1}.

\subsection{Sensitivity of the BiPo-3 detector}

\begin{figure}[h]
  \centering
  \includegraphics[scale=0.37]{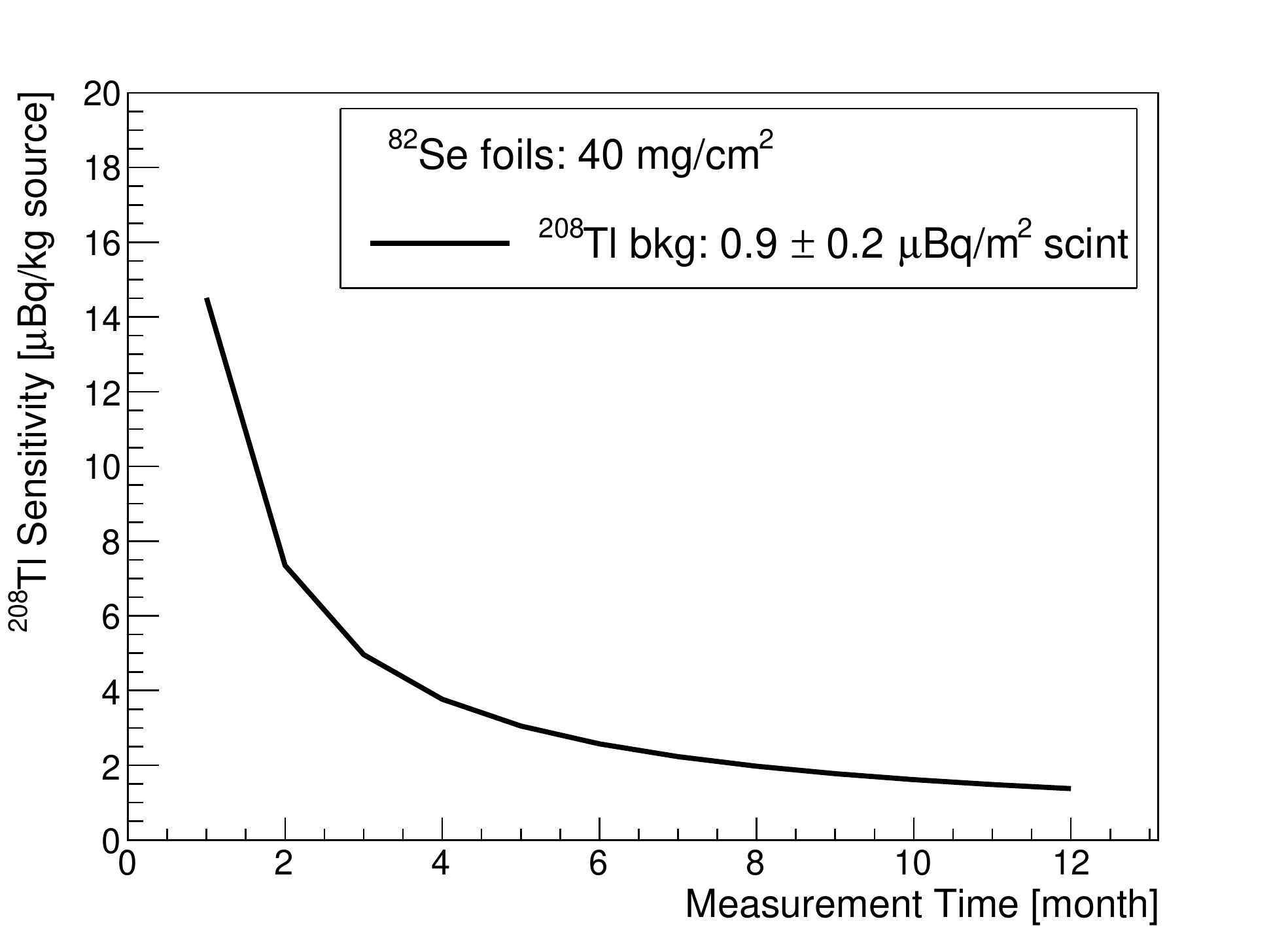}
  \includegraphics[scale=0.37]{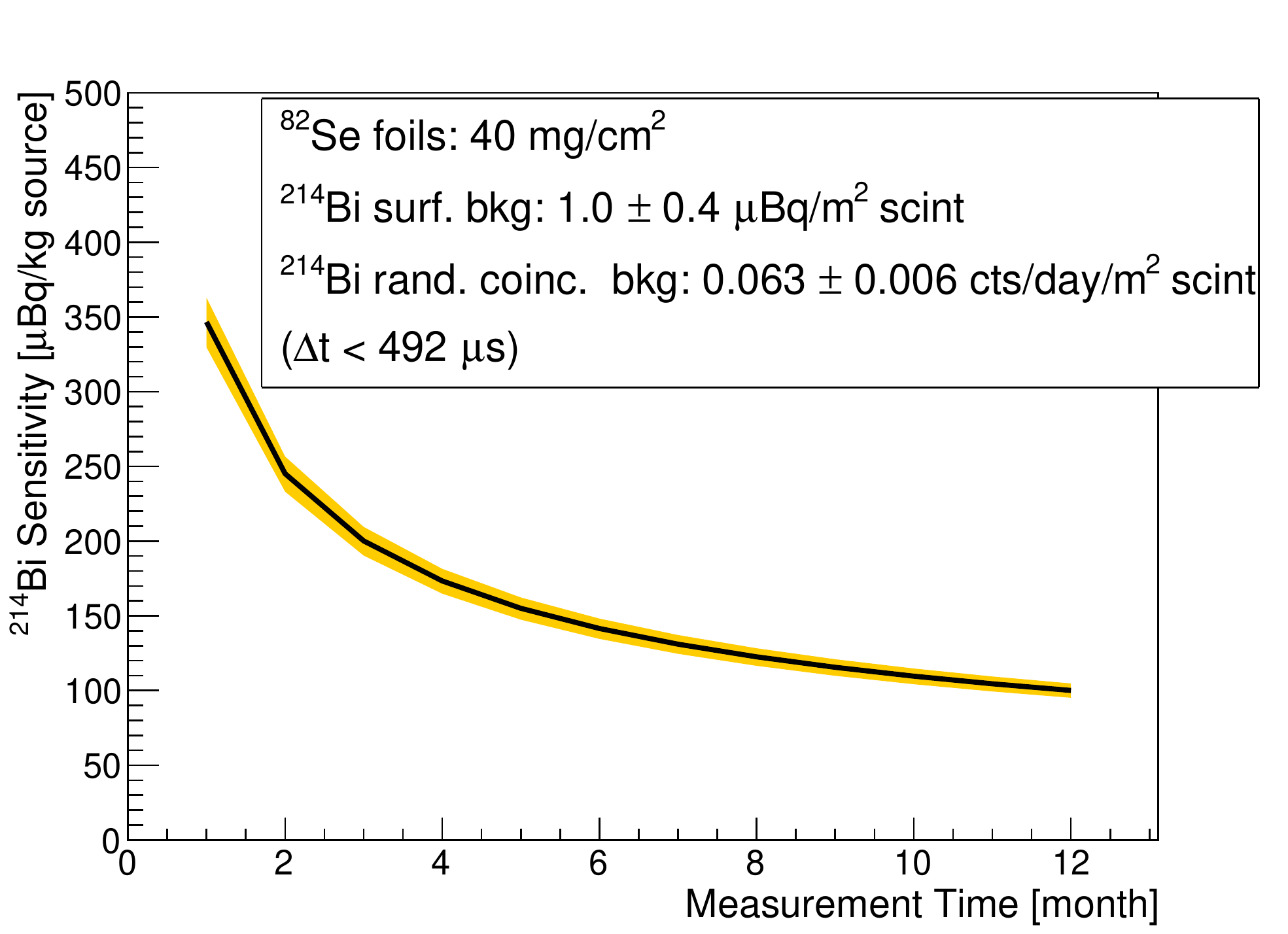}
  \caption{Sensitivity of the BiPo-3 detector for the measurement of a SuperNEMO $^{82}$Se foil, in $^{208}$Tl (left) and in $^{214}$Bi (right).}
  \label{fig:bipo3-sensitivity}
\end{figure}

The  sensitivities of the BiPo-3 detector in $^{208}$Tl and $^{214}$Bi are presented in Figure~\ref{fig:bipo3-sensitivity} for the measurement of the SuperNEMO $^{82}$Se foils. 
The sensitivities are calculated assuming that the composition of the foils is the same as those measured previously and that the total sensitive area of the BiPo-3 detector (3.6~m$^2$) is used. 
We assume that the Mylar is ideally pure and contains no bismuth ($^{212}$Bi and $^{214}$Bi) contamination. 
Efficiencies are calculated assuming only bismuth contaminations inside the $^{82}$Se+PVA mixture. 
We also take into account the surface background and the random coincidences. Their activities are set to the combined measured values reported in Tables~\ref{tab:bkg-bipo212} and \ref{tab:bkg-bipo214}.
The requirement of a delay time lower than 492~$\mu$s is added for the $^{214}$BiPo events.
Sensitivities of $\mathcal{A}$($^{208}$Tl)~$< 2$~$\mu$Bq/kg (90\%~C.L.) and $\mathcal{A}$($^{214}$Bi)~$< 140$~$\mu$Bq/kg (90\%~C.L.) will be reached in 6 months of measurement.
The sensitivity in $^{208}$Tl is in line with SuperNEMO $^{208}$Tl requirements. However the SuperNEMO required $^{214}$Bi activity is about 10 times smaller than the BiPo-3 $^{214}$Bi sensitivity which is limited by the random coincidence background. An electron$/\alpha$ discrimination based on a pulse shape analysis is under development in order to reduce the random coincidence background.

\section{Conclusion}

The BiPo-3 detector is a new generation of low-radioactivity detectors, dedicated to the measurement of ultra low $^{208}$Tl and $^{214}$Bi contaminations in thin materials with a surface area of measurement of 3.6~m$^2$.
It is fully operational since 2013 in Canfranc Underground Laboratory. 
Long periods have been dedicated to the background measurement.
The first source of background is the scintillator surface contamination. 
Surface activities of $\mathcal{A}(^{208}\mathrm{Tl})=0.9 \pm 0.2 \ \mathrm{\mu Bq/m^2}$ and $\mathcal{A}(^{214}\mathrm{Bi}) = 1.0 \pm 0.4 \ \mathrm{\mu Bq/m^2}$ have been measured. It has been verified that this background is stable in time, and unaffected by the opening of the detector. Moreover, it has been shown that this background component can be strongly suppressed by analysing the delayed $\alpha$ energy spectrum. 
The second source of background is the random coincidences between opposite scintillators. The coincidence rate is measured independently and with consistent results. Although the coincidences are negligible for the $^{208}$Tl measurement, they represent the dominant component of background for the $^{214}$Bi measurement, with a level of 0.13 counts/day/m$^2$ of scintillator surface area for the first BiPo-3 module and 0.10~counts/days/m$^2$ for the second module.
An electron$/\alpha$ discrimination based on a pulse shape analysis is under development in order to reduce even further the random coincidence background for the $^{214}$Bi measurement.
The detection efficiency has been experimentally verified by measuring a calibrated aluminium foil with BiPo-3 and HPGe detectors, which showed for the $^{208}$Tl and $^{214}$Bi activities equivalent results within uncertainties.

In the last two years, several samples have been measured with the BiPo-3 detector to select materials for the production of the enriched $^{82}$Se foils for the SuperNEMO experiment. The measurement of the first four foils (with a total $^{82}$Se mass of 0.52~kg and a total surface area of 1.4~m$^2$) shows a low $^{208}$Tl contamination with an activity $\mathcal{A}(^{208}\mathrm{Tl}) = 24^{+20}_{-15} \ \mathrm{\mu Bq/kg} \  \mathrm{(90\% \ C.L.)}$. No significant excess has been observed in $^{214}$Bi and an upper limit (at 90\%~C.L.) is set to the $^{214}$Bi contamination inside the $^{82}$Se mixture $ \mathcal{A}(^{214}\mathrm{Bi}) < 300 \ \mathrm{\mu Bq/kg} \  \mathrm{(90\% \ C.L.)} $.
The sensitivity of the BiPo-3 detector for the measurement of the SuperNEMO $^{82}$Se foils is $\mathcal{A}$($^{208}$Tl)~$<2$~$\mu$Bq/kg (90\%~C.L.) and $\mathcal{A}$($^{214}$Bi)~$<140$~$\mu$Bq/kg (90\%~C.L.) after 6 months of measurement.

At the time of this paper, 11 enriched $^{82}$Se foils (for a total $^{82}$Se mass of 1.43~kg and a total surface area of 3.85~m$^2$) have been produced and are under measurement in the BiPo-3 detector.
Other enriched $^{82}$Se foils are under development, with a different purification method and a different production technology. Part of these foils will also be measured in the BiPo-3 detector. 
More samples used by fundamental physics experiments (reflecting foils and plastic film for scintillating bolometer and readouts used in Micromegas TPC) have been measured with the BiPo-3 detector. 
Results of these measurements will be given in a future dedicated paper.
The BiPo-3 detector has become a generic detector and will be available in 2017 to measure samples for various purposes.

\acknowledgments

The authors would like to thank  all LSC staff for providing their support to run the BiPo-3 detector. We acknowledge support by the grant agencies of the Czech Republic, CNRS/IN2P3 in France, RFBR 16-52-16018 in Russia, APVV in Slovakia, STFC in the U.K. and NSF in the U.S.

\end{document}